\documentclass[onecolumn,secnumarabic,amssymb, nobibnotes, aps, prd]{revtex4-2}

\usepackage{graphicx}
\usepackage{booktabs}
\usepackage{amssymb,bm,mathrsfs,bbm,amscd}
\usepackage[tbtags]{amsmath}

\begin{document}

\title{Constraints on violation of Lorentz symmetry with clock-comparison redshift experiments}
\thanks{This manuscript has been accepted for publication in Physical Review D.111.055008.(2025)}

\author{Cheng-Gang Qin$^{1}$} \email{cgqin@hust.edu.cn}
\author{Yu-Jie Tan$^{1}$} \email{yjtan@hust.edu.cn}
\author{Xiao-Yu Lu$^{2}$} \email{xiaoyulu1993@163.com}
\author{Tong Liu$^{3}$}
\author{Yan-Rui Yang$^{1}$}
\author{Qin Li$^{1}$}
\author{Cheng-Gang Shao$^1$}\email[E-mail:]{cgshao@hust.edu.cn}

\affiliation
{$^{1}$MOE Key Laboratory of Fundamental Physical Quantities
Measurement $\&$ Hubei Key Laboratory of Gravitation and Quantum Physics, PGMF and School of Physics, Huazhong University of Science and Technology, Wuhan 430074, People's Republic of China\\
$^2$Henan International Joint Laboratory of MXene Materials Microstructure, College of Physics and Electronic Engineering, Nanyang Normal University, Nanyang 473061, China\\
$^{3}$ Key Laboratory Of Space Utilization, Technology And Engineering Center For Space Utilization, Chinese Academy Of Sciences, Beijing 100094, China}
\date{\today}

\begin{abstract}
Lorentz symmetry is a cornerstone of both the General relativity and Standard Model and its experimental verification deepens our understanding of nature. This paper focuses on the investigation of Lorentz violations with the context of clock comparison experiments in the framework of Standard Model Extension (SME). Considering matter-gravity coupling sector, we provide a generic frame to study the sensitivities of Lorentz-violating coefficients for three distinct types of clock redshift tests, including the traditional gravitational redshift test, null-redshift test I and null-redshift test II. Each of these tests is sensitivity to different combinations of Lorentz-violating coefficients. By using the current clock comparison results, we estimate the limits of SME coefficients at level of parts in $10^{4}$ down to parts in $10^{7}$. Better sensitivity may be achieved in the clock comparisons by using the state-of-the-art optical clocks. Additionally considering relativistic factors in null-redshift I, the frequency comparison result of E2 and E3 transitions of Yb$^{+}$ can set the limit $c^{e}_{00}=(7.4\pm9.3)\times10^{-9}$ in the electron sector. Our analysis demonstrates that clock-comparison redshift experiments may contribute to explore the vast parameters space on searching for the Lorentz violation.
\end{abstract}

\maketitle

\section{Introduction}

Lorentz symmetry is at the heart of general relativity (GR), which is known for its remarkable description of various gravitational phenomena. However, it is believed that GR may be the low-energy limit of a more fundamental theory that unifies gravity and quantum physics, such as string theory. Many attempts to unifies gravity and quantum physics predict a tiny violation of lorentz symmetry at the Planck scale \cite{kostelecky1989spontaneous}. Consequently, the experimental tests of Lorentz symmetry can detect the fundamental structure of GR and potentially provide new insights at the quantum gravity scale. These concepts have led to the development of a comprehensive framework based on effective field theory for testing Lorentz symmetry, which is used in many modern searches for Lorentz violations. While direct experiments at Planck energy remain unfeasible in the foreseeable future, the Planck suppression effects may be detected with experiments at currently achievable energies, which may shed on light on the nature of fundamental theories \cite{Mattingly2005Modern,Tasson2014What}.

The Standard Model Extension (SME) represents a generic effective filed theory describing the Lorentz violation in an underlying theory of quantum gravity. By encompassing all possible Lorentz-violating operators in all sectors of physics \cite{PhysRevD.58.116002,PhysRevD.51.3923,PhysRevD.69.105009}, the SME contains all possible Lorentz-violating terms in both the general relativity and the standard model. This theoretical framework provides a valuable approach for exploring the experimental signals of Lorentz violations at currently attainable energies. Various experimental predictions have been proposed with attainable sensitivities in terms of SME by specifying the SME coefficient values \cite{PhysRevD.58.116002,PhysRevD.51.3923,PhysRevD.69.105009,PhysRevD.74.045001}. The Minkowski-space constraints of the minimal SME have been scrutinized in the diverse experiments, which includes the photon sector \cite{PhysRevLett.90.060403,PhysRevLett.90.211601,PhysRevD.71.025004,PhysRevD.87.125028,PhysRevD.89.105019,PhysRevLett.91.020401,PhysRevD.97.115043,PhysRevD.100.055036}, electron sector \cite{PhysRevLett.84.1381,PhysRevLett.90.201101,PhysRevD.68.116006,PhysRevD.70.076004,PhysRevD.71.045004}, and nucleon sector \cite{PhysRevLett.86.3228,PhysRevD.64.076001,PhysRevD.72.087702}. Additionally, the Lorentz violation within post-newtonian gravity has been investigated in the lunar and satellite laser ranging \cite{PhysRevLett.99.241103,PhysRevD.103.064055,PhysRevLett.117.241301,PhysRevLett.119.201102,PhysRevD.103.064055}, short-range gravity experiments \cite{PhysRevLett.122.011102,PhysRevD.91.022006,PhysRevD.91.092003,PhysRevD.91.102007,kostelecky2017testing}, laboratory measurements with gravimeters \cite{PhysRevLett.100.031101,PhysRevLett.119.201101,PhysRevD.80.016002,PhysRevD.97.024019}, clock-comparison experiments \cite{sanner2019optical,PhysRevD.98.036003,qin2021test,PhysRevD.79.061702}, etc. A summarized account of experimental constraints on SME coefficients is referenced in \cite{RevModPhys.83.11}. To date, no significant evidence has emerged for non-zero coefficients indicating Lorentz violations in existing experiments and observations.

Amongst the diverse Lorentz symmetry tests, the clock-comparison experiments have recently garnered substantial attention. Some of the significant laboratory results of testing Lorentz symmetry are given by clock comparison experiments \cite{sanner2019optical,dreissen2022improved,PhysRevD.98.036003,qin2021test,PhysRevD.79.061702,PhysRevLett.118.142501,PhysRevD.95.075026,PhysRevLett.105.151604,PhysRevLett.93.230801,PhysRevLett.103.081602,PhysRevLett.111.050401,pruttivarasin2015michelson}. This is motivated by the latest advancements in time and frequency metrology, enabling unparalleled sensitivity to the Lorentz violation coefficients. Within the clock experiments, the Lorentz-violating effects can be predicted by the effective field theory. In the SME, anticipated signals include the variations in the frequency of clock, which are dependent on the electron and nucleon compositions of the clocks. The frequency comparisons between different kinds of clock can sever as a probe for the potential Lorentz-violating signals. Some null results of the clock-comparison experiments can be reinterpreted as constraints on Lorentz-violating coefficients. Numerous coefficients governing Lorentz violation have been bounded through the clock-comparison experiments, e.g. the frequency comparisons of Cs and Rb fountain clocks \cite{PhysRevLett.96.060801}, of trapped ultracold neutrons and Hg atoms \cite{PhysRevLett.103.081602}, of two Yb$^{+}$ clocks with different quantization axes \cite{sanner2019optical}, of entangled states of Ca$^+$ \cite{pruttivarasin2015michelson}. The bounds on protons and neutrons have been set using atomic fountain clocks \cite{PhysRevLett.96.060801,PhysRevD.95.075026}. In term of electron-photon sector for Lorentz violations, strong constraints have been set with trapped Ca$^{+}$ ions \cite{pruttivarasin2015michelson,PhysRevLett.122.123605}. The higher sensitivity can be achieved by using Yb$^{+}$ with 1.6 years radiative lifetime \cite{PhysRevLett.120.103202,dzuba2016strongly,PhysRevLett.127.213001}. Sanner $et$ $al$.\cite{sanner2019optical} reported stringent limits by using Yb$^{+}$ clocks with a fractional uncertainty at the $10^{-18}$ level. With extending the coherence time, Dreissen $et$ $al$.\cite{dreissen2022improved} further improved the previous bounds by nearly an order of magnitude reaching the level of $10^{-21}$. More clock experiments also have the potential to investigate the tests of Lorentz symmetry \cite{PhysRevLett.131.083002,roberts2020search,robinson2024direct,zhang2024frequency,king2022optical,PhysRevApplied.21.044017,li2024strontium,PhysRevApplied.19.064004}.
Amongst the constrains on the matter-gravity coupling sector, the limit on the $\alpha(\bar{a})_{0}$ term and $\bar{c}_{00}$ term is relatively poor. The laboratory experiments with systems of Cs interferometer and nuclear binding energy reported constraints at the level of $10^{-6}$ to $10^{-8}$ \cite{PhysRevLett.111.151102,PhysRevLett.106.151102}. In addition, by introducing the relativistic factor for atomic transitions, the electron-sector coefficient is limited to the level of $10^{-7}$ \cite{PhysRevD.95.015019}. With ongoing enhancements in clock technology, the clock-comparison experiments are expected to set better constraint on the matter-gravity coefficients for Lorentz violation, specially on the coefficients $\alpha(\bar{a})_{0}$ and $\bar{c}_{00}$.

In this work, we mainly study the Lorentz-violating signal arising from the matter-gravity coupling sector of the SME in the clock comparison experiments. Considering the SME's matter-gravity coupling, the transitions between Bohr levels in the atomic clocks are dependent on the atomic component of atomic clock. The matter-gravity coupling coefficients of SME can be directly constrained through clock experiments involving redshift ``tests''.  Under the classification of ``redshift tests", three types of gravitational tests namely, the traditional gravitational redshift test, null-redshift test I, and null-redshift test II, yield distinct sensitivities for SME coefficients. Some current results of clock experiments can be reinterpreted the bounds of SME coefficients.

The rest of this paper is organized as follows. In Sec. \ref{ii}, we present the theoretical foundations of this work and discuss the key information of matter-gravity coupling sector of the SME. In Sec.\ref{iii}, we discuss the three types of redshift tests and Lorentz violation, including the traditional gravitational redshift test, null-redshift test I and null-redshift test II. The conclusion is given in Sec.\ref{iv}.

\section{The framework of SME}\label{ii}

SME is a general effective field theory describing the violations of Lorentz invariance and others tenets of Einstein equivalence principle (EEP). It is constructed by the standard model, general relativity and all possible Lorentz-violating terms, including a large number of coefficients to be limited experimentally. We focus on the SME coefficients of matter-gravity coupling sector. In the SME framework, the Lorentz violations for the spin-independent fermion sector have been systematically discussed in Ref. \cite{PhysRevD.83.016013}.
In the minimal SME, the action of the point particle of mass $m^{k}$ can be written as \cite{PhysRevD.83.016013,PhysRevLett.111.151102}
\begin{equation}\label{sve1}
  S_{c}=-\int d\lambda  m^{k} c \left( \frac{\sqrt{-(\textsl{g}_{\mu\nu}+2(\bar{c}^{k})_{\mu\nu})u^{\mu}u^{\nu}}}{1+(5/3)(\bar{c}^{k})_{00}} +\frac{({a}^{k}_{\text{eff}})_{\mu}  u^{\mu}}{m^{k}}  \right),
\end{equation}
where the superscript $k=p,n,\text{or}~e$ indicates the type of particle for proton, neutron or electron, $\textsl{g}_{\mu\nu}$ is the metric tensor, $c$ is the speed of light, and $u^{\mu}=dx^{\mu}/d\lambda$ is the four-velocity of the particle $k$ with the particle path $x^{\mu}=x^{\mu}(\lambda)$ parametrized by $\lambda$. The coefficients $(\bar{c}^{k})_{\mu\nu}$ and $({a}^{k}_{\text{eff}})_{\mu}$ control the Lorentz violation for matter. The $(\bar{c}^{k})_{\mu\nu}$ tensor describes the fixed background field which modified the effective metric. The $(a^{k}_{\text{eff}})_{\mu}$ is given by $(a^{k}_{\text{eff}})_{\mu}=
\{ (1-U\alpha)(\bar{a}^{k}_{\text{eff}})_{0},(\bar{a}^{k}_{\text{eff}})_{i} \}$ ($U$ is the Newtonian potential), which represents the coupling of the particle to a field with a non-metric interaction $\alpha$ with gravity. Clearly, these coefficients are particle-species dependent and so they can lead to the violations of the EEP. In the general relativity case, both $(\bar{c}^{k})_{\mu\nu}$ and $(a^{k}_{\text{eff}})_{\mu}$ vanish.
We focus on the isotropic subset in this model \cite{PhysRevD.83.016013} and thereby upon the most poorly constrained $(\bar{c}^{k})_{00}$ and $\alpha (\bar{a}^{k}_{\text{eff}})_{0}$ coefficients. These coefficients can be measured only by gravitational experiments \cite{PhysRevD.55.6760,PhysRevD.58.116002,PhysRevLett.111.151102}. Notice that the action (\ref{sve1}) includes an unobservable scaling of particle mass by factor $1+(5/3)(\bar{c}^{k})_{00}$. In the non-relativistic Newtonian limit and redefining $m^{k}\rightarrow m^{k}[1+(5/3)(\bar{c}^{k})_{00}]$, the Hamiltonian of a single particle $m^{k}$ is given by the action (\ref{sve1}), as follows
\begin{equation}\label{sve2}
  H=\frac{1}{2}m^{k} v^{2}-m^{k}_{\text{G}}U,
\end{equation}
where $v$ is the velocity of point particle $k$, $m^{k}_{\text{G}}$ is the effective gravitational mass that is given by
\begin{equation}\label{sve3}
  {m^{k}_{\text{G}}}=\left( 1-\frac{2}{3}(\bar{c}^{k})_{00}+\frac{2}{m^{k}}\alpha(\bar{a}^{k}_{\text{eff}})_{0}\right)m^{k}  \equiv \left(  1 + \beta^{k}\right)m^{k},
\end{equation}
where the particle-dependent parameter $\beta^{k}$ is defined by $2 \alpha (\bar{a}^{k}_{\text{eff}})_{0}/m^{k}-2(\bar{c}^{k})_{00}/3$. The parameter $\beta^{k}$ indicates a particle-species dependence for the ratio between gravitational mass and inertial mass, which is responsible for violations of the WEP or EEP. For a free particle, the violations of EEP or WEP are not apparent on the nonrelativistic motion if $m^{k}(\bar{c}^{k})_{00}=3 \alpha(\bar{a}^{k}_{\text{eff}})_{0}$ \cite{PhysRevLett.111.151102}. However, the violations are maintained for the motion of antiparticle $m^{\tilde{k}}$ since the corresponding parameter $\beta^{\tilde{k}}$ is given by $-2 \alpha (\bar{a}^{\tilde{k}}_{\text{eff}})_{0}/m^{\tilde{k}}-2(\bar{c}^{\tilde{k}})_{00}/3$.

For a charge-neutral test mass $A$ made of elementary particles $k$, its ratio between gravitational mass and inertial mass can be derived in a similar way as above. From the total Hamiltonian, the violation parameter $\beta_{A}$ of test body $A$ is
\begin{equation}\label{sve30}
  \beta_{A}=
  \frac{1}{m^{A}}\sum_{k}N^{k}m^{k} \left( \frac{2 \alpha}{m^{k}}(\bar{a}^{k}_{\text{eff}})_{0}-\frac{2}{3}(\bar{c}^{k})_{00} \right),
\end{equation}
where $m_{A}$ is the total mass, and $N^{k}$ is the number of particle $m^{k}$. The parameter $\beta_{A}$ depends on the composition of body $A$, therefore a classical test of WEP can be performed by comparing accelerations of two bodies of different compositions in an external gravitational field. The classical experimental tests of the university of free fall can be reinterpreted as the upper limit on the SME coefficients.

In addition, the violations of Lorentz symmetry produce a shift $\delta H$ in the effective Hamiltonian for the proton, neutron, and electron. In the nonrelativistic limit, they can be seen as a potential-dependent rescaling of electron, proton and neutron inertial masses, leading to modified binding energy, thus to a modified clock gravitational redshift. For the atomic clock based on the transition between two states, this manifests as an anomalous gravitational redshift by a clock-dependent factor of $(1+\xi_{\text{clock}})$. The parameter $\xi_{\text{clock}}$ for a realistic clock can be calculated if the Hamiltonian that describes the clock is given. Thus modified gravitational redshift is given by
\begin{equation}\label{kekeke}
  \frac{\delta f}{f}=\left( 1+\xi_{\text{clock}}\right)\frac{\Delta U}{c^{2}},
\end{equation}
where $\Delta U$ is the difference of gravitational potential between two clocks. This equation can be used to describe the violation of the local position invariance or the universality of gravitational redshift. As the analysis in Ref.\cite{PhysRevD.83.016013}, we focus on the kinetic portion of the Hamiltonian to calculate the clock parameter $\xi_{\text{clock}}$ for the purpose of simplification since the accurate calculation of the clock-dependent parameter $\xi_{\text{clock}}$ is not the purpose of this work. 
For determining the parameter $\xi_{\text{clock}}$, the zeroth-order approximation of kinetic portion of the Hamiltonian can be obtained by simple replacements for the proton, neutron and electron masses
\begin{eqnarray}
  \frac{1}{m^{k}}  &\rightarrow& \frac{1}{m^{k}} \left( 1-3U+\frac{5}{3}(\bar{c}^{k})_{00}+\frac{13}{3}(\bar{c}^{k})_{00}U \right).
\end{eqnarray}
Considering the hyperfine transition with the Bohr levels of H clock, the hyperfine splitting scales with the electron mass $m^e$ and proton mass $m^p$ as $(m^e m^p)^2/(m^{e}+m^p)^3$. As the previous treatment in Ref. \cite{PhysRevD.83.016013,PhysRevLett.111.151102}, the parameter $\xi_{\text{H,Bohr}}$ can be calculated by implemented the above replacements in the standard result of Bohr energy levels
\begin{equation}\label{ham4}
  \xi_{\text{H,Bohr}}=-\frac{2}{3}\frac{m^{p}(2(\bar{c}^{e})_{00}-(\bar{c}^{p})_{00})+m^{e}((2\bar{c}^{p})_{00}-(\bar{c}^{e})_{00})}{m^{p}+m^{e}}.
\end{equation}

The classical experiments of gravitational redshift have been performed with H clock in the GPA experiment and Galileo satellites. The experimental results can be reinterpreted as the bounds on the SME coefficients.
For other types of atomic clock $O$, we use a rough hydrogenic model to estimate $\xi_{O}$ with replacements in Eq.(\ref{ham4}) from $(\bar{c}^{p})_{00}$ to $(\bar{c}^{O})_{00}$ and the proton mass $m^{p}$ to $O$ atom mass $m^{O}$
\begin{equation}\label{ham5}
  \xi_{O}=-\frac{2}{3}\frac{m^{O}(2(\bar{c}^{e})_{00}-(\bar{c}^{O})_{00})+m^{e}((2\bar{c}^{O})_{00}-(\bar{c}^{e})_{00})}{m^{O}},
\end{equation}
where
\begin{equation}\label{ham6}
  (\bar{c}^{O})_{00}=\frac{1}{m^{O}}\sum_{k}N^{k}m^{k}(\bar{c}^{k})_{00}.
\end{equation}

\section{Lorentz violation and clock-comparison experiments}\label{iii}

We discuss the Lorentz-violating contributions to the gravitational redshift in the clock-comparison experiments. It is noted that the time delay and Doppler shift are the subdominated effects since the experimental time scale is slow. In dedicated redshift experiments, the Lorentz-violating gravitational redshift can appear as the dominant effects. We discuss the Lorentz-violating modifications to the ``gravitational redshift tests" and ``null-redshift tests".

The clock-comparison experiments can be split into three distinct types of gravitational tests. The first type is the traditional gravitational redshift, which involves two identical clocks held at different gravitational potentials and compares their frequencies by using the electromagnetic-wave signals. The other two types are called ``null-redshift test", which involves two clocks placed in a freely falling ``elevator".

\subsection{The traditional gravitational redshift test}

In this subsection, we discuss the traditional gravitational redshift tests and clock-comparison experiments. In the presence of Lorentz violation with matter-gravity coupling, the clock frequency is dependent on its composition or structure. Considering traditional gravitational redshift measurement with two identical clock, we can express the gravitational redshift as the form
\begin{equation}\label{gr1}
  \left(\frac{\delta  f}{f}\right)_{\text{gr}}=\left(1+\xi_{\text{clock}}\right) \frac{\Delta U}{c^{2}},
\end{equation}
where $\xi_{\text{clock}}$ represents the function of the Lorentz-violating coefficients associated with the clocks used in the experiment, and $\Delta U$ is the difference of gravitational potential between two clocks. In general, the coefficient $(1+\xi_{\text{clock}})$ is an unobservable scaling in a gravitational redshift test since it can be scaled by a redefined gravitational mass. For measuring coefficient $\xi_{\text{clock}}$, it is necessary to introduce the additional measurements.

For the gravitational redshift test experiment in the laboratory, it is essentially to compare the frequency difference of clocks $(\delta f/f)_{\text{gr}}$ with the measured gravitational potential difference $\Delta U'$. The frequency difference is given by the frequency comparison between two identical clocks. The measured gravitational potential difference can be expressed as $\Delta U'=\emph{\textbf{g}}_{m}\cdot\emph{\textbf{h}}$ (this is only valid for small separation of the clocks) that is determined by the gravimeter measurement, in which $\emph{\textbf{g}}_{m}$ is the average value of gravitational acceleration measured by the gravimeter, and $\emph{\textbf{h}}$ is the separation vector of the two clocks. For a particular gravimeter, the measured gravitational acceleration is given by
\begin{equation}\label{gram}
  \emph{\textbf{g}}_{m}=(1+\beta_{\text{gravimeter}})\emph{\textbf{g}},
\end{equation}
where $\beta_{\text{gravimeter}}$ is the parameter specific to the gravimeter, and $\emph{\textbf{g}}$ is the gravitational acceleration. Therefore, we can obtain an observable scaling $1+\xi_{\text{clock}}-\beta_{\text{gravimeter}}$ in the gravitational redshift test of the laboratory experiment with
\begin{equation}\label{gr2}
  |\xi_{\text{clock}}-\beta_{\text{gravimeter}}|\leq |\alpha|,
\end{equation}
where $\alpha$ is the violation parameter in the traditional test of gravitational redshift.

The results of gravitational redshift laboratory test can be used to set limit on $ \xi_{\text{clock}}-\beta_{\text{gravimeter}}$, which provides sensitivity to combinations of coefficients for Lorentz violation. The RIKEN researchers reported the test of gravitational redshift at $(1.4\pm9.1)\times10^{-5}$ with a pair of Sr optical clocks \cite{takamoto2020test}. The measurement of gravitational acceleration was provided with the relative gravimeters and FG5 absolute gravimeter. For the sake of simplicity, the test mass is assumed as the silicon dioxide. The RIKEN test obtained the sensitivity $|\xi_{\text{Sr}}-\beta_{\text{SiO}_{2}}|\leq 9.1\times10^{-5}$,
where parameter $\xi_{\text{Sr}}$ is given by Eq.(\ref{ham5}) with replacement $O$ to $\text{Sr}$, and parameter $\beta_{\text{SiO}_{2}}$ is expressed by Eq.(\ref{sve30}) with the silicon-dioxide point mass.

\begin{table}
\caption{\label{tab1} The estimated sensitivity of SME coefficients with the traditional gravitational redshift test. For the constraints on SME coefficients, the index $T$ replacing 0 indicates these limits hold in the Sun-centered celestial equatorial frame. The differences between the parameters $\bar{c}^{k}_{TT}$, $\alpha(\bar{a}^{k}_{\text{eff}})_{T}$ and parameters $\bar{c}^{k}_{00}$, $\alpha(\bar{a}^{k}_{\text{eff}})_{0}$ are small and neglected. The results $x(y)$ represents $x\pm y$. Values in brackets represent the estimate of sensitivities attainable in the future tests.}
\begin{ruledtabular}
\begin{tabular}{ccccccc}
{SME coefficients} &$\xi_{\text{clock}}-\beta_{\text{T}}$  &$\alpha(\bar{a}^{e+p}_{\text{eff}})_{T}$ (GeV)  &$\alpha(\bar{a}^{n}_{\text{eff}})_{T}$ (GeV)    & $\bar{c}^{p}_{TT}$  & $\bar{c}^{n}_{TT}$  & $\bar{c}^{e}_{TT}$ \\
\hline
{RIKEN \cite{takamoto2020test}} & $1.4(9.1)\times10^{-5}$  &$-1.3(8.5)\times10^{-5}$ &$-1.3(8.5)\times10^{-5}$  &$2.3(14.8)\times10^{-5}$ & $2.0(12.7)\times10^{-5}$ &$-1.1(6.8)\times10^{-5}$\\
{Galileo \cite{PhysRevLett.121.231101,*herrmann2018test}} & $0.19(2.48)\times10^{-5}$  &$-0.18(2.32)\times10^{-5} $  &$-0.18(2.32)\times10^{-5}$  &$0.19(2.48)\times10^{-5}$  &$0.57(7.44)\times10^{-5}$  &$-0.14(1.86)\times10^{-5}$\\
{ACES \cite{savalle2019gravitational}} & $[2-3\times10^{-6}]$   &$[10^{-6}]$   &$[10^{-6}]$   &$[10^{-6}]$ &$[10^{-6}]$  &$[10^{-6}]$\\
{CLEP \cite{qin2024preliminary}} & $[\sim4\times10^{-6}]$   &$[10^{-6}]$   &$[10^{-6}]$  &$[10^{-6}]$ &$[10^{-6}]$  &$[10^{-6}]$\\
{CSS \cite{PhysRevD.108.064031}} & $[\sim 5\times10^{-6}]$  & $[10^{-6}]$   & $[10^{-6}]$  &$[10^{-6}]$ &$[10^{-6}]$  &$[10^{-6}]$\\
{FOCOS \cite{derevianko2022fundamental}} & $[\sim 10^{-9}]$   & $[10^{-9}]$  &$[10^{-9}]$  &$[10^{-9}]$ &$[10^{-9}]$  &$[10^{-9}]$
\end{tabular}
\end{ruledtabular}
\end{table}

Considering the traditional gravitational redshift test with eccentric satellites, this type of clock experiment compared a clock on the satellite to an identical clock on the ground. A first influence of the Lorentz violation comes from the anomalous redshift by a clock-dependent factor $1+\xi_{\text{clock}}$. Another influence is the change in the motion of the test mass or the satellite. The Earth gravitational potential is based on the motion of the test mass. The effective gravitational potential is mapped by test mass as $U'=(1+\beta_{\text{T}})U$, where $\beta_{\text{T}}$ is Lorentz-violating parameter of test mass. In the satellite-type experiments, the gravitational redshift test is to compare the clock's gravitational redshift with the effective gravitational potential measured by test mass. The anomalies in the motion of satellite need not to be considered since these are removed by continuous monitoring in the satellite's trajectory. The gravitational redshift test with eccentric satellites is given by
\begin{equation}\label{grtt1}
  \left(\frac{\delta  f}{f}\right)_{\text{gr}}=(1+\alpha)\frac{ {\Delta U'}}{c^2}=(1+\xi_{\text{clock}}-\beta_{\text{T}}) \frac{{\Delta U'}}{c^2}.
\end{equation}
The Gravity probe A (GPA) experiment tested the gravitational redshift with H clocks reaching an uncertainty of $1.4\times10^{-4}$, and Galileo satellites improved the limits to $(0.19\pm2.48)\times10^{-5}$ \cite{PhysRevLett.121.231101,*herrmann2018test}, which yields the measurement $|\xi_{\text{H}}-\beta_{\text{SiO}_{2}}|\leq 2.48\times10^{-5}$,
where for simplicity, we assume that the test mass mapping potentials is the silicon dioxide. All these results can be used to set constraints on the SME coefficients, and as shown in the TABLE.\ref{tab1} the estimated constraints are expressed in the Sun-centered inertial reference frame where $\bar{c}^{k}_{TT}=\bar{c}^{k}_{00}$, $\alpha(\bar{a}^{k}_{\text{eff}})_{T}=\alpha(\bar{a}^{k}_{\text{eff}})_{0}$, and $\alpha(\bar{a}^{e+p}_{\text{eff}})_{T}=\alpha(\bar{a}^{e}_{\text{eff}})_{T}+\alpha(\bar{a}^{p}_{\text{eff}})_{T}$.
The future Atomic Clock Ensemble in Space (ACES) planed to used cold atom Cesium clock to perform a gravitational redshift test at uncertainty of $(2-3)\times 10^{-6}$ \cite{savalle2019gravitational}. The China's Lunar Exploration Project (CLEP) has a potential to test gravitational redshift with the uncertainty $4\times 10^{-6}$ by the onboard H clock \cite{qin2024preliminary}. The China Space Station (CSS) was equipped with a Sr optical clock that may test gravitational redshift at the accuracy of $5\times10^{-7}$ \cite{PhysRevD.108.064031}. The fundamental physics with a state-of-the-art optical clock in space (FOCOS) would placed optical atomic clock in an eccentric orbit to test gravitational redshift with the accuracy of $10^{-9}$ \cite{derevianko2022fundamental}. These missions provides the clock-experiment attainable sensitivities to $\xi_{\text{clock}}-\beta_{\text{T}}$ in the next few years. We summarize the limits on the SME coefficients from the traditional gravitational redshift with maximum reach analysis in Table.\ref{tab1} (For the constraints on SME coefficients, the index $T$ replacing 0 indicates these limits hold in the Sun-centered celestial equatorial frame. The differences between the parameters $\bar{c}^{k}_{TT}$, $\alpha(\bar{a}^{k}_{\text{eff}})_{T}$ and parameters $\bar{c}^{k}_{00}$,  $\alpha(\bar{a}^{k}_{\text{eff}})_{0}$ are small and neglected.). In the maximum reach analysis, each SME coefficient is estimated separately assuming that all coefficients of other sectors remain equal to zero. Thus, we can limit one nonvanishing coefficient at one time. Values without brackets represent our estimate of sensitivities from currently published experimental results. Values in brackets represent our estimate of sensitivities attainable in the future tests.

\subsection{Null-redshift test I}

\begin{figure}
\includegraphics[width=0.35\textwidth]{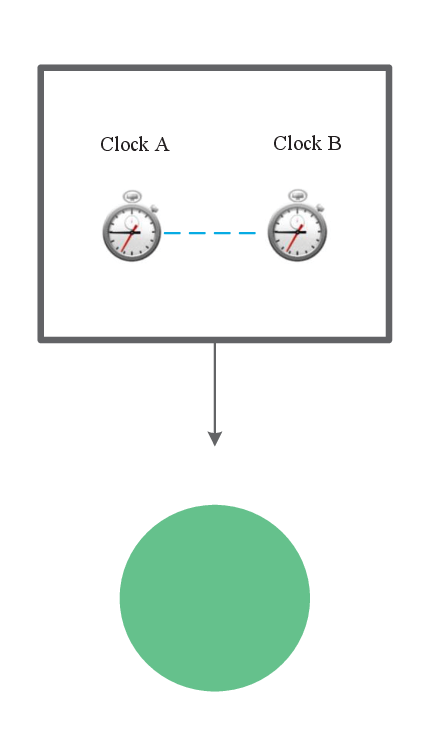}
\caption{\label{fig1} A schematic picture of the null-redshift test I. Two different clocks $A$ and $B$ are placed on same position in the Einstein elevator and experience the same changes in the gravitational potential when the elevator moves. If Lorentz Symmetry is violated, their frequencies ratio dependent on position of elevator.}
\end{figure}

Another type is called the null-redshift test where the gravitational redshift between two clocks should be zero in the general relativity. Considering the configuration of the clocks, the null-redshift test can be split into two types as null-redshift test I and null-redshift test II. We discuss the null-reshift test I in this subsetion.
As shown in Ref.\cite{PhysRevLett.109.080801,PhysRevA.87.010102,ashby2018null,PhysRevResearch.2.033242}, the null-redshift test I involves monitoring frequencies of two different clocks held at same position when they move through the gravitational potential. We can use the Einstein elevator to illustrate this test. In a free-falling Einstein elevator, two different types of atomic clocks are fixed at the same position in the elevator, as shown in FIG.\ref{fig1}. Noted that in the elevator, two clocks are different types of clock, and two clocks are located in the same location. If Lorentz symmetry is valid, the frequency difference between the two clocks is zero. In the case of Lorentz violation, two clocks have different couplings to the external gravitational potential, and thus the frequency difference between the two clocks changes as the elevator moves in the gravitational field.
Considering frequency comparison of clocks $A$ and $B$, we can monitor the change in frequency differences to perform null-redshift test I
\begin{equation}\label{nr1}
   \left(\frac{\delta  f}{f}\right)_{\text{nrI}}=\left(\frac{\delta  f}{f}\right)_{A}-\left(\frac{\delta  f}{f}\right)_{B}=\left(\xi_{A}  - \xi_{B}\right) \frac{\Delta U}{c^{2}},
\end{equation}
where $\Delta U$ represents the changes in gravitational potential, $\xi_{A}$ and $\xi_{B}$ are the parameters measuring the Lorentz violation of clocks $A$ and $B$, respectively. This frequency shift of Lorentz violation varies with spacetime position, and it exhibits characteristics similar to the violation of Local Position Invariance. The results of testing LPI can be reinterpreted as the measurements for null-redshift test I. Based on Earth-based LPI tests comparing a Rb clock and a Cs clock \cite{PhysRevLett.109.080801}, a Rb clock with a H clock \cite{PhysRevA.87.010102}, a H clock with a Cs clock \cite{ashby2018null}, a Sr clock with a Cs clock \cite{PhysRevResearch.2.033242}, a Hg$^{+}$ clock and a Cs clock\cite{PhysRevLett.98.070801}, a Yb clock with a Sr clock\cite{goti2023absolute,mcgrew2019towards,PhysRevLett.113.210802}, and a Hg$^{+}$ clock with a Al$^{+}$ clock \cite{rosenband2008frequency}, the sensitivities to SME coefficients can be obtained. These experiments involve different types of clocks, providing different sensitivity to SME coefficients (as shown in TABLE \ref{tab2}). Based on the clock-dependence parameter Eq.(\ref{ham4}) and the bounds of null-redshift test I Eq.(\ref{nr1}), Table.\ref{tab2} presents the limits on the SME coefficients where values without brackets are the estimate of sensitivities from published experimental results and values in brackets represent the estimate of attainable sensitivities with clocks with the accuracy of $10^{-18}$.
The bounds on the $\bar{c}^{p}_{TT}$  and $\bar{c}^{n}_{TT}$ can reach a level comparable to the experiments with systems of Cs interferometer and nuclear binding energy \cite{PhysRevLett.111.151102,PhysRevLett.106.151102}.
By using the results of other clock comparison experiments, it is possible to obtain limits on other SME coefficients.

From Eq.(\ref{nr1}), the null-redshift test I is sensitive to the changes of external gravitational potential. With the highly eccentric orbits, the onboard clock of FOCOS undergoes a large change in the Earth's gravitational potential, which corresponds to gravitational redshift of several $10^{-10}$ \cite{derevianko2022fundamental}. Thus by carrying multiple optical clocks, FOCOS has the potential to test the null-redshift test I to the level of $10^{-9}$.

\begin{table}
\caption{\label{tab2} The estimated sensitivity of SME coefficients with the null-redshift test I. For the constraints on SME coefficients, the index $T$ replacing 0 indicates these limits hold in the Sun-centered celestial equatorial frame. Values in brackets represent the estimate of sensitivities attainable in the future tests.}
\begin{ruledtabular}
\begin{tabular}{cccccccc}
{SME coefficients} &$\xi_{\text{clock,A}}-\xi_{\text{clock,B}}$  &Clocks A,B & $\bar{c}^{p}_{TT}$  & $\bar{c}^{n}_{TT}$  & $\bar{c}^{e}_{TT}$ \\
\hline
{Gu$\acute{e}$na et al.\cite{PhysRevLett.109.080801}} & $0.11(1.04)\times10^{-6}$  &Rb, Cs  &$1.4(13)\times10^{-5}$ & $1.4(13)\times10^{-5}$ &$-0.2(1.8)\times10^{-1}$ \\
{Goti et al. \cite{goti2023absolute}} & $-0.0(1.2)\times10^{-6}$ &Yb, Cs  &$0.0(4.3)\times10^{-4}$  &$0.0(4.3)\times10^{-4}$ &$0.0(5.7)\times10^{-1}$\\
{Fortier et al. \cite{PhysRevLett.98.070801}} & $2.0(3.5)\times10^{-6}$ &Hg$^{+}$, Cs   &$-2.6(4.6)\times10^{-4}$  &$2.6(4.6)\times10^{-4}$ &$-3.9(6.9)\times10^{-1}$\\
{Schwarz et al. \cite{PhysRevResearch.2.033242}} & $-1.1(5.2)\times10^{-7}$  &Sr, Cs  &$-0.7(3.4)\times10^{-5}$  &$0.7(3.4)\times10^{-5}$ &$-1.1(5.3)\times10^{-2}$\\
{Rosenband et al. \cite{rosenband2008frequency}}  & $1.6(3.0)\times10^{-7}$ &Al$^+$, Hg$^+$  &$3.0(5.7)\times10^{-6}$  &$-3.0(5.7)\times10^{-6}$ &$4.0(7.4)\times10^{-3}$\\
{Peil et al. \cite{PhysRevA.87.010102}} & $-2.7(4.9)\times10^{-7}$ &Rb, H  &$7.1(12.8)\times10^{-7}$  &$-7.0(12.8)\times10^{-7}$  &$4.8(8.6)\times10^{-4}$\\
{Ashby et al. \cite{ashby2018null}} & $2.24(2.48)\times10^{-7}$ &H, Cs  &$5.73(6.36)\times10^{-7}$  &$-5.73(6.34)\times10^{-7}$ &$3.91(4.33)\times10^{-4}$\\
{Current clock} &$[\sim 10^{-8}]$ &$-$ &$[10^{-8}]$ &$[10^{-8}]$  &$[10^{-5}]$
\end{tabular}
\end{ruledtabular}
\end{table}

The above constraints are based on the atomic transition model with nonrelativistic limits, which considers clock comparison experiments with different atomic species. This model cannot be used to limit SME coefficients with clock comparison configurations where the two clocks are based on the same atomic species but different quantum transitions. However, such clock comparisons can also be utilized for analyzing the null-redshift test II. In this clock comparison configurations with different transitions, the parameter $\xi_{\text{clock}}$ in Eq.(\ref{nr1}) should performed a relativistic calculation introducing relativistic factors $R$ in the electron sector \cite{PhysRevD.95.015019}. The Lorentz violation of the frequency ratio is dominated by the difference in the values of relativistic factors $R$. The relativistic factor $R$ leads to a modification factor $(2/3)Rc^{e}_{00}$ in gravitational redshift \cite{PhysRevD.95.015019}. Then, the null-redshift test I becomes
\begin{equation}\label{nr2}
   \left(\frac{\delta  f}{f}\right)_{\text{nrI}}=\Delta \xi   \frac{\Delta U}{c^{2}}=\Delta R \frac{2c^{e}_{00}}{3}   \frac{\Delta U}{c^{2}},
\end{equation}
where $\Delta \xi$ and $\Delta R$ describe Lorentz violation between two transitions, $R$ can be calculated by using Eq.(5) in Ref.\cite{PhysRevD.95.015019}. For the atomic clock $O$, the long-term comparison of different transition frequencies or tests of local position invariance can be used to set constraint on SME coefficient $c^{e}_{00}$. For example, the long-term frequency comparisons of the electric-quadrupole (E2) and electron-octupole (E3) transitions of Yb$^{+}$ \cite{PhysRevLett.126.011102,PhysRevLett.130.253001,PhysRevLett.113.210801} have high sensitive to the coefficient $c^{e}_{00}$. Generally, the experimental data of frequency ratio are fitted by the Sun's gravitational potential on Earth with the nonlinear least square fitting of $C\cos[2\pi(t-t_{0})/T_{0}]+D$, where $t_{0}$ is the perihelion time of the year in which the experiment began, $T_{0}$ is the duration of the anomalistic year, and $C$ and $D$ are free fitting parameters. From the fitting function and Eq(\ref{nr2}), we obtain the constraint on the SME coefficient as $c^{e}_{00}=3C/2\Delta R({\Delta U}/{c^{2}})$. Recently, Lange $et$ $al$.\cite{PhysRevLett.126.011102} and Filzinger $et$ $al$.\cite{PhysRevLett.130.253001} have reported the high-precision values of free parameter $C$ by using precise comparisons of the frequencies of the E2 and E3 transitions of Yb$^{+}$: $C=(-16\pm13)\times10^{-18}$ \cite{PhysRevLett.126.011102} and $C=(2.6\pm3.2)\times10^{-18}$ \cite{PhysRevLett.130.253001}. Using $C$ values and $\Delta R$=3.38 \cite{PhysRevD.95.015019} for E2 and E3 transitions of Yb$^{+}$, we get the limits
\begin{equation}\label{nr3}
   c^{e}_{00}= (-4.3\pm3.5)\times10^{-8}, \quad c^{e}_{00}=(7.4\pm9.3)\times10^{-9}
\end{equation}
correspondingly. The results improve the uncertainty of the previous limits \cite{PhysRevD.95.015019} by about a factor of $60$, and this improvement is primarily due to the high-precision frequency comparison results and the exceptionally high sensitivity $\Delta R$. To set constraints on the SME coefficients of other sector, a more comprehensive model of atomic transition may need to incorporate relativistic calculations for electron, proton, and neutron sectors.

\subsection{Null-redshift test II}
\begin{figure}
\includegraphics[width=0.35\textwidth]{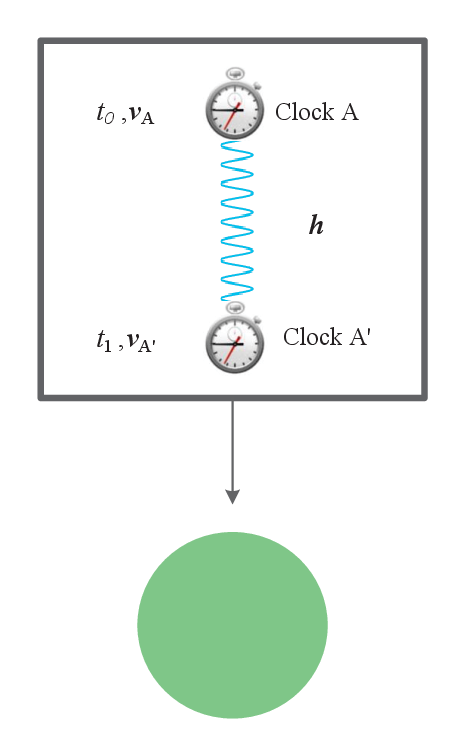}
\caption{\label{fig2} A schematic picture of the null-redshift test II. The Einstein elevator is freely falling in an external gravitational field. In elevator, two identical clocks $A$ and $A'$ are placed on different positions with separation $\textbf{\emph{h}}$ and experience the different external gravitational potentials. The dependence of their frequency difference on the difference in the external gravitational potentials reveals a sensitivity to Lorentz violation. }
\end{figure}

In this subsection, we discuss the null-redshift test II with clock experiments. As shown in Fig.\ref{fig2}, the null-redshift test II involves monitoring the frequency difference of two identical clocks at different positions when they are in free fall together in the external gravitational field. Different from the null-redshift test I where two clocks are of different types and both are located at the same position (or gravitational potential), the clock configuration of the null-redshift test II is that the two clocks are identical and are situated at different locations (or gravitational potentials). Note that null-redshift test I: same location and different types of clocks; null-redshift test II: different locations and same type of clocks. Following section is dedicated to the analysis of null-redshift test II.

The null-redshift test II can be demonstrated by the Einstein elevator experiment. As shown with the Einstein elevator in Fig.\ref{fig2}, an elevator is freely falling in an external gravitational field, two identical clocks are fixed at the different positions in the elevator, and their frequencies are compared through a light signal. In this elevator, two identical clocks are fixed at different heights and their frequencies are compared through light signals. Note that unlike null-redshift test I, the two clocks are same type of clock for null-redshift test II, and they are located at different locations with different external gravitational potentials. In the case of Lorentz symmetry or equivalence principle being valid, the frequencies of the two clocks are the same and their frequency difference is zero. However, in the case of Lorentz violation, the frequency comparison of two clocks may measure an additional frequency shift, which changes with the motion of the elevator. To study the Lorentz-violating effect in test II, we consider a scenario that in the external gravitational field $\textbf{\emph{g}}$ the Einstein elevator is freely falling with the acceleration $\textbf{\emph{a}}$ and the separation of two clocks is $\textbf{\emph{h}}$. At the time $t_0$, clock $A$ emits a light signal with frequency $f_{A}$ to clock $A'$. After traveling a distance $h$, this light signal is received by clock $A'$ with the frequency of $f_{A'}$ at the time $t_1$. For this clock comparison process, the frequency ratio of clocks $A$ and $A'$ can be expressed as
\begin{eqnarray}\label{nrii1}
  \frac{f_{A'}}{f_{A}}&=& 1-\frac{1}{c}\textbf{\emph{N}}\cdot\left(\textbf{\emph{v}}_{A'}-\textbf{\emph{v}}_{A}\right)+\frac{U_{A'}-U_{A}}{c^{2}}\nonumber\\
  &+&\frac{1}{c^{2}}\left(\frac{\textbf{\emph{v}}_{A'}^{2}}{2}-\frac{\textbf{\emph{v}}_{A}^2}{2}
  -\left[\textbf{\emph{N}}\cdot\left(\textbf{\emph{v}}_{A'}-\textbf{\emph{v}}_{A}\right)\right]\cdot(\textbf{\emph{N}}\cdot\textbf{\emph{v}}_{A})\right),
\end{eqnarray}
where $\textbf{\emph{v}}_{A}$ is the velocity of clock $A$ at time $t_0$, $\textbf{\emph{v}}_{A'}$ is the velocity of clock $A'$ at time $t_1$, $\textbf{\emph{N}}$ is the unit vector from $A$ to $A'$, the second term represents the doppler effect, and the third term is the gravitational redshift.

Between clocks $A$ and $A'$, the travel time of light signal can be written as $\Delta t=t_1-t_0=h/c$, thus the difference between velocities becomes $\textbf{\emph{v}}_{A'}-\textbf{\emph{v}}_{A}=\textbf{\emph{a}}\Delta t=(h\textbf{\emph{a}})/c$, where $\textbf{\emph{a}}$ is the acceleration of the clocks.
Then, the doppler effect can be expressed as
\begin{equation}\label{nrii04}
  -\frac{1}{c}\textbf{\emph{N}}\cdot\left(\textbf{\emph{v}}_{A'}-\textbf{\emph{v}}_{A}\right)=\frac{\textbf{\emph{a}}\cdot\textbf{\emph{h}}}{c^2}.
\end{equation}
The gravitational redshift can be written in the form of the gravitational acceleration
\begin{equation}\label{nrii05}
  \frac{U_{A'}-U_{A}}{c^{2}}=-\frac{\textbf{\emph{g}}\cdot\textbf{\emph{h}}}{c^2},
\end{equation}
where the separation is small so that the inhomogeneity of the gravitational field can be ignored. Considering the velocity difference between $\textbf{\emph{v}}_{A}$ and $\textbf{\emph{v}}_{A'}$, the fourth term is the order of $c^{-3}$, which does not contribute to order of $c^{-2}$. The frequency ratio Eq.(\ref{nrii1}) of clocks $A$ and $A'$ becomes
\begin{equation}\label{nrii06}
  \frac{f_{A'}}{f_{A}}= 1+\frac{\textbf{\emph{a}}\cdot\textbf{\emph{h}}}{c^2}-\frac{\textbf{\emph{g}}\cdot\textbf{\emph{h}}}{c^2}.
\end{equation}
In general relativity, equivalence principle implies that $\textbf{\emph{a}}$ and $\textbf{\emph{g}}$ are equivalent and the second and third terms in Eq.(\ref{nrii06}) cancel each other out. This frequency ratio ${f_{A'}}/{f_{A}}$ is equal to 1 and is independent on the external gravitational field.

In the case of Lorentz violation, there may exist the violation of the weak equivalence principle related to the Einstein elevator (or clock) and the violation of gravitational redshift related to the type of atomic clock. For the clock comparison in Fig.\ref{fig2}, we need to reanalyse the Doppler shift term and gravitational redshift term. From Eqs.(\ref{sve3}) and (\ref{sve30}), the acceleration of the Einstein elevator (or clock) is given by $\textbf{\emph{a}}=(1+\beta_{T})\textbf{\emph{g}}$. The Doppler shift term induced by Lorentz violation becomes
\begin{equation}\label{nrii07}
  -\frac{1}{c}\textbf{\emph{N}}\cdot\left(\textbf{\emph{v}}_{A'}-\textbf{\emph{v}}_{A}\right)=(1+\beta_{T})\frac{\textbf{\emph{g}}\cdot\textbf{\emph{h}}}{c^2},
\end{equation}
where $\beta_{T}$ describes the Lorentz violation for the elevator or clock given by Eq.(\ref{sve30}).
From Eq.(\ref{kekeke}), the gravitational redshift term induced by Lorentz violation is modified by a clock-dependent parameter $\xi_{\text{clock}}$
\begin{equation}\label{nrii08}
  \xi_{\text{clock}}\frac{U_{A'}-U_{A}}{c^{2}}=-\xi_{\text{clock}}\frac{\textbf{\emph{g}}\cdot\textbf{\emph{h}}}{c^2}.
\end{equation}
Therefore, the null-redshift test II can be written as
\begin{equation}\label{nrii2}
   \left(\frac{  f_{A'}}{f_{A}}\right)_{\text{nrII}}=1+  \beta_{T}\frac{\textbf{\emph{g}}\cdot\textbf{\emph{h}}}{c^{2}} -\xi_{A} \frac{\textbf{\emph{g}}\cdot\textbf{\emph{h}}}{c^{2}} = 1+\left( \xi_{A}  -\beta_{T} \right) \frac{\Delta U}{c^{2}},
\end{equation}
where $\xi_{A}$ is the Lorentz-violating parameter of clock $A$, $\Delta U$ is the difference of gravitational potential between two clocks and $\left( \xi_{A}  -\beta_{T} \right) {\Delta U}/{c^{2}}$ is defined as $\left({\delta  f}/{f}\right)_{\text{nrII}}$. This anomalous frequency shift is dependent on the difference of gravitational potential and the spacetime position.
It is worth noting that test II in Eq.(\ref{nrii2}) is different from the traditional gravitational redshift test Eq.(\ref{gr2}). The parameter $\beta_{T}$ in Eq.(\ref{nrii2}) represents the Lorentz violation of the elevator, while the parameter $\beta_{\text{gravimeter}}$ in Eq.(\ref{gr2}) represents the Lorentz violation related to the gravimeter. In the null-redshift test II the clock configuration is in free fall in an external gravitational field, while in the traditional redshift test the clock configuration is stationary in a gravitational field. Two redshift tests are sensitivity to different combinations of SME coefficients.

For an experimental example of the test II, we consider that the Earth can be treated as the Einstein elevator, the Sun's gravitational field is the external gravitational field, and the same type of clocks in different laboratories can be considered as the clocks $A$ and $A'$ in the Einstein elevator. The laboratories at different positions provides the potential difference with respect to the Sun. It is therefore possible to constrain the Lorentz-violating parameter by monitoring the variations in frequency differences between remote clocks, located at different longitudes or different latitudes. The ground clock networks of the same type of clock linked by the optical fiber links can provide the experimental test for the null-redshift test II. The Sr optical clocks in the PTB of Germany, NPL of United Kingdom, and LNE-SYRTE of France constitute a clock network \cite{le2013experimental,falke2014strontium,hill2016low,lodewyck2016optical}, which can used to place the bounds on the SME coefficients. In this Sr clock network, the null-redshift test II between two Sr clocks is given by
\begin{equation}\label{nrii3}
   \left(\frac{\delta  f}{f}\right)_{\text{nrII,Sr}}=  \left( \xi_{Sr}  -\beta_{E} \right) \frac{\Delta U_{S}}{c^{2}},
\end{equation}
where $\xi_{Sr}$ is the Lorentz-violating parameter of Sr clock, $\beta_{E}$ is the Lorentz-violating parameter of the Earth, and $\Delta U_{S}$ is the difference of Sun gravitational potential between two Sr clocks. The rotation of the Earth will cause changes in the difference of gravitational potential, which leads to variations in the frequency differences with a sidereal day period.
A search for a daily variations in frequency difference of Sr clock comparisons obtained a constraint on the Robertson-Mansouri-Sexl parameter \cite{PhysRevLett.118.221102}, which can be viewed as the null-redshift test II, corresponding to the bound on parameters $ |\xi_{Sr}  -\beta_{E}|\leq 9\times10^{-5}$. From Eq.(\ref{nrii3}), the violated effects are sensitivity to the gravitational potential difference and the clock-comparison experiments with more distances have more potential to test Lorentz symmetry, such as the frequency comparison of 2220 km with optical fibre \cite{schioppo2022comparing,roberts2024ultralight}. Based on the worldwide network of the state-of-the-art optical clocks, the parameter $\xi-\beta$ may be limited to the level of $10^{-6}$ or better. TABLE.\ref{tab3} presents the limits on the SME coefficients with the null-redshift test II, in which values without brackets are the estimate from the Sr clock network and values in brackets represent the estimate of attainable sensitivities from the worldwide clock network with the accuracy of $10^{-18}$.

\begin{table}
\caption{\label{tab3} The estimated sensitivity of SME coefficients with the null-redshift test II. For the constraints on SME coefficients, the index $T$ replacing 0 indicates these limits hold in the Sun-centered celestial equatorial frame. Values in brackets represent the estimate of sensitivities attainable in the future tests.}
\begin{ruledtabular}
\begin{tabular}{cccccccc}
{SME coefficients} &$\xi_{\text{clock}}-\beta_{\text{T}}$   &$\Delta U_S$ &$\alpha(\bar{a}^{e+p}_{\text{eff}})_{T}$ (GeV)  &$\alpha(\bar{a}^{n}_{\text{eff}})_{T}$ (GeV) & $\bar{c}^{p}_{TT}$  &  $\bar{c}^{n}_{TT}$  & $\bar{c}^{e}_{TT}$ \\
\hline
{Sr clock network \cite{PhysRevLett.118.221102}} & $3(9)\times10^{-5}$  &$4\times10^{-14}$ &$-3(9)\times10^{-5}$   &$-3(8)\times10^{-5}$ &$5(15)\times10^{-5}$ & $3(12)\times10^{-5}$ &$-2(7)\times10^{-5}$\\
{Worldwide clock network} & $[\sim 10^{-6}]$ &$[8\times10^{-13}]$ &$[10^{-6}]$ &$[10^{-6}]$ &$[10^{-6}]$ &$[10^{-6}]$  &$[10^{-6}]$
\end{tabular}
\end{ruledtabular}
\end{table}

\begin{figure}
\includegraphics[width=0.7\textwidth]{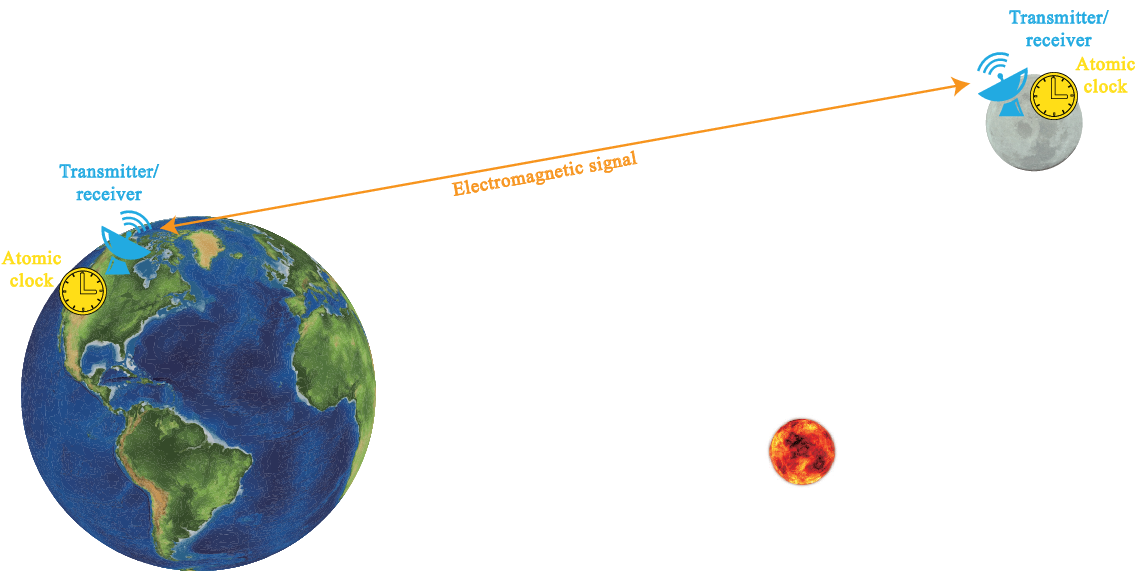}
\caption{\label{fig3} Sketch of the clock comparison of Earth-Moon system. The atomic clocks are deployed on the lunar station. Earth-based station clocks and lunar station clocks exchange electromagnetic pulses via a transmitter/receiver link system.  }
\end{figure}

A better test of null-redshift test II can be achieved by the clock comparison of Earth-Moon system. With recent implementation of various lunar exploration programs and study of clock rates on the Moon \cite{PhysRevD.110.084047,ashby2024relativistic}, there is a potential for atomic clocks to be deployed on the Moon in the future. As shown in Fig.\ref{fig3}, through microwave or laser links, the clocks at Earth-based stations can undergo frequency comparisons with those deployed at lunar stations. Such clock comparisons within the Earth-Moon system can be applied to the null-redshift test II, where the clocks at the Earth station and lunar station correspond to clocks $A$ and $A'$, respectively, and the Earth-Moon system is corresponding to a freely falling elevator. The clock comparisons in this system can be utilized to constrain parameter combination $\xi_{\text{clock}}-\beta_{\text{EM}}$, where $\beta_{\text{EM}}$ is related to the Earth-Moon system. The great distance between the Earth and the Moon offers a large gravitational potential difference for the Sun, $\Delta U_{S}/c^{2}\approx 2 \times10^{-11}$. Considering a clock comparison accuracy at the level of $1\times10^{-18}$, the parameters $\xi_{\text{clock}}-\beta_{\text{EM}}$ can be constrained to a level of $10^{-8}$.

Additionally, we can also consider clock comparisons using ion clocks. In the International Space Station or China Space Station, the clock comparisons can be performed by using two identical ion clocks comparison, where these clocks are deployed at different positions or gravitational potentials. Considering that the ion clocks are in free fall, such clock comparisons also may be applied to the null-redshift test II, constraining parameter combination $\xi_{\text{clock}} - \beta_{\text{ion}} $, where $\beta_{\text{ion}}$ is the Lorentz-violating parameter of the ion clock. Under ideal conditions, the $\beta_{\text{ion}}$ can be regarded as the Lorentz-violating parameter for ions. For a charged test mass or an ion, the number of electrons is not equal to the number of protons as $N^{e}\neq N^{p}$. Then, from Eq.(\ref{sve30}), by using the results from multiple ion clock comparisons, one can independently constrain the coefficients $\alpha(\bar{a}^{e}_{\text{eff}})_{T}$ and $\alpha(\bar{a}^{p}_{\text{eff}})_{T}$ (other clock-comparison experiments are sensitive to the parameter $\alpha(\bar{a}^{e+p}_{\text{eff}})_{T}$). This clock configuration provides a method disentangling electron from proton contributions.

\section{conclusion}\label{iv}

Lorentz symmetry is one of the most important foundations in physics. This symmetry is predicted to be broken in many scenarios of unification and of quantum gravity, and experimental searching of Lorentz violation is a powerful method to test the fundamental physics.  Atomic clocks serve as indispensable tools in probing Lorentz symmetry due to their unparalleled precision in timekeeping. This work studies the tests of Lorentz symmetry through clock comparison experiments involving the redshift ``test". Based on the matter-gravity coupling sector in Standard Model Extension, we investigate the matter-dependence violation of the weak equivalence principle and the clock-dependence violation of gravitational redshift. The parameter $\beta$ describing the weak equivalence principle is dependent on the composition of test mass. The clock-dependence parameter $\xi$ describing the gravitational redshift violation is given by a rough hydrogenic model.

Considering the experimental clock configurations, we introduce three distinct types of clock redshift experiments for testing Lorentz symmetry, namely traditional gravitational redshift experiments, null-redshift test I and null-redshift test II. These experiments exhibit varying sensitivities to different combinations of the SME coefficients, thereby offering a generic framework for constraining SME coefficients within clock-comparison experiments involving redshift tests.
By considering current clock comparison results, we estimate the upper limits on the SME coefficients with maximum reach analysis, reaching levels of $10^{-7}$ to $10^{-4}$ (TABLEs.\ref{tab1}-\ref{tab3}). Especially, considering relativistic factors in null-redshift test I, the frequency comparisons of E2 and E3 transitions of Yb$^{+}$ can set limits on parameter $c^{e}_{00}=(7.4\pm9.3)\times10^{-9}$, improving the previous bounds more than an order of magnitude. In null-redshift test II, a proposal of ion clock comparisons helps disentangling electron from proton contributions. Our analysis demonstrates that clock redshift experiments are competitive on improving the partial SME coefficients.

Continued advancements in scientific and technological domains are expected to enhance the accuracy and sensitivity of atomic clocks and other precision measurement instruments. One can expect more experimental methods and technologies to be used to test Lorentz symmetry. Meanwhile, same proposed and ongoing clock-comparison experiments, such as ACES, CSS, CLEP, FOCOS, and laboratory clock experiments, may allow to yield heightened precision in constraining gravitational redshift and Lorentz-violating coefficients. Furthermore, it is important to develop a more comprehensive model for atomic transition that includes relativistic calculations for electron, proton, and neutron sectors.

\section{Acknowledgment}
The authors thank the anonymous referee for useful comments and constructive suggestion on improving this work. This work is supported by the National Natural Science Foundation of China (Grants No.12305062, No.12247150, No.12175076 and No.11925503), the Post-doctoral Science Foundation of China (Grant No.2022M721257), and the Guangdong Major Project of Basic and Applied Basic Research (Grant No. 2019B030302001).

\bibliographystyle{apsrev4-2}

\begin{thebibliography}{0}%
\makeatletter
\providecommand \@ifxundefined [1]{%
 \@ifx{#1\undefined}
}%
\providecommand \@ifnum [1]{%
 \ifnum #1\expandafter \@firstoftwo
 \else \expandafter \@secondoftwo
 \fi
}%
\providecommand \@ifx [1]{%
 \ifx #1\expandafter \@firstoftwo
 \else \expandafter \@secondoftwo
 \fi
}%
\providecommand \natexlab [1]{#1}%
\providecommand \enquote  [1]{``#1''}%
\providecommand \bibnamefont  [1]{#1}%
\providecommand \bibfnamefont [1]{#1}%
\providecommand \citenamefont [1]{#1}%
\providecommand \href@noop [0]{\@secondoftwo}%
\providecommand \href [0]{\begingroup \@sanitize@url \@href}%
\providecommand \@href[1]{\@@startlink{#1}\@@href}%
\providecommand \@@href[1]{\endgroup#1\@@endlink}%
\providecommand \@sanitize@url [0]{\catcode `\\12\catcode `\$12\catcode
  `\&12\catcode `\#12\catcode `\^12\catcode `\_12\catcode `\%12\relax}%
\providecommand \@@startlink[1]{}%
\providecommand \@@endlink[0]{}%
\providecommand \url  [0]{\begingroup\@sanitize@url \@url }%
\providecommand \@url [1]{\endgroup\@href {#1}{\urlprefix }}%
\providecommand \urlprefix  [0]{URL }%
\providecommand \Eprint [0]{\href }%
\providecommand \doibase [0]{https://doi.org/}%
\providecommand \selectlanguage [0]{\@gobble}%
\providecommand \bibinfo  [0]{\@secondoftwo}%
\providecommand \bibfield  [0]{\@secondoftwo}%
\providecommand \translation [1]{[#1]}%
\providecommand \BibitemOpen [0]{}%
\providecommand \bibitemStop [0]{}%
\providecommand \bibitemNoStop [0]{.\EOS\space}%
\providecommand \EOS [0]{\spacefactor3000\relax}%
\providecommand \BibitemShut  [1]{\csname bibitem#1\endcsname}%
\let\auto@bib@innerbib\@empty
\end{thebibliography}%


\begin{thebibliography}{95}%
\makeatletter
\providecommand \@ifxundefined [1]{%
 \@ifx{#1\undefined}
}%
\providecommand \@ifnum [1]{%
 \ifnum #1\expandafter \@firstoftwo
 \else \expandafter \@secondoftwo
 \fi
}%
\providecommand \@ifx [1]{%
 \ifx #1\expandafter \@firstoftwo
 \else \expandafter \@secondoftwo
 \fi
}%
\providecommand \natexlab [1]{#1}%
\providecommand \enquote  [1]{``#1''}%
\providecommand \bibnamefont  [1]{#1}%
\providecommand \bibfnamefont [1]{#1}%
\providecommand \citenamefont [1]{#1}%
\providecommand \href@noop [0]{\@secondoftwo}%
\providecommand \href [0]{\begingroup \@sanitize@url \@href}%
\providecommand \@href[1]{\@@startlink{#1}\@@href}%
\providecommand \@@href[1]{\endgroup#1\@@endlink}%
\providecommand \@sanitize@url [0]{\catcode `\\12\catcode `\$12\catcode
  `\&12\catcode `\#12\catcode `\^12\catcode `\_12\catcode `\%12\relax}%
\providecommand \@@startlink[1]{}%
\providecommand \@@endlink[0]{}%
\providecommand \url  [0]{\begingroup\@sanitize@url \@url }%
\providecommand \@url [1]{\endgroup\@href {#1}{\urlprefix }}%
\providecommand \urlprefix  [0]{URL }%
\providecommand \Eprint [0]{\href }%
\providecommand \doibase [0]{https://doi.org/}%
\providecommand \selectlanguage [0]{\@gobble}%
\providecommand \bibinfo  [0]{\@secondoftwo}%
\providecommand \bibfield  [0]{\@secondoftwo}%
\providecommand \translation [1]{[#1]}%
\providecommand \BibitemOpen [0]{}%
\providecommand \bibitemStop [0]{}%
\providecommand \bibitemNoStop [0]{.\EOS\space}%
\providecommand \EOS [0]{\spacefactor3000\relax}%
\providecommand \BibitemShut  [1]{\csname bibitem#1\endcsname}%
\let\auto@bib@innerbib\@empty
\bibitem [{\citenamefont {Kosteleck{\`y}}\ and\ \citenamefont
  {Samuel}(1989)}]{kostelecky1989spontaneous}%
  \BibitemOpen
  \bibfield  {author} {\bibinfo {author} {\bibfnamefont {V.~A.}\ \bibnamefont
  {Kosteleck{\`y}}}\ and\ \bibinfo {author} {\bibfnamefont {S.}~\bibnamefont
  {Samuel}},\ }\bibfield  {title} {\bibinfo {title} {Spontaneous breaking of
  lorentz symmetry in string theory},\ }\href@noop {} {\bibfield  {journal}
  {\bibinfo  {journal} {Physical Review D}\ }\textbf {\bibinfo {volume} {39}},\
  \bibinfo {pages} {683} (\bibinfo {year} {1989})}\BibitemShut {NoStop}%
\bibitem [{\citenamefont {Mattingly}\ and\ \citenamefont
  {David}(2005)}]{Mattingly2005Modern}%
  \BibitemOpen
  \bibfield  {author} {\bibinfo {author} {\bibnamefont {Mattingly}}\ and\
  \bibinfo {author} {\bibnamefont {David}},\ }\bibfield  {title} {\bibinfo
  {title} {Modern tests of lorentz invariance},\ }\href@noop {} {\bibfield
  {journal} {\bibinfo  {journal} {Living Reviews in Relativity}\ }\textbf
  {\bibinfo {volume} {8}} (\bibinfo {year} {2005})}\BibitemShut {NoStop}%
\bibitem [{\citenamefont {Tasson}\ and\ \citenamefont
  {Jay}(2014)}]{Tasson2014What}%
  \BibitemOpen
  \bibfield  {author} {\bibinfo {author} {\bibnamefont {Tasson}}\ and\ \bibinfo
  {author} {\bibfnamefont {D.}~\bibnamefont {Jay}},\ }\bibfield  {title}
  {\bibinfo {title} {What do we know about lorentz invariance?},\ }\href@noop
  {} {\bibfield  {journal} {\bibinfo  {journal} {Reports on Progress in Physics
  Physical Society}\ }\textbf {\bibinfo {volume} {77}},\ \bibinfo {pages}
  {062901} (\bibinfo {year} {2014})}\BibitemShut {NoStop}%
\bibitem [{\citenamefont {Colladay}\ and\ \citenamefont
  {Kosteleck\'y}(1998)}]{PhysRevD.58.116002}%
  \BibitemOpen
  \bibfield  {author} {\bibinfo {author} {\bibfnamefont {D.}~\bibnamefont
  {Colladay}}\ and\ \bibinfo {author} {\bibfnamefont {V.~A.}\ \bibnamefont
  {Kosteleck\'y}},\ }\bibfield  {title} {\bibinfo {title} {Lorentz-violating
  extension of the standard model},\ }\href
  {https://doi.org/10.1103/PhysRevD.58.116002} {\bibfield  {journal} {\bibinfo
  {journal} {Phys. Rev. D}\ }\textbf {\bibinfo {volume} {58}},\ \bibinfo
  {pages} {116002} (\bibinfo {year} {1998})}\BibitemShut {NoStop}%
\bibitem [{\citenamefont {Kosteleck\'y}\ and\ \citenamefont
  {Potting}(1995)}]{PhysRevD.51.3923}%
  \BibitemOpen
  \bibfield  {author} {\bibinfo {author} {\bibfnamefont {V.~A.}\ \bibnamefont
  {Kosteleck\'y}}\ and\ \bibinfo {author} {\bibfnamefont {R.}~\bibnamefont
  {Potting}},\ }\bibfield  {title} {\bibinfo {title} {Cpt, strings, and meson
  factories},\ }\href {https://doi.org/10.1103/PhysRevD.51.3923} {\bibfield
  {journal} {\bibinfo  {journal} {Phys. Rev. D}\ }\textbf {\bibinfo {volume}
  {51}},\ \bibinfo {pages} {3923} (\bibinfo {year} {1995})}\BibitemShut
  {NoStop}%
\bibitem [{\citenamefont {Kosteleck\'y}(2004)}]{PhysRevD.69.105009}%
  \BibitemOpen
  \bibfield  {author} {\bibinfo {author} {\bibfnamefont {V.~A.}\ \bibnamefont
  {Kosteleck\'y}},\ }\bibfield  {title} {\bibinfo {title} {Gravity, lorentz
  violation, and the standard model},\ }\href
  {https://doi.org/10.1103/PhysRevD.69.105009} {\bibfield  {journal} {\bibinfo
  {journal} {Phys. Rev. D}\ }\textbf {\bibinfo {volume} {69}},\ \bibinfo
  {pages} {105009} (\bibinfo {year} {2004})}\BibitemShut {NoStop}%
\bibitem [{\citenamefont {Bailey}\ and\ \citenamefont
  {Kosteleck\'y}(2006)}]{PhysRevD.74.045001}%
  \BibitemOpen
  \bibfield  {author} {\bibinfo {author} {\bibfnamefont {Q.~G.}\ \bibnamefont
  {Bailey}}\ and\ \bibinfo {author} {\bibfnamefont {V.~A.}\ \bibnamefont
  {Kosteleck\'y}},\ }\bibfield  {title} {\bibinfo {title} {Signals for lorentz
  violation in post-newtonian gravity},\ }\href
  {https://doi.org/10.1103/PhysRevD.74.045001} {\bibfield  {journal} {\bibinfo
  {journal} {Phys. Rev. D}\ }\textbf {\bibinfo {volume} {74}},\ \bibinfo
  {pages} {045001} (\bibinfo {year} {2006})}\BibitemShut {NoStop}%
\bibitem [{\citenamefont {Lipa}\ \emph {et~al.}(2003)\citenamefont {Lipa},
  \citenamefont {Nissen}, \citenamefont {Wang}, \citenamefont {Stricker},\ and\
  \citenamefont {Avaloff}}]{PhysRevLett.90.060403}%
  \BibitemOpen
  \bibfield  {author} {\bibinfo {author} {\bibfnamefont {J.~A.}\ \bibnamefont
  {Lipa}}, \bibinfo {author} {\bibfnamefont {J.~A.}\ \bibnamefont {Nissen}},
  \bibinfo {author} {\bibfnamefont {S.}~\bibnamefont {Wang}}, \bibinfo {author}
  {\bibfnamefont {D.~A.}\ \bibnamefont {Stricker}},\ and\ \bibinfo {author}
  {\bibfnamefont {D.}~\bibnamefont {Avaloff}},\ }\bibfield  {title} {\bibinfo
  {title} {New limit on signals of lorentz violation in electrodynamics},\
  }\href {https://doi.org/10.1103/PhysRevLett.90.060403} {\bibfield  {journal}
  {\bibinfo  {journal} {Phys. Rev. Lett.}\ }\textbf {\bibinfo {volume} {90}},\
  \bibinfo {pages} {060403} (\bibinfo {year} {2003})}\BibitemShut {NoStop}%
\bibitem [{\citenamefont {Myers}\ and\ \citenamefont
  {Pospelov}(2003)}]{PhysRevLett.90.211601}%
  \BibitemOpen
  \bibfield  {author} {\bibinfo {author} {\bibfnamefont {R.~C.}\ \bibnamefont
  {Myers}}\ and\ \bibinfo {author} {\bibfnamefont {M.}~\bibnamefont
  {Pospelov}},\ }\bibfield  {title} {\bibinfo {title} {Ultraviolet
  modifications of dispersion relations in effective field theory},\ }\href
  {https://doi.org/10.1103/PhysRevLett.90.211601} {\bibfield  {journal}
  {\bibinfo  {journal} {Phys. Rev. Lett.}\ }\textbf {\bibinfo {volume} {90}},\
  \bibinfo {pages} {211601} (\bibinfo {year} {2003})}\BibitemShut {NoStop}%
\bibitem [{\citenamefont {Tobar}\ \emph {et~al.}(2005)\citenamefont {Tobar},
  \citenamefont {Wolf}, \citenamefont {Fowler},\ and\ \citenamefont
  {Hartnett}}]{PhysRevD.71.025004}%
  \BibitemOpen
  \bibfield  {author} {\bibinfo {author} {\bibfnamefont {M.~E.}\ \bibnamefont
  {Tobar}}, \bibinfo {author} {\bibfnamefont {P.}~\bibnamefont {Wolf}},
  \bibinfo {author} {\bibfnamefont {A.}~\bibnamefont {Fowler}},\ and\ \bibinfo
  {author} {\bibfnamefont {J.~G.}\ \bibnamefont {Hartnett}},\ }\bibfield
  {title} {\bibinfo {title} {New methods of testing lorentz violation in
  electrodynamics},\ }\href {https://doi.org/10.1103/PhysRevD.71.025004}
  {\bibfield  {journal} {\bibinfo  {journal} {Phys. Rev. D}\ }\textbf {\bibinfo
  {volume} {71}},\ \bibinfo {pages} {025004} (\bibinfo {year}
  {2005})}\BibitemShut {NoStop}%
\bibitem [{\citenamefont {Reyes}(2013)}]{PhysRevD.87.125028}%
  \BibitemOpen
  \bibfield  {author} {\bibinfo {author} {\bibfnamefont {C.~M.}\ \bibnamefont
  {Reyes}},\ }\bibfield  {title} {\bibinfo {title} {Unitarity in higher-order
  lorentz-invariance violating qed},\ }\href
  {https://doi.org/10.1103/PhysRevD.87.125028} {\bibfield  {journal} {\bibinfo
  {journal} {Phys. Rev. D}\ }\textbf {\bibinfo {volume} {87}},\ \bibinfo
  {pages} {125028} (\bibinfo {year} {2013})}\BibitemShut {NoStop}%
\bibitem [{\citenamefont {Schreck}(2014)}]{PhysRevD.89.105019}%
  \BibitemOpen
  \bibfield  {author} {\bibinfo {author} {\bibfnamefont {M.}~\bibnamefont
  {Schreck}},\ }\bibfield  {title} {\bibinfo {title} {Quantum field theoretic
  properties of lorentz-violating operators of nonrenormalizable dimension in
  the photon sector},\ }\href {https://doi.org/10.1103/PhysRevD.89.105019}
  {\bibfield  {journal} {\bibinfo  {journal} {Phys. Rev. D}\ }\textbf {\bibinfo
  {volume} {89}},\ \bibinfo {pages} {105019} (\bibinfo {year}
  {2014})}\BibitemShut {NoStop}%
\bibitem [{\citenamefont {M\"uller}\ \emph
  {et~al.}(2003{\natexlab{a}})\citenamefont {M\"uller}, \citenamefont
  {Herrmann}, \citenamefont {Braxmaier}, \citenamefont {Schiller},\ and\
  \citenamefont {Peters}}]{PhysRevLett.91.020401}%
  \BibitemOpen
  \bibfield  {author} {\bibinfo {author} {\bibfnamefont {H.}~\bibnamefont
  {M\"uller}}, \bibinfo {author} {\bibfnamefont {S.}~\bibnamefont {Herrmann}},
  \bibinfo {author} {\bibfnamefont {C.}~\bibnamefont {Braxmaier}}, \bibinfo
  {author} {\bibfnamefont {S.}~\bibnamefont {Schiller}},\ and\ \bibinfo
  {author} {\bibfnamefont {A.}~\bibnamefont {Peters}},\ }\bibfield  {title}
  {\bibinfo {title} {Modern michelson-morley experiment using cryogenic optical
  resonators},\ }\href {https://doi.org/10.1103/PhysRevLett.91.020401}
  {\bibfield  {journal} {\bibinfo  {journal} {Phys. Rev. Lett.}\ }\textbf
  {\bibinfo {volume} {91}},\ \bibinfo {pages} {020401} (\bibinfo {year}
  {2003}{\natexlab{a}})}\BibitemShut {NoStop}%
\bibitem [{\citenamefont {Casana}\ \emph {et~al.}(2018)\citenamefont {Casana},
  \citenamefont {Ferreira}, \citenamefont {Lisboa-Santos}, \citenamefont {dos
  Santos},\ and\ \citenamefont {Schreck}}]{PhysRevD.97.115043}%
  \BibitemOpen
  \bibfield  {author} {\bibinfo {author} {\bibfnamefont {R.}~\bibnamefont
  {Casana}}, \bibinfo {author} {\bibfnamefont {M.~M.}\ \bibnamefont
  {Ferreira}}, \bibinfo {author} {\bibfnamefont {L.}~\bibnamefont
  {Lisboa-Santos}}, \bibinfo {author} {\bibfnamefont {F.~E.~P.}\ \bibnamefont
  {dos Santos}},\ and\ \bibinfo {author} {\bibfnamefont {M.}~\bibnamefont
  {Schreck}},\ }\bibfield  {title} {\bibinfo {title} {Maxwell electrodynamics
  modified by $cpt$-even and lorentz-violating dimension-6 higher-derivative
  terms},\ }\href {https://doi.org/10.1103/PhysRevD.97.115043} {\bibfield
  {journal} {\bibinfo  {journal} {Phys. Rev. D}\ }\textbf {\bibinfo {volume}
  {97}},\ \bibinfo {pages} {115043} (\bibinfo {year} {2018})}\BibitemShut
  {NoStop}%
\bibitem [{\citenamefont {Ferreira}\ \emph {et~al.}(2019)\citenamefont
  {Ferreira}, \citenamefont {Lisboa-Santos}, \citenamefont {Maluf},\ and\
  \citenamefont {Schreck}}]{PhysRevD.100.055036}%
  \BibitemOpen
  \bibfield  {author} {\bibinfo {author} {\bibfnamefont {M.~M.}\ \bibnamefont
  {Ferreira}}, \bibinfo {author} {\bibfnamefont {L.}~\bibnamefont
  {Lisboa-Santos}}, \bibinfo {author} {\bibfnamefont {R.~V.}\ \bibnamefont
  {Maluf}},\ and\ \bibinfo {author} {\bibfnamefont {M.}~\bibnamefont
  {Schreck}},\ }\bibfield  {title} {\bibinfo {title} {Maxwell electrodynamics
  modified by a $cpt$-odd dimension-five higher-derivative term},\ }\href
  {https://doi.org/10.1103/PhysRevD.100.055036} {\bibfield  {journal} {\bibinfo
   {journal} {Phys. Rev. D}\ }\textbf {\bibinfo {volume} {100}},\ \bibinfo
  {pages} {055036} (\bibinfo {year} {2019})}\BibitemShut {NoStop}%
\bibitem [{\citenamefont {Bluhm}\ and\ \citenamefont
  {Kosteleck\'y}(2000)}]{PhysRevLett.84.1381}%
  \BibitemOpen
  \bibfield  {author} {\bibinfo {author} {\bibfnamefont {R.}~\bibnamefont
  {Bluhm}}\ and\ \bibinfo {author} {\bibfnamefont {V.~A.}\ \bibnamefont
  {Kosteleck\'y}},\ }\bibfield  {title} {\bibinfo {title} {Lorentz and
  $\mathit{CPT}$ tests with spin-polarized solids},\ }\href
  {https://doi.org/10.1103/PhysRevLett.84.1381} {\bibfield  {journal} {\bibinfo
   {journal} {Phys. Rev. Lett.}\ }\textbf {\bibinfo {volume} {84}},\ \bibinfo
  {pages} {1381} (\bibinfo {year} {2000})}\BibitemShut {NoStop}%
\bibitem [{\citenamefont {Hou}\ \emph {et~al.}(2003)\citenamefont {Hou},
  \citenamefont {Ni},\ and\ \citenamefont {Li}}]{PhysRevLett.90.201101}%
  \BibitemOpen
  \bibfield  {author} {\bibinfo {author} {\bibfnamefont {L.-S.}\ \bibnamefont
  {Hou}}, \bibinfo {author} {\bibfnamefont {W.-T.}\ \bibnamefont {Ni}},\ and\
  \bibinfo {author} {\bibfnamefont {Y.-C.~M.}\ \bibnamefont {Li}},\ }\bibfield
  {title} {\bibinfo {title} {Test of cosmic spatial isotropy for polarized
  electrons using a rotatable torsion balance},\ }\href
  {https://doi.org/10.1103/PhysRevLett.90.201101} {\bibfield  {journal}
  {\bibinfo  {journal} {Phys. Rev. Lett.}\ }\textbf {\bibinfo {volume} {90}},\
  \bibinfo {pages} {201101} (\bibinfo {year} {2003})}\BibitemShut {NoStop}%
\bibitem [{\citenamefont {M\"uller}\ \emph
  {et~al.}(2003{\natexlab{b}})\citenamefont {M\"uller}, \citenamefont
  {Herrmann}, \citenamefont {Saenz}, \citenamefont {Peters},\ and\
  \citenamefont {L\"ammerzahl}}]{PhysRevD.68.116006}%
  \BibitemOpen
  \bibfield  {author} {\bibinfo {author} {\bibfnamefont {H.}~\bibnamefont
  {M\"uller}}, \bibinfo {author} {\bibfnamefont {S.}~\bibnamefont {Herrmann}},
  \bibinfo {author} {\bibfnamefont {A.}~\bibnamefont {Saenz}}, \bibinfo
  {author} {\bibfnamefont {A.}~\bibnamefont {Peters}},\ and\ \bibinfo {author}
  {\bibfnamefont {C.}~\bibnamefont {L\"ammerzahl}},\ }\bibfield  {title}
  {\bibinfo {title} {Optical cavity tests of lorentz invariance for the
  electron},\ }\href {https://doi.org/10.1103/PhysRevD.68.116006} {\bibfield
  {journal} {\bibinfo  {journal} {Phys. Rev. D}\ }\textbf {\bibinfo {volume}
  {68}},\ \bibinfo {pages} {116006} (\bibinfo {year}
  {2003}{\natexlab{b}})}\BibitemShut {NoStop}%
\bibitem [{\citenamefont {M\"uller}\ \emph {et~al.}(2004)\citenamefont
  {M\"uller}, \citenamefont {Herrmann}, \citenamefont {Saenz}, \citenamefont
  {Peters},\ and\ \citenamefont {L\"ammerzahl}}]{PhysRevD.70.076004}%
  \BibitemOpen
  \bibfield  {author} {\bibinfo {author} {\bibfnamefont {H.}~\bibnamefont
  {M\"uller}}, \bibinfo {author} {\bibfnamefont {S.}~\bibnamefont {Herrmann}},
  \bibinfo {author} {\bibfnamefont {A.}~\bibnamefont {Saenz}}, \bibinfo
  {author} {\bibfnamefont {A.}~\bibnamefont {Peters}},\ and\ \bibinfo {author}
  {\bibfnamefont {C.}~\bibnamefont {L\"ammerzahl}},\ }\bibfield  {title}
  {\bibinfo {title} {Tests of lorentz invariance using hydrogen molecules},\
  }\href {https://doi.org/10.1103/PhysRevD.70.076004} {\bibfield  {journal}
  {\bibinfo  {journal} {Phys. Rev. D}\ }\textbf {\bibinfo {volume} {70}},\
  \bibinfo {pages} {076004} (\bibinfo {year} {2004})}\BibitemShut {NoStop}%
\bibitem [{\citenamefont {M\"uller}(2005)}]{PhysRevD.71.045004}%
  \BibitemOpen
  \bibfield  {author} {\bibinfo {author} {\bibfnamefont {H.}~\bibnamefont
  {M\"uller}},\ }\bibfield  {title} {\bibinfo {title} {Testing lorentz
  invariance by the use of vacuum and matter filled cavity resonators},\ }\href
  {https://doi.org/10.1103/PhysRevD.71.045004} {\bibfield  {journal} {\bibinfo
  {journal} {Phys. Rev. D}\ }\textbf {\bibinfo {volume} {71}},\ \bibinfo
  {pages} {045004} (\bibinfo {year} {2005})}\BibitemShut {NoStop}%
\bibitem [{\citenamefont {Abe}\ \emph {et~al.}(2001)\citenamefont {Abe},
  \citenamefont {Abe}, \citenamefont {Adachi}, \citenamefont {Ahn},
  \citenamefont {Aihara}, \citenamefont {Akatsu}, \citenamefont {Alimonti},
  \citenamefont {Aoki}, \citenamefont {Asai},\ and\ \citenamefont
  {Asai}}]{PhysRevLett.86.3228}%
  \BibitemOpen
  \bibfield  {author} {\bibinfo {author} {\bibfnamefont {K.}~\bibnamefont
  {Abe}}, \bibinfo {author} {\bibfnamefont {K.}~\bibnamefont {Abe}}, \bibinfo
  {author} {\bibfnamefont {I.}~\bibnamefont {Adachi}}, \bibinfo {author}
  {\bibfnamefont {B.~S.}\ \bibnamefont {Ahn}}, \bibinfo {author} {\bibfnamefont
  {H.}~\bibnamefont {Aihara}}, \bibinfo {author} {\bibfnamefont
  {M.}~\bibnamefont {Akatsu}}, \bibinfo {author} {\bibfnamefont
  {G.}~\bibnamefont {Alimonti}}, \bibinfo {author} {\bibfnamefont
  {K.}~\bibnamefont {Aoki}}, \bibinfo {author} {\bibfnamefont {K.}~\bibnamefont
  {Asai}},\ and\ \bibinfo {author} {\bibfnamefont {M.~e.~a.}\ \bibnamefont
  {Asai}} (\bibinfo {collaboration} {Belle Collaboration}),\ }\bibfield
  {title} {\bibinfo {title} {Measurement of
  ${B}_{d}^{0}\ensuremath{-}{\overline{b}}_{d}^{0}$ mixing rate from the time
  evolution of dilepton events at the
  $\mathit{\ensuremath{\Upsilon}}(4\mathit{S})$},\ }\href
  {https://doi.org/10.1103/PhysRevLett.86.3228} {\bibfield  {journal} {\bibinfo
   {journal} {Phys. Rev. Lett.}\ }\textbf {\bibinfo {volume} {86}},\ \bibinfo
  {pages} {3228} (\bibinfo {year} {2001})}\BibitemShut {NoStop}%
\bibitem [{\citenamefont {Kosteleck\'y}(2001)}]{PhysRevD.64.076001}%
  \BibitemOpen
  \bibfield  {author} {\bibinfo {author} {\bibfnamefont {V.~A.}\ \bibnamefont
  {Kosteleck\'y}},\ }\bibfield  {title} {\bibinfo {title} {Cpt, t, and lorentz
  violation in neutral-meson oscillations},\ }\href
  {https://doi.org/10.1103/PhysRevD.64.076001} {\bibfield  {journal} {\bibinfo
  {journal} {Phys. Rev. D}\ }\textbf {\bibinfo {volume} {64}},\ \bibinfo
  {pages} {076001} (\bibinfo {year} {2001})}\BibitemShut {NoStop}%
\bibitem [{\citenamefont {Lambiase}(2005)}]{PhysRevD.72.087702}%
  \BibitemOpen
  \bibfield  {author} {\bibinfo {author} {\bibfnamefont {G.}~\bibnamefont
  {Lambiase}},\ }\bibfield  {title} {\bibinfo {title} {Lorentz invariance
  breakdown and constraints from big-bang nucleosynthesis},\ }\href
  {https://doi.org/10.1103/PhysRevD.72.087702} {\bibfield  {journal} {\bibinfo
  {journal} {Phys. Rev. D}\ }\textbf {\bibinfo {volume} {72}},\ \bibinfo
  {pages} {087702} (\bibinfo {year} {2005})}\BibitemShut {NoStop}%
\bibitem [{\citenamefont {Battat}\ \emph {et~al.}(2007)\citenamefont {Battat},
  \citenamefont {Chandler},\ and\ \citenamefont
  {Stubbs}}]{PhysRevLett.99.241103}%
  \BibitemOpen
  \bibfield  {author} {\bibinfo {author} {\bibfnamefont {J.~B.~R.}\
  \bibnamefont {Battat}}, \bibinfo {author} {\bibfnamefont {J.~F.}\
  \bibnamefont {Chandler}},\ and\ \bibinfo {author} {\bibfnamefont {C.~W.}\
  \bibnamefont {Stubbs}},\ }\bibfield  {title} {\bibinfo {title} {Testing for
  lorentz violation: Constraints on standard-model-extension parameters via
  lunar laser ranging},\ }\href {https://doi.org/10.1103/PhysRevLett.99.241103}
  {\bibfield  {journal} {\bibinfo  {journal} {Phys. Rev. Lett.}\ }\textbf
  {\bibinfo {volume} {99}},\ \bibinfo {pages} {241103} (\bibinfo {year}
  {2007})}\BibitemShut {NoStop}%
\bibitem [{\citenamefont {Bourgoin}\ \emph {et~al.}(2021)\citenamefont
  {Bourgoin}, \citenamefont {Bouquillon}, \citenamefont {Hees}, \citenamefont
  {Le~Poncin-Lafitte}, \citenamefont {Bailey}, \citenamefont {Howard},
  \citenamefont {Angonin}, \citenamefont {Francou}, \citenamefont {Chab\'e},
  \citenamefont {Courde},\ and\ \citenamefont {Torre}}]{PhysRevD.103.064055}%
  \BibitemOpen
  \bibfield  {author} {\bibinfo {author} {\bibfnamefont {A.}~\bibnamefont
  {Bourgoin}}, \bibinfo {author} {\bibfnamefont {S.}~\bibnamefont
  {Bouquillon}}, \bibinfo {author} {\bibfnamefont {A.}~\bibnamefont {Hees}},
  \bibinfo {author} {\bibfnamefont {C.}~\bibnamefont {Le~Poncin-Lafitte}},
  \bibinfo {author} {\bibfnamefont {Q.~G.}\ \bibnamefont {Bailey}}, \bibinfo
  {author} {\bibfnamefont {J.~J.}\ \bibnamefont {Howard}}, \bibinfo {author}
  {\bibfnamefont {M.-C.}\ \bibnamefont {Angonin}}, \bibinfo {author}
  {\bibfnamefont {G.}~\bibnamefont {Francou}}, \bibinfo {author} {\bibfnamefont
  {J.}~\bibnamefont {Chab\'e}}, \bibinfo {author} {\bibfnamefont
  {C.}~\bibnamefont {Courde}},\ and\ \bibinfo {author} {\bibfnamefont {J.-M.}\
  \bibnamefont {Torre}},\ }\bibfield  {title} {\bibinfo {title} {Constraining
  velocity-dependent lorentz and $cpt$ violations using lunar laser ranging},\
  }\href {https://doi.org/10.1103/PhysRevD.103.064055} {\bibfield  {journal}
  {\bibinfo  {journal} {Phys. Rev. D}\ }\textbf {\bibinfo {volume} {103}},\
  \bibinfo {pages} {064055} (\bibinfo {year} {2021})}\BibitemShut {NoStop}%
\bibitem [{\citenamefont {Bourgoin}\ \emph {et~al.}(2016)\citenamefont
  {Bourgoin}, \citenamefont {Hees}, \citenamefont {Bouquillon}, \citenamefont
  {Le~Poncin-Lafitte}, \citenamefont {Francou},\ and\ \citenamefont
  {Angonin}}]{PhysRevLett.117.241301}%
  \BibitemOpen
  \bibfield  {author} {\bibinfo {author} {\bibfnamefont {A.}~\bibnamefont
  {Bourgoin}}, \bibinfo {author} {\bibfnamefont {A.}~\bibnamefont {Hees}},
  \bibinfo {author} {\bibfnamefont {S.}~\bibnamefont {Bouquillon}}, \bibinfo
  {author} {\bibfnamefont {C.}~\bibnamefont {Le~Poncin-Lafitte}}, \bibinfo
  {author} {\bibfnamefont {G.}~\bibnamefont {Francou}},\ and\ \bibinfo {author}
  {\bibfnamefont {M.~C.}\ \bibnamefont {Angonin}},\ }\bibfield  {title}
  {\bibinfo {title} {Testing lorentz symmetry with lunar laser ranging},\
  }\href {https://doi.org/10.1103/PhysRevLett.117.241301} {\bibfield  {journal}
  {\bibinfo  {journal} {Phys. Rev. Lett.}\ }\textbf {\bibinfo {volume} {117}},\
  \bibinfo {pages} {241301} (\bibinfo {year} {2016})}\BibitemShut {NoStop}%
\bibitem [{\citenamefont {Bourgoin}\ \emph {et~al.}(2017)\citenamefont
  {Bourgoin}, \citenamefont {Le~Poncin-Lafitte}, \citenamefont {Hees},
  \citenamefont {Bouquillon}, \citenamefont {Francou},\ and\ \citenamefont
  {Angonin}}]{PhysRevLett.119.201102}%
  \BibitemOpen
  \bibfield  {author} {\bibinfo {author} {\bibfnamefont {A.}~\bibnamefont
  {Bourgoin}}, \bibinfo {author} {\bibfnamefont {C.}~\bibnamefont
  {Le~Poncin-Lafitte}}, \bibinfo {author} {\bibfnamefont {A.}~\bibnamefont
  {Hees}}, \bibinfo {author} {\bibfnamefont {S.}~\bibnamefont {Bouquillon}},
  \bibinfo {author} {\bibfnamefont {G.}~\bibnamefont {Francou}},\ and\ \bibinfo
  {author} {\bibfnamefont {M.-C.}\ \bibnamefont {Angonin}},\ }\bibfield
  {title} {\bibinfo {title} {Lorentz symmetry violations from matter-gravity
  couplings with lunar laser ranging},\ }\href
  {https://doi.org/10.1103/PhysRevLett.119.201102} {\bibfield  {journal}
  {\bibinfo  {journal} {Phys. Rev. Lett.}\ }\textbf {\bibinfo {volume} {119}},\
  \bibinfo {pages} {201102} (\bibinfo {year} {2017})}\BibitemShut {NoStop}%
\bibitem [{\citenamefont {Shao}\ \emph {et~al.}(2019)\citenamefont {Shao},
  \citenamefont {Chen}, \citenamefont {Tan}, \citenamefont {Yang},
  \citenamefont {Luo}, \citenamefont {Tobar}, \citenamefont {Long},
  \citenamefont {Weisman},\ and\ \citenamefont
  {Kosteleck\'y}}]{PhysRevLett.122.011102}%
  \BibitemOpen
  \bibfield  {author} {\bibinfo {author} {\bibfnamefont {C.-G.}\ \bibnamefont
  {Shao}}, \bibinfo {author} {\bibfnamefont {Y.-F.}\ \bibnamefont {Chen}},
  \bibinfo {author} {\bibfnamefont {Y.-J.}\ \bibnamefont {Tan}}, \bibinfo
  {author} {\bibfnamefont {S.-Q.}\ \bibnamefont {Yang}}, \bibinfo {author}
  {\bibfnamefont {J.}~\bibnamefont {Luo}}, \bibinfo {author} {\bibfnamefont
  {M.~E.}\ \bibnamefont {Tobar}}, \bibinfo {author} {\bibfnamefont {J.~C.}\
  \bibnamefont {Long}}, \bibinfo {author} {\bibfnamefont {E.}~\bibnamefont
  {Weisman}},\ and\ \bibinfo {author} {\bibfnamefont {V.~A.}\ \bibnamefont
  {Kosteleck\'y}},\ }\bibfield  {title} {\bibinfo {title} {Combined search for
  a lorentz-violating force in short-range gravity varying as the inverse sixth
  power of distance},\ }\href {https://doi.org/10.1103/PhysRevLett.122.011102}
  {\bibfield  {journal} {\bibinfo  {journal} {Phys. Rev. Lett.}\ }\textbf
  {\bibinfo {volume} {122}},\ \bibinfo {pages} {011102} (\bibinfo {year}
  {2019})}\BibitemShut {NoStop}%
\bibitem [{\citenamefont {Bailey}\ \emph {et~al.}(2015)\citenamefont {Bailey},
  \citenamefont {Kosteleck\'y},\ and\ \citenamefont {Xu}}]{PhysRevD.91.022006}%
  \BibitemOpen
  \bibfield  {author} {\bibinfo {author} {\bibfnamefont {Q.~G.}\ \bibnamefont
  {Bailey}}, \bibinfo {author} {\bibfnamefont {V.~A.}\ \bibnamefont
  {Kosteleck\'y}},\ and\ \bibinfo {author} {\bibfnamefont {R.}~\bibnamefont
  {Xu}},\ }\bibfield  {title} {\bibinfo {title} {Short-range gravity and
  lorentz violation},\ }\href {https://doi.org/10.1103/PhysRevD.91.022006}
  {\bibfield  {journal} {\bibinfo  {journal} {Phys. Rev. D}\ }\textbf {\bibinfo
  {volume} {91}},\ \bibinfo {pages} {022006} (\bibinfo {year}
  {2015})}\BibitemShut {NoStop}%
\bibitem [{\citenamefont {Long}\ and\ \citenamefont
  {Kosteleck\'y}(2015)}]{PhysRevD.91.092003}%
  \BibitemOpen
  \bibfield  {author} {\bibinfo {author} {\bibfnamefont {J.~C.}\ \bibnamefont
  {Long}}\ and\ \bibinfo {author} {\bibfnamefont {V.~A.}\ \bibnamefont
  {Kosteleck\'y}},\ }\bibfield  {title} {\bibinfo {title} {Search for lorentz
  violation in short-range gravity},\ }\href
  {https://doi.org/10.1103/PhysRevD.91.092003} {\bibfield  {journal} {\bibinfo
  {journal} {Phys. Rev. D}\ }\textbf {\bibinfo {volume} {91}},\ \bibinfo
  {pages} {092003} (\bibinfo {year} {2015})}\BibitemShut {NoStop}%
\bibitem [{\citenamefont {Shao}\ \emph {et~al.}(2015)\citenamefont {Shao},
  \citenamefont {Tan}, \citenamefont {Tan}, \citenamefont {Yang}, \citenamefont
  {Luo},\ and\ \citenamefont {Tobar}}]{PhysRevD.91.102007}%
  \BibitemOpen
  \bibfield  {author} {\bibinfo {author} {\bibfnamefont {C.-G.}\ \bibnamefont
  {Shao}}, \bibinfo {author} {\bibfnamefont {Y.-J.}\ \bibnamefont {Tan}},
  \bibinfo {author} {\bibfnamefont {W.-H.}\ \bibnamefont {Tan}}, \bibinfo
  {author} {\bibfnamefont {S.-Q.}\ \bibnamefont {Yang}}, \bibinfo {author}
  {\bibfnamefont {J.}~\bibnamefont {Luo}},\ and\ \bibinfo {author}
  {\bibfnamefont {M.~E.}\ \bibnamefont {Tobar}},\ }\bibfield  {title} {\bibinfo
  {title} {Search for lorentz invariance violation through tests of the
  gravitational inverse square law at short ranges},\ }\href
  {https://doi.org/10.1103/PhysRevD.91.102007} {\bibfield  {journal} {\bibinfo
  {journal} {Phys. Rev. D}\ }\textbf {\bibinfo {volume} {91}},\ \bibinfo
  {pages} {102007} (\bibinfo {year} {2015})}\BibitemShut {NoStop}%
\bibitem [{\citenamefont {Kosteleck{\`y}}\ and\ \citenamefont
  {Mewes}(2017)}]{kostelecky2017testing}%
  \BibitemOpen
  \bibfield  {author} {\bibinfo {author} {\bibfnamefont {V.~A.}\ \bibnamefont
  {Kosteleck{\`y}}}\ and\ \bibinfo {author} {\bibfnamefont {M.}~\bibnamefont
  {Mewes}},\ }\bibfield  {title} {\bibinfo {title} {Testing local lorentz
  invariance with short-range gravity},\ }\href@noop {} {\bibfield  {journal}
  {\bibinfo  {journal} {Physics Letters B}\ }\textbf {\bibinfo {volume}
  {766}},\ \bibinfo {pages} {137} (\bibinfo {year} {2017})}\BibitemShut
  {NoStop}%
\bibitem [{\citenamefont {M\"uller}\ \emph {et~al.}(2008)\citenamefont
  {M\"uller}, \citenamefont {Chiow}, \citenamefont {Herrmann}, \citenamefont
  {Chu},\ and\ \citenamefont {Chung}}]{PhysRevLett.100.031101}%
  \BibitemOpen
  \bibfield  {author} {\bibinfo {author} {\bibfnamefont {H.}~\bibnamefont
  {M\"uller}}, \bibinfo {author} {\bibfnamefont {S.-w.}\ \bibnamefont {Chiow}},
  \bibinfo {author} {\bibfnamefont {S.}~\bibnamefont {Herrmann}}, \bibinfo
  {author} {\bibfnamefont {S.}~\bibnamefont {Chu}},\ and\ \bibinfo {author}
  {\bibfnamefont {K.-Y.}\ \bibnamefont {Chung}},\ }\bibfield  {title} {\bibinfo
  {title} {Atom-interferometry tests of the isotropy of post-newtonian
  gravity},\ }\href {https://doi.org/10.1103/PhysRevLett.100.031101} {\bibfield
   {journal} {\bibinfo  {journal} {Phys. Rev. Lett.}\ }\textbf {\bibinfo
  {volume} {100}},\ \bibinfo {pages} {031101} (\bibinfo {year}
  {2008})}\BibitemShut {NoStop}%
\bibitem [{\citenamefont {Flowers}\ \emph {et~al.}(2017)\citenamefont
  {Flowers}, \citenamefont {Goodge},\ and\ \citenamefont
  {Tasson}}]{PhysRevLett.119.201101}%
  \BibitemOpen
  \bibfield  {author} {\bibinfo {author} {\bibfnamefont {N.~A.}\ \bibnamefont
  {Flowers}}, \bibinfo {author} {\bibfnamefont {C.}~\bibnamefont {Goodge}},\
  and\ \bibinfo {author} {\bibfnamefont {J.~D.}\ \bibnamefont {Tasson}},\
  }\bibfield  {title} {\bibinfo {title} {Superconducting-gravimeter tests of
  local lorentz invariance},\ }\href
  {https://doi.org/10.1103/PhysRevLett.119.201101} {\bibfield  {journal}
  {\bibinfo  {journal} {Phys. Rev. Lett.}\ }\textbf {\bibinfo {volume} {119}},\
  \bibinfo {pages} {201101} (\bibinfo {year} {2017})}\BibitemShut {NoStop}%
\bibitem [{\citenamefont {Chung}\ \emph {et~al.}(2009)\citenamefont {Chung},
  \citenamefont {Chiow}, \citenamefont {Herrmann}, \citenamefont {Chu},\ and\
  \citenamefont {M\"uller}}]{PhysRevD.80.016002}%
  \BibitemOpen
  \bibfield  {author} {\bibinfo {author} {\bibfnamefont {K.-Y.}\ \bibnamefont
  {Chung}}, \bibinfo {author} {\bibfnamefont {S.-W.}\ \bibnamefont {Chiow}},
  \bibinfo {author} {\bibfnamefont {S.}~\bibnamefont {Herrmann}}, \bibinfo
  {author} {\bibfnamefont {S.}~\bibnamefont {Chu}},\ and\ \bibinfo {author}
  {\bibfnamefont {H.}~\bibnamefont {M\"uller}},\ }\bibfield  {title} {\bibinfo
  {title} {Atom interferometry tests of local lorentz invariance in gravity and
  electrodynamics},\ }\href {https://doi.org/10.1103/PhysRevD.80.016002}
  {\bibfield  {journal} {\bibinfo  {journal} {Phys. Rev. D}\ }\textbf {\bibinfo
  {volume} {80}},\ \bibinfo {pages} {016002} (\bibinfo {year}
  {2009})}\BibitemShut {NoStop}%
\bibitem [{\citenamefont {Shao}\ \emph {et~al.}(2018)\citenamefont {Shao},
  \citenamefont {Chen}, \citenamefont {Sun}, \citenamefont {Cao}, \citenamefont
  {Zhou}, \citenamefont {Hu}, \citenamefont {Yu},\ and\ \citenamefont
  {M\"uller}}]{PhysRevD.97.024019}%
  \BibitemOpen
  \bibfield  {author} {\bibinfo {author} {\bibfnamefont {C.-G.}\ \bibnamefont
  {Shao}}, \bibinfo {author} {\bibfnamefont {Y.-F.}\ \bibnamefont {Chen}},
  \bibinfo {author} {\bibfnamefont {R.}~\bibnamefont {Sun}}, \bibinfo {author}
  {\bibfnamefont {L.-S.}\ \bibnamefont {Cao}}, \bibinfo {author} {\bibfnamefont
  {M.-K.}\ \bibnamefont {Zhou}}, \bibinfo {author} {\bibfnamefont {Z.-K.}\
  \bibnamefont {Hu}}, \bibinfo {author} {\bibfnamefont {C.}~\bibnamefont
  {Yu}},\ and\ \bibinfo {author} {\bibfnamefont {H.}~\bibnamefont {M\"uller}},\
  }\bibfield  {title} {\bibinfo {title} {Limits on lorentz violation in gravity
  from worldwide superconducting gravimeters},\ }\href
  {https://doi.org/10.1103/PhysRevD.97.024019} {\bibfield  {journal} {\bibinfo
  {journal} {Phys. Rev. D}\ }\textbf {\bibinfo {volume} {97}},\ \bibinfo
  {pages} {024019} (\bibinfo {year} {2018})}\BibitemShut {NoStop}%
\bibitem [{\citenamefont {Sanner}\ \emph {et~al.}(2019)\citenamefont {Sanner},
  \citenamefont {Huntemann}, \citenamefont {Lange}, \citenamefont {Tamm},
  \citenamefont {Peik}, \citenamefont {Safronova},\ and\ \citenamefont
  {Porsev}}]{sanner2019optical}%
  \BibitemOpen
  \bibfield  {author} {\bibinfo {author} {\bibfnamefont {C.}~\bibnamefont
  {Sanner}}, \bibinfo {author} {\bibfnamefont {N.}~\bibnamefont {Huntemann}},
  \bibinfo {author} {\bibfnamefont {R.}~\bibnamefont {Lange}}, \bibinfo
  {author} {\bibfnamefont {C.}~\bibnamefont {Tamm}}, \bibinfo {author}
  {\bibfnamefont {E.}~\bibnamefont {Peik}}, \bibinfo {author} {\bibfnamefont
  {M.~S.}\ \bibnamefont {Safronova}},\ and\ \bibinfo {author} {\bibfnamefont
  {S.~G.}\ \bibnamefont {Porsev}},\ }\bibfield  {title} {\bibinfo {title}
  {Optical clock comparison for lorentz symmetry testing},\ }\href@noop {}
  {\bibfield  {journal} {\bibinfo  {journal} {Nature}\ }\textbf {\bibinfo
  {volume} {567}},\ \bibinfo {pages} {204} (\bibinfo {year}
  {2019})}\BibitemShut {NoStop}%
\bibitem [{\citenamefont {Kosteleck\'y}\ and\ \citenamefont
  {Vargas}(2018)}]{PhysRevD.98.036003}%
  \BibitemOpen
  \bibfield  {author} {\bibinfo {author} {\bibfnamefont {V.~A.}\ \bibnamefont
  {Kosteleck\'y}}\ and\ \bibinfo {author} {\bibfnamefont {A.~J.}\ \bibnamefont
  {Vargas}},\ }\bibfield  {title} {\bibinfo {title} {Lorentz and $cpt$ tests
  with clock-comparison experiments},\ }\href
  {https://doi.org/10.1103/PhysRevD.98.036003} {\bibfield  {journal} {\bibinfo
  {journal} {Phys. Rev. D}\ }\textbf {\bibinfo {volume} {98}},\ \bibinfo
  {pages} {036003} (\bibinfo {year} {2018})}\BibitemShut {NoStop}%
\bibitem [{\citenamefont {Qin}\ \emph {et~al.}(2021)\citenamefont {Qin},
  \citenamefont {Tan},\ and\ \citenamefont {Shao}}]{qin2021test}%
  \BibitemOpen
  \bibfield  {author} {\bibinfo {author} {\bibfnamefont {C.}~\bibnamefont
  {Qin}}, \bibinfo {author} {\bibfnamefont {Y.}~\bibnamefont {Tan}},\ and\
  \bibinfo {author} {\bibfnamefont {C.}~\bibnamefont {Shao}},\ }\bibfield
  {title} {\bibinfo {title} {Test of einstein equivalence principle by
  frequency comparisons of optical clocks},\ }\href@noop {} {\bibfield
  {journal} {\bibinfo  {journal} {Physics Letters B}\ }\textbf {\bibinfo
  {volume} {820}},\ \bibinfo {pages} {136471} (\bibinfo {year}
  {2021})}\BibitemShut {NoStop}%
\bibitem [{\citenamefont {Altschul}(2009)}]{PhysRevD.79.061702}%
  \BibitemOpen
  \bibfield  {author} {\bibinfo {author} {\bibfnamefont {B.}~\bibnamefont
  {Altschul}},\ }\bibfield  {title} {\bibinfo {title} {Disentangling forms of
  lorentz violation with complementary clock comparison experiments},\ }\href
  {https://doi.org/10.1103/PhysRevD.79.061702} {\bibfield  {journal} {\bibinfo
  {journal} {Phys. Rev. D}\ }\textbf {\bibinfo {volume} {79}},\ \bibinfo
  {pages} {061702(R)} (\bibinfo {year} {2009})}\BibitemShut {NoStop}%
\bibitem [{\citenamefont {Kosteleck\'y}\ and\ \citenamefont
  {Russell}(2011)}]{RevModPhys.83.11}%
  \BibitemOpen
  \bibfield  {author} {\bibinfo {author} {\bibfnamefont {V.~A.}\ \bibnamefont
  {Kosteleck\'y}}\ and\ \bibinfo {author} {\bibfnamefont {N.}~\bibnamefont
  {Russell}},\ }\bibfield  {title} {\bibinfo {title} {Data tables for lorentz
  and $cpt$ violation},\ }\href {https://doi.org/10.1103/RevModPhys.83.11}
  {\bibfield  {journal} {\bibinfo  {journal} {Rev. Mod. Phys.}\ }\textbf
  {\bibinfo {volume} {83}},\ \bibinfo {pages} {11} (\bibinfo {year}
  {2011})}\BibitemShut {NoStop}%
\bibitem [{\citenamefont {Dreissen}\ \emph {et~al.}(2022)\citenamefont
  {Dreissen}, \citenamefont {Yeh}, \citenamefont {F{\"u}rst}, \citenamefont
  {Grensemann},\ and\ \citenamefont {Mehlst{\"a}ubler}}]{dreissen2022improved}%
  \BibitemOpen
  \bibfield  {author} {\bibinfo {author} {\bibfnamefont {L.~S.}\ \bibnamefont
  {Dreissen}}, \bibinfo {author} {\bibfnamefont {C.-H.}\ \bibnamefont {Yeh}},
  \bibinfo {author} {\bibfnamefont {H.~A.}\ \bibnamefont {F{\"u}rst}}, \bibinfo
  {author} {\bibfnamefont {K.~C.}\ \bibnamefont {Grensemann}},\ and\ \bibinfo
  {author} {\bibfnamefont {T.~E.}\ \bibnamefont {Mehlst{\"a}ubler}},\
  }\bibfield  {title} {\bibinfo {title} {Improved bounds on lorentz violation
  from composite pulse ramsey spectroscopy in a trapped ion},\ }\href@noop {}
  {\bibfield  {journal} {\bibinfo  {journal} {Nature communications}\ }\textbf
  {\bibinfo {volume} {13}},\ \bibinfo {pages} {7314} (\bibinfo {year}
  {2022})}\BibitemShut {NoStop}%
\bibitem [{\citenamefont {Flambaum}\ and\ \citenamefont
  {Romalis}(2017)}]{PhysRevLett.118.142501}%
  \BibitemOpen
  \bibfield  {author} {\bibinfo {author} {\bibfnamefont {V.~V.}\ \bibnamefont
  {Flambaum}}\ and\ \bibinfo {author} {\bibfnamefont {M.~V.}\ \bibnamefont
  {Romalis}},\ }\bibfield  {title} {\bibinfo {title} {Limits on lorentz
  invariance violation from coulomb interactions in nuclei and atoms},\ }\href
  {https://doi.org/10.1103/PhysRevLett.118.142501} {\bibfield  {journal}
  {\bibinfo  {journal} {Phys. Rev. Lett.}\ }\textbf {\bibinfo {volume} {118}},\
  \bibinfo {pages} {142501} (\bibinfo {year} {2017})}\BibitemShut {NoStop}%
\bibitem [{\citenamefont {Pihan-Le~Bars}\ \emph {et~al.}(2017)\citenamefont
  {Pihan-Le~Bars}, \citenamefont {Guerlin}, \citenamefont {Lasseri},
  \citenamefont {Ebran}, \citenamefont {Bailey}, \citenamefont {Bize},
  \citenamefont {Khan},\ and\ \citenamefont {Wolf}}]{PhysRevD.95.075026}%
  \BibitemOpen
  \bibfield  {author} {\bibinfo {author} {\bibfnamefont {H.}~\bibnamefont
  {Pihan-Le~Bars}}, \bibinfo {author} {\bibfnamefont {C.}~\bibnamefont
  {Guerlin}}, \bibinfo {author} {\bibfnamefont {R.-D.}\ \bibnamefont
  {Lasseri}}, \bibinfo {author} {\bibfnamefont {J.-P.}\ \bibnamefont {Ebran}},
  \bibinfo {author} {\bibfnamefont {Q.~G.}\ \bibnamefont {Bailey}}, \bibinfo
  {author} {\bibfnamefont {S.}~\bibnamefont {Bize}}, \bibinfo {author}
  {\bibfnamefont {E.}~\bibnamefont {Khan}},\ and\ \bibinfo {author}
  {\bibfnamefont {P.}~\bibnamefont {Wolf}},\ }\bibfield  {title} {\bibinfo
  {title} {Lorentz-symmetry test at planck-scale suppression with nucleons in a
  spin-polarized $^{133}\mathrm{Cs}$ cold atom clock},\ }\href
  {https://doi.org/10.1103/PhysRevD.95.075026} {\bibfield  {journal} {\bibinfo
  {journal} {Phys. Rev. D}\ }\textbf {\bibinfo {volume} {95}},\ \bibinfo
  {pages} {075026} (\bibinfo {year} {2017})}\BibitemShut {NoStop}%
\bibitem [{\citenamefont {Brown}\ \emph {et~al.}(2010)\citenamefont {Brown},
  \citenamefont {Smullin}, \citenamefont {Kornack},\ and\ \citenamefont
  {Romalis}}]{PhysRevLett.105.151604}%
  \BibitemOpen
  \bibfield  {author} {\bibinfo {author} {\bibfnamefont {J.~M.}\ \bibnamefont
  {Brown}}, \bibinfo {author} {\bibfnamefont {S.~J.}\ \bibnamefont {Smullin}},
  \bibinfo {author} {\bibfnamefont {T.~W.}\ \bibnamefont {Kornack}},\ and\
  \bibinfo {author} {\bibfnamefont {M.~V.}\ \bibnamefont {Romalis}},\
  }\bibfield  {title} {\bibinfo {title} {New limit on lorentz- and
  $cpt$-violating neutron spin interactions},\ }\href
  {https://doi.org/10.1103/PhysRevLett.105.151604} {\bibfield  {journal}
  {\bibinfo  {journal} {Phys. Rev. Lett.}\ }\textbf {\bibinfo {volume} {105}},\
  \bibinfo {pages} {151604} (\bibinfo {year} {2010})}\BibitemShut {NoStop}%
\bibitem [{\citenamefont {Can\`e}\ \emph {et~al.}(2004)\citenamefont {Can\`e},
  \citenamefont {Bear}, \citenamefont {Phillips}, \citenamefont {Rosen},
  \citenamefont {Smallwood}, \citenamefont {Stoner}, \citenamefont
  {Walsworth},\ and\ \citenamefont {Kosteleck\'y}}]{PhysRevLett.93.230801}%
  \BibitemOpen
  \bibfield  {author} {\bibinfo {author} {\bibfnamefont {F.}~\bibnamefont
  {Can\`e}}, \bibinfo {author} {\bibfnamefont {D.}~\bibnamefont {Bear}},
  \bibinfo {author} {\bibfnamefont {D.~F.}\ \bibnamefont {Phillips}}, \bibinfo
  {author} {\bibfnamefont {M.~S.}\ \bibnamefont {Rosen}}, \bibinfo {author}
  {\bibfnamefont {C.~L.}\ \bibnamefont {Smallwood}}, \bibinfo {author}
  {\bibfnamefont {R.~E.}\ \bibnamefont {Stoner}}, \bibinfo {author}
  {\bibfnamefont {R.~L.}\ \bibnamefont {Walsworth}},\ and\ \bibinfo {author}
  {\bibfnamefont {V.~A.}\ \bibnamefont {Kosteleck\'y}},\ }\bibfield  {title}
  {\bibinfo {title} {Bound on lorentz and $cpt$ violating boost effects for the
  neutron},\ }\href {https://doi.org/10.1103/PhysRevLett.93.230801} {\bibfield
  {journal} {\bibinfo  {journal} {Phys. Rev. Lett.}\ }\textbf {\bibinfo
  {volume} {93}},\ \bibinfo {pages} {230801} (\bibinfo {year}
  {2004})}\BibitemShut {NoStop}%
\bibitem [{\citenamefont {Altarev}\ \emph {et~al.}(2009)\citenamefont
  {Altarev}, \citenamefont {Baker}, \citenamefont {Ban}, \citenamefont {Bison},
  \citenamefont {Bodek}, \citenamefont {Daum}, \citenamefont {Fierlinger},
  \citenamefont {Geltenbort}, \citenamefont {Green}, \citenamefont {van~der
  Grinten}, \citenamefont {Gutsmiedl}, \citenamefont {Harris}, \citenamefont
  {Heil}, \citenamefont {Henneck}, \citenamefont {Horras}, \citenamefont
  {Iaydjiev}, \citenamefont {Ivanov}, \citenamefont {Khomutov}, \citenamefont
  {Kirch}, \citenamefont {Kistryn}, \citenamefont {Knecht}, \citenamefont
  {Knowles}, \citenamefont {Kozela}, \citenamefont {Kuchler}, \citenamefont
  {Ku\ifmmode~\acute{z}\else \'{z}\fi{}niak}, \citenamefont {Lauer},
  \citenamefont {Lauss}, \citenamefont {Lefort}, \citenamefont
  {Mtchedlishvili}, \citenamefont {Naviliat-Cuncic}, \citenamefont {Pazgalev},
  \citenamefont {Pendlebury}, \citenamefont {Petzoldt}, \citenamefont {Pierre},
  \citenamefont {Pignol}, \citenamefont {Qu\'em\'ener}, \citenamefont
  {Rebetez}, \citenamefont {Rebreyend}, \citenamefont {Roccia}, \citenamefont
  {Rogel}, \citenamefont {Severijns}, \citenamefont {Shiers}, \citenamefont
  {Sobolev}, \citenamefont {Weis}, \citenamefont {Zejma},\ and\ \citenamefont
  {Zsigmond}}]{PhysRevLett.103.081602}%
  \BibitemOpen
  \bibfield  {author} {\bibinfo {author} {\bibfnamefont {I.}~\bibnamefont
  {Altarev}}, \bibinfo {author} {\bibfnamefont {C.~A.}\ \bibnamefont {Baker}},
  \bibinfo {author} {\bibfnamefont {G.}~\bibnamefont {Ban}}, \bibinfo {author}
  {\bibfnamefont {G.}~\bibnamefont {Bison}}, \bibinfo {author} {\bibfnamefont
  {K.}~\bibnamefont {Bodek}}, \bibinfo {author} {\bibfnamefont
  {M.}~\bibnamefont {Daum}}, \bibinfo {author} {\bibfnamefont {P.}~\bibnamefont
  {Fierlinger}}, \bibinfo {author} {\bibfnamefont {P.}~\bibnamefont
  {Geltenbort}}, \bibinfo {author} {\bibfnamefont {K.}~\bibnamefont {Green}},
  \bibinfo {author} {\bibfnamefont {M.~G.~D.}\ \bibnamefont {van~der Grinten}},
  \bibinfo {author} {\bibfnamefont {E.}~\bibnamefont {Gutsmiedl}}, \bibinfo
  {author} {\bibfnamefont {P.~G.}\ \bibnamefont {Harris}}, \bibinfo {author}
  {\bibfnamefont {W.}~\bibnamefont {Heil}}, \bibinfo {author} {\bibfnamefont
  {R.}~\bibnamefont {Henneck}}, \bibinfo {author} {\bibfnamefont
  {M.}~\bibnamefont {Horras}}, \bibinfo {author} {\bibfnamefont
  {P.}~\bibnamefont {Iaydjiev}}, \bibinfo {author} {\bibfnamefont {S.~N.}\
  \bibnamefont {Ivanov}}, \bibinfo {author} {\bibfnamefont {N.}~\bibnamefont
  {Khomutov}}, \bibinfo {author} {\bibfnamefont {K.}~\bibnamefont {Kirch}},
  \bibinfo {author} {\bibfnamefont {S.}~\bibnamefont {Kistryn}}, \bibinfo
  {author} {\bibfnamefont {A.}~\bibnamefont {Knecht}}, \bibinfo {author}
  {\bibfnamefont {P.}~\bibnamefont {Knowles}}, \bibinfo {author} {\bibfnamefont
  {A.}~\bibnamefont {Kozela}}, \bibinfo {author} {\bibfnamefont
  {F.}~\bibnamefont {Kuchler}}, \bibinfo {author} {\bibfnamefont
  {M.}~\bibnamefont {Ku\ifmmode~\acute{z}\else \'{z}\fi{}niak}}, \bibinfo
  {author} {\bibfnamefont {T.}~\bibnamefont {Lauer}}, \bibinfo {author}
  {\bibfnamefont {B.}~\bibnamefont {Lauss}}, \bibinfo {author} {\bibfnamefont
  {T.}~\bibnamefont {Lefort}}, \bibinfo {author} {\bibfnamefont
  {A.}~\bibnamefont {Mtchedlishvili}}, \bibinfo {author} {\bibfnamefont
  {O.}~\bibnamefont {Naviliat-Cuncic}}, \bibinfo {author} {\bibfnamefont
  {A.}~\bibnamefont {Pazgalev}}, \bibinfo {author} {\bibfnamefont {J.~M.}\
  \bibnamefont {Pendlebury}}, \bibinfo {author} {\bibfnamefont
  {G.}~\bibnamefont {Petzoldt}}, \bibinfo {author} {\bibfnamefont
  {E.}~\bibnamefont {Pierre}}, \bibinfo {author} {\bibfnamefont
  {G.}~\bibnamefont {Pignol}}, \bibinfo {author} {\bibfnamefont
  {G.}~\bibnamefont {Qu\'em\'ener}}, \bibinfo {author} {\bibfnamefont
  {M.}~\bibnamefont {Rebetez}}, \bibinfo {author} {\bibfnamefont
  {D.}~\bibnamefont {Rebreyend}}, \bibinfo {author} {\bibfnamefont
  {S.}~\bibnamefont {Roccia}}, \bibinfo {author} {\bibfnamefont
  {G.}~\bibnamefont {Rogel}}, \bibinfo {author} {\bibfnamefont
  {N.}~\bibnamefont {Severijns}}, \bibinfo {author} {\bibfnamefont
  {D.}~\bibnamefont {Shiers}}, \bibinfo {author} {\bibfnamefont
  {Y.}~\bibnamefont {Sobolev}}, \bibinfo {author} {\bibfnamefont
  {A.}~\bibnamefont {Weis}}, \bibinfo {author} {\bibfnamefont {J.}~\bibnamefont
  {Zejma}},\ and\ \bibinfo {author} {\bibfnamefont {G.}~\bibnamefont
  {Zsigmond}},\ }\bibfield  {title} {\bibinfo {title} {Test of lorentz
  invariance with spin precession of ultracold neutrons},\ }\href
  {https://doi.org/10.1103/PhysRevLett.103.081602} {\bibfield  {journal}
  {\bibinfo  {journal} {Phys. Rev. Lett.}\ }\textbf {\bibinfo {volume} {103}},\
  \bibinfo {pages} {081602} (\bibinfo {year} {2009})}\BibitemShut {NoStop}%
\bibitem [{\citenamefont {Hohensee}\ \emph
  {et~al.}(2013{\natexlab{a}})\citenamefont {Hohensee}, \citenamefont {Leefer},
  \citenamefont {Budker}, \citenamefont {Harabati}, \citenamefont {Dzuba},\
  and\ \citenamefont {Flambaum}}]{PhysRevLett.111.050401}%
  \BibitemOpen
  \bibfield  {author} {\bibinfo {author} {\bibfnamefont {M.~A.}\ \bibnamefont
  {Hohensee}}, \bibinfo {author} {\bibfnamefont {N.}~\bibnamefont {Leefer}},
  \bibinfo {author} {\bibfnamefont {D.}~\bibnamefont {Budker}}, \bibinfo
  {author} {\bibfnamefont {C.}~\bibnamefont {Harabati}}, \bibinfo {author}
  {\bibfnamefont {V.~A.}\ \bibnamefont {Dzuba}},\ and\ \bibinfo {author}
  {\bibfnamefont {V.~V.}\ \bibnamefont {Flambaum}},\ }\bibfield  {title}
  {\bibinfo {title} {Limits on violations of lorentz symmetry and the einstein
  equivalence principle using radio-frequency spectroscopy of atomic
  dysprosium},\ }\href {https://doi.org/10.1103/PhysRevLett.111.050401}
  {\bibfield  {journal} {\bibinfo  {journal} {Phys. Rev. Lett.}\ }\textbf
  {\bibinfo {volume} {111}},\ \bibinfo {pages} {050401} (\bibinfo {year}
  {2013}{\natexlab{a}})}\BibitemShut {NoStop}%
\bibitem [{\citenamefont {Pruttivarasin}\ \emph {et~al.}(2015)\citenamefont
  {Pruttivarasin}, \citenamefont {Ramm}, \citenamefont {Porsev}, \citenamefont
  {Tupitsyn}, \citenamefont {Safronova}, \citenamefont {Hohensee},\ and\
  \citenamefont {H{\"a}ffner}}]{pruttivarasin2015michelson}%
  \BibitemOpen
  \bibfield  {author} {\bibinfo {author} {\bibfnamefont {T.}~\bibnamefont
  {Pruttivarasin}}, \bibinfo {author} {\bibfnamefont {M.}~\bibnamefont {Ramm}},
  \bibinfo {author} {\bibfnamefont {S.}~\bibnamefont {Porsev}}, \bibinfo
  {author} {\bibfnamefont {I.}~\bibnamefont {Tupitsyn}}, \bibinfo {author}
  {\bibfnamefont {M.}~\bibnamefont {Safronova}}, \bibinfo {author}
  {\bibfnamefont {M.}~\bibnamefont {Hohensee}},\ and\ \bibinfo {author}
  {\bibfnamefont {H.}~\bibnamefont {H{\"a}ffner}},\ }\bibfield  {title}
  {\bibinfo {title} {Michelson--morley analogue for electrons using trapped
  ions to test lorentz symmetry},\ }\href@noop {} {\bibfield  {journal}
  {\bibinfo  {journal} {Nature}\ }\textbf {\bibinfo {volume} {517}},\ \bibinfo
  {pages} {592} (\bibinfo {year} {2015})}\BibitemShut {NoStop}%
\bibitem [{\citenamefont {Wolf}\ \emph {et~al.}(2006)\citenamefont {Wolf},
  \citenamefont {Chapelet}, \citenamefont {Bize},\ and\ \citenamefont
  {Clairon}}]{PhysRevLett.96.060801}%
  \BibitemOpen
  \bibfield  {author} {\bibinfo {author} {\bibfnamefont {P.}~\bibnamefont
  {Wolf}}, \bibinfo {author} {\bibfnamefont {F.}~\bibnamefont {Chapelet}},
  \bibinfo {author} {\bibfnamefont {S.}~\bibnamefont {Bize}},\ and\ \bibinfo
  {author} {\bibfnamefont {A.}~\bibnamefont {Clairon}},\ }\bibfield  {title}
  {\bibinfo {title} {Cold atom clock test of lorentz invariance in the matter
  sector},\ }\href {https://doi.org/10.1103/PhysRevLett.96.060801} {\bibfield
  {journal} {\bibinfo  {journal} {Phys. Rev. Lett.}\ }\textbf {\bibinfo
  {volume} {96}},\ \bibinfo {pages} {060801} (\bibinfo {year}
  {2006})}\BibitemShut {NoStop}%
\bibitem [{\citenamefont {Megidish}\ \emph {et~al.}(2019)\citenamefont
  {Megidish}, \citenamefont {Broz}, \citenamefont {Greene},\ and\ \citenamefont
  {H\"affner}}]{PhysRevLett.122.123605}%
  \BibitemOpen
  \bibfield  {author} {\bibinfo {author} {\bibfnamefont {E.}~\bibnamefont
  {Megidish}}, \bibinfo {author} {\bibfnamefont {J.}~\bibnamefont {Broz}},
  \bibinfo {author} {\bibfnamefont {N.}~\bibnamefont {Greene}},\ and\ \bibinfo
  {author} {\bibfnamefont {H.}~\bibnamefont {H\"affner}},\ }\bibfield  {title}
  {\bibinfo {title} {Improved test of local lorentz invariance from a
  deterministic preparation of entangled states},\ }\href
  {https://doi.org/10.1103/PhysRevLett.122.123605} {\bibfield  {journal}
  {\bibinfo  {journal} {Phys. Rev. Lett.}\ }\textbf {\bibinfo {volume} {122}},\
  \bibinfo {pages} {123605} (\bibinfo {year} {2019})}\BibitemShut {NoStop}%
\bibitem [{\citenamefont {Shaniv}\ \emph {et~al.}(2018)\citenamefont {Shaniv},
  \citenamefont {Ozeri}, \citenamefont {Safronova}, \citenamefont {Porsev},
  \citenamefont {Dzuba}, \citenamefont {Flambaum},\ and\ \citenamefont
  {H\"affner}}]{PhysRevLett.120.103202}%
  \BibitemOpen
  \bibfield  {author} {\bibinfo {author} {\bibfnamefont {R.}~\bibnamefont
  {Shaniv}}, \bibinfo {author} {\bibfnamefont {R.}~\bibnamefont {Ozeri}},
  \bibinfo {author} {\bibfnamefont {M.~S.}\ \bibnamefont {Safronova}}, \bibinfo
  {author} {\bibfnamefont {S.~G.}\ \bibnamefont {Porsev}}, \bibinfo {author}
  {\bibfnamefont {V.~A.}\ \bibnamefont {Dzuba}}, \bibinfo {author}
  {\bibfnamefont {V.~V.}\ \bibnamefont {Flambaum}},\ and\ \bibinfo {author}
  {\bibfnamefont {H.}~\bibnamefont {H\"affner}},\ }\bibfield  {title} {\bibinfo
  {title} {New methods for testing lorentz invariance with atomic systems},\
  }\href {https://doi.org/10.1103/PhysRevLett.120.103202} {\bibfield  {journal}
  {\bibinfo  {journal} {Phys. Rev. Lett.}\ }\textbf {\bibinfo {volume} {120}},\
  \bibinfo {pages} {103202} (\bibinfo {year} {2018})}\BibitemShut {NoStop}%
\bibitem [{\citenamefont {Dzuba}\ \emph {et~al.}(2016)\citenamefont {Dzuba},
  \citenamefont {Flambaum}, \citenamefont {Safronova}, \citenamefont {Porsev},
  \citenamefont {Pruttivarasin}, \citenamefont {Hohensee},\ and\ \citenamefont
  {H{\"a}ffner}}]{dzuba2016strongly}%
  \BibitemOpen
  \bibfield  {author} {\bibinfo {author} {\bibfnamefont {V.}~\bibnamefont
  {Dzuba}}, \bibinfo {author} {\bibfnamefont {V.}~\bibnamefont {Flambaum}},
  \bibinfo {author} {\bibfnamefont {M.}~\bibnamefont {Safronova}}, \bibinfo
  {author} {\bibfnamefont {S.}~\bibnamefont {Porsev}}, \bibinfo {author}
  {\bibfnamefont {T.}~\bibnamefont {Pruttivarasin}}, \bibinfo {author}
  {\bibfnamefont {M.}~\bibnamefont {Hohensee}},\ and\ \bibinfo {author}
  {\bibfnamefont {H.}~\bibnamefont {H{\"a}ffner}},\ }\bibfield  {title}
  {\bibinfo {title} {Strongly enhanced effects of lorentz symmetry violation in
  entangled yb+ ions},\ }\href@noop {} {\bibfield  {journal} {\bibinfo
  {journal} {Nature Physics}\ }\textbf {\bibinfo {volume} {12}},\ \bibinfo
  {pages} {465} (\bibinfo {year} {2016})}\BibitemShut {NoStop}%
\bibitem [{\citenamefont {Lange}\ \emph
  {et~al.}(2021{\natexlab{a}})\citenamefont {Lange}, \citenamefont {Peshkov},
  \citenamefont {Huntemann}, \citenamefont {Tamm}, \citenamefont {Surzhykov},\
  and\ \citenamefont {Peik}}]{PhysRevLett.127.213001}%
  \BibitemOpen
  \bibfield  {author} {\bibinfo {author} {\bibfnamefont {R.}~\bibnamefont
  {Lange}}, \bibinfo {author} {\bibfnamefont {A.~A.}\ \bibnamefont {Peshkov}},
  \bibinfo {author} {\bibfnamefont {N.}~\bibnamefont {Huntemann}}, \bibinfo
  {author} {\bibfnamefont {C.}~\bibnamefont {Tamm}}, \bibinfo {author}
  {\bibfnamefont {A.}~\bibnamefont {Surzhykov}},\ and\ \bibinfo {author}
  {\bibfnamefont {E.}~\bibnamefont {Peik}},\ }\bibfield  {title} {\bibinfo
  {title} {Lifetime of the $^{2}{F}_{7/2}$ level in ${\mathrm{yb}}^{+}$ for
  spontaneous emission of electric octupole radiation},\ }\href
  {https://doi.org/10.1103/PhysRevLett.127.213001} {\bibfield  {journal}
  {\bibinfo  {journal} {Phys. Rev. Lett.}\ }\textbf {\bibinfo {volume} {127}},\
  \bibinfo {pages} {213001} (\bibinfo {year} {2021}{\natexlab{a}})}\BibitemShut
  {NoStop}%
\bibitem [{\citenamefont {Steinel}\ \emph {et~al.}(2023)\citenamefont
  {Steinel}, \citenamefont {Shao}, \citenamefont {Filzinger}, \citenamefont
  {Lipphardt}, \citenamefont {Brinkmann}, \citenamefont {Didier}, \citenamefont
  {Mehlst\"aubler}, \citenamefont {Lindvall}, \citenamefont {Peik},\ and\
  \citenamefont {Huntemann}}]{PhysRevLett.131.083002}%
  \BibitemOpen
  \bibfield  {author} {\bibinfo {author} {\bibfnamefont {M.}~\bibnamefont
  {Steinel}}, \bibinfo {author} {\bibfnamefont {H.}~\bibnamefont {Shao}},
  \bibinfo {author} {\bibfnamefont {M.}~\bibnamefont {Filzinger}}, \bibinfo
  {author} {\bibfnamefont {B.}~\bibnamefont {Lipphardt}}, \bibinfo {author}
  {\bibfnamefont {M.}~\bibnamefont {Brinkmann}}, \bibinfo {author}
  {\bibfnamefont {A.}~\bibnamefont {Didier}}, \bibinfo {author} {\bibfnamefont
  {T.~E.}\ \bibnamefont {Mehlst\"aubler}}, \bibinfo {author} {\bibfnamefont
  {T.}~\bibnamefont {Lindvall}}, \bibinfo {author} {\bibfnamefont
  {E.}~\bibnamefont {Peik}},\ and\ \bibinfo {author} {\bibfnamefont
  {N.}~\bibnamefont {Huntemann}},\ }\bibfield  {title} {\bibinfo {title}
  {Evaluation of a $^{88}{\mathrm{sr}}^{+}$ optical clock with a direct
  measurement of the blackbody radiation shift and determination of the clock
  frequency},\ }\href {https://doi.org/10.1103/PhysRevLett.131.083002}
  {\bibfield  {journal} {\bibinfo  {journal} {Phys. Rev. Lett.}\ }\textbf
  {\bibinfo {volume} {131}},\ \bibinfo {pages} {083002} (\bibinfo {year}
  {2023})}\BibitemShut {NoStop}%
\bibitem [{\citenamefont {Roberts}\ \emph {et~al.}(2020)\citenamefont
  {Roberts}, \citenamefont {Delva}, \citenamefont {Al-Masoudi}, \citenamefont
  {Amy-Klein}, \citenamefont {Baerentsen}, \citenamefont {Baynham},
  \citenamefont {Benkler}, \citenamefont {Bilicki}, \citenamefont {Bize},
  \citenamefont {Bowden} \emph {et~al.}}]{roberts2020search}%
  \BibitemOpen
  \bibfield  {author} {\bibinfo {author} {\bibfnamefont {B.~M.}\ \bibnamefont
  {Roberts}}, \bibinfo {author} {\bibfnamefont {P.}~\bibnamefont {Delva}},
  \bibinfo {author} {\bibfnamefont {A.}~\bibnamefont {Al-Masoudi}}, \bibinfo
  {author} {\bibfnamefont {A.}~\bibnamefont {Amy-Klein}}, \bibinfo {author}
  {\bibfnamefont {C.}~\bibnamefont {Baerentsen}}, \bibinfo {author}
  {\bibfnamefont {C.}~\bibnamefont {Baynham}}, \bibinfo {author} {\bibfnamefont
  {E.}~\bibnamefont {Benkler}}, \bibinfo {author} {\bibfnamefont
  {S.}~\bibnamefont {Bilicki}}, \bibinfo {author} {\bibfnamefont
  {S.}~\bibnamefont {Bize}}, \bibinfo {author} {\bibfnamefont {W.}~\bibnamefont
  {Bowden}}, \emph {et~al.},\ }\bibfield  {title} {\bibinfo {title} {Search for
  transient variations of the fine structure constant and dark matter using
  fiber-linked optical atomic clocks},\ }\href@noop {} {\bibfield  {journal}
  {\bibinfo  {journal} {New Journal of Physics}\ }\textbf {\bibinfo {volume}
  {22}},\ \bibinfo {pages} {093010} (\bibinfo {year} {2020})}\BibitemShut
  {NoStop}%
\bibitem [{\citenamefont {Robinson}\ \emph {et~al.}(2024)\citenamefont
  {Robinson}, \citenamefont {Miklos}, \citenamefont {Tso}, \citenamefont
  {Kennedy}, \citenamefont {Bothwell}, \citenamefont {Kedar}, \citenamefont
  {Thompson},\ and\ \citenamefont {Ye}}]{robinson2024direct}%
  \BibitemOpen
  \bibfield  {author} {\bibinfo {author} {\bibfnamefont {J.~M.}\ \bibnamefont
  {Robinson}}, \bibinfo {author} {\bibfnamefont {M.}~\bibnamefont {Miklos}},
  \bibinfo {author} {\bibfnamefont {Y.~M.}\ \bibnamefont {Tso}}, \bibinfo
  {author} {\bibfnamefont {C.~J.}\ \bibnamefont {Kennedy}}, \bibinfo {author}
  {\bibfnamefont {T.}~\bibnamefont {Bothwell}}, \bibinfo {author}
  {\bibfnamefont {D.}~\bibnamefont {Kedar}}, \bibinfo {author} {\bibfnamefont
  {J.~K.}\ \bibnamefont {Thompson}},\ and\ \bibinfo {author} {\bibfnamefont
  {J.}~\bibnamefont {Ye}},\ }\bibfield  {title} {\bibinfo {title} {Direct
  comparison of two spin-squeezed optical clock ensembles at the 10- 17
  level},\ }\href@noop {} {\bibfield  {journal} {\bibinfo  {journal} {Nature
  Physics}\ }\textbf {\bibinfo {volume} {20}},\ \bibinfo {pages} {208}
  (\bibinfo {year} {2024})}\BibitemShut {NoStop}%
\bibitem [{\citenamefont {Zhang}\ \emph {et~al.}(2024)\citenamefont {Zhang},
  \citenamefont {Ooi}, \citenamefont {Higgins}, \citenamefont {Doyle},
  \citenamefont {von~der Wense}, \citenamefont {Beeks}, \citenamefont
  {Leitner}, \citenamefont {Kazakov}, \citenamefont {Li}, \citenamefont
  {Thirolf} \emph {et~al.}}]{zhang2024frequency}%
  \BibitemOpen
  \bibfield  {author} {\bibinfo {author} {\bibfnamefont {C.}~\bibnamefont
  {Zhang}}, \bibinfo {author} {\bibfnamefont {T.}~\bibnamefont {Ooi}}, \bibinfo
  {author} {\bibfnamefont {J.~S.}\ \bibnamefont {Higgins}}, \bibinfo {author}
  {\bibfnamefont {J.~F.}\ \bibnamefont {Doyle}}, \bibinfo {author}
  {\bibfnamefont {L.}~\bibnamefont {von~der Wense}}, \bibinfo {author}
  {\bibfnamefont {K.}~\bibnamefont {Beeks}}, \bibinfo {author} {\bibfnamefont
  {A.}~\bibnamefont {Leitner}}, \bibinfo {author} {\bibfnamefont {G.~A.}\
  \bibnamefont {Kazakov}}, \bibinfo {author} {\bibfnamefont {P.}~\bibnamefont
  {Li}}, \bibinfo {author} {\bibfnamefont {P.~G.}\ \bibnamefont {Thirolf}},
  \emph {et~al.},\ }\bibfield  {title} {\bibinfo {title} {Frequency ratio of
  the 229mth nuclear isomeric transition and the 87sr atomic clock},\
  }\href@noop {} {\bibfield  {journal} {\bibinfo  {journal} {Nature}\ }\textbf
  {\bibinfo {volume} {633}},\ \bibinfo {pages} {63} (\bibinfo {year}
  {2024})}\BibitemShut {NoStop}%
\bibitem [{\citenamefont {King}\ \emph {et~al.}(2022)\citenamefont {King},
  \citenamefont {Spie{\ss}}, \citenamefont {Micke}, \citenamefont {Wilzewski},
  \citenamefont {Leopold}, \citenamefont {Benkler}, \citenamefont {Lange},
  \citenamefont {Huntemann}, \citenamefont {Surzhykov}, \citenamefont
  {Yerokhin} \emph {et~al.}}]{king2022optical}%
  \BibitemOpen
  \bibfield  {author} {\bibinfo {author} {\bibfnamefont {S.~A.}\ \bibnamefont
  {King}}, \bibinfo {author} {\bibfnamefont {L.~J.}\ \bibnamefont {Spie{\ss}}},
  \bibinfo {author} {\bibfnamefont {P.}~\bibnamefont {Micke}}, \bibinfo
  {author} {\bibfnamefont {A.}~\bibnamefont {Wilzewski}}, \bibinfo {author}
  {\bibfnamefont {T.}~\bibnamefont {Leopold}}, \bibinfo {author} {\bibfnamefont
  {E.}~\bibnamefont {Benkler}}, \bibinfo {author} {\bibfnamefont
  {R.}~\bibnamefont {Lange}}, \bibinfo {author} {\bibfnamefont
  {N.}~\bibnamefont {Huntemann}}, \bibinfo {author} {\bibfnamefont
  {A.}~\bibnamefont {Surzhykov}}, \bibinfo {author} {\bibfnamefont {V.~A.}\
  \bibnamefont {Yerokhin}}, \emph {et~al.},\ }\bibfield  {title} {\bibinfo
  {title} {An optical atomic clock based on a highly charged ion},\ }\href@noop
  {} {\bibfield  {journal} {\bibinfo  {journal} {Nature}\ }\textbf {\bibinfo
  {volume} {611}},\ \bibinfo {pages} {43} (\bibinfo {year} {2022})}\BibitemShut
  {NoStop}%
\bibitem [{\citenamefont {Ma}\ \emph {et~al.}(2024)\citenamefont {Ma},
  \citenamefont {Deng}, \citenamefont {Wang}, \citenamefont {Wei},
  \citenamefont {Hao}, \citenamefont {Zhang}, \citenamefont {Pang},
  \citenamefont {Wang}, \citenamefont {Wu}, \citenamefont {Liu}, \citenamefont
  {Yuan}, \citenamefont {Chang}, \citenamefont {Zhang}, \citenamefont {Wu},
  \citenamefont {Zhang},\ and\ \citenamefont {Lu}}]{PhysRevApplied.21.044017}%
  \BibitemOpen
  \bibfield  {author} {\bibinfo {author} {\bibfnamefont {Z.~Y.}\ \bibnamefont
  {Ma}}, \bibinfo {author} {\bibfnamefont {K.}~\bibnamefont {Deng}}, \bibinfo
  {author} {\bibfnamefont {Z.~Y.}\ \bibnamefont {Wang}}, \bibinfo {author}
  {\bibfnamefont {W.~Z.}\ \bibnamefont {Wei}}, \bibinfo {author} {\bibfnamefont
  {P.}~\bibnamefont {Hao}}, \bibinfo {author} {\bibfnamefont {H.~X.}\
  \bibnamefont {Zhang}}, \bibinfo {author} {\bibfnamefont {L.~R.}\ \bibnamefont
  {Pang}}, \bibinfo {author} {\bibfnamefont {B.}~\bibnamefont {Wang}}, \bibinfo
  {author} {\bibfnamefont {F.~F.}\ \bibnamefont {Wu}}, \bibinfo {author}
  {\bibfnamefont {H.~L.}\ \bibnamefont {Liu}}, \bibinfo {author} {\bibfnamefont
  {W.~H.}\ \bibnamefont {Yuan}}, \bibinfo {author} {\bibfnamefont {J.~L.}\
  \bibnamefont {Chang}}, \bibinfo {author} {\bibfnamefont {J.~X.}\ \bibnamefont
  {Zhang}}, \bibinfo {author} {\bibfnamefont {Q.~Y.}\ \bibnamefont {Wu}},
  \bibinfo {author} {\bibfnamefont {J.}~\bibnamefont {Zhang}},\ and\ \bibinfo
  {author} {\bibfnamefont {Z.~H.}\ \bibnamefont {Lu}},\ }\bibfield  {title}
  {\bibinfo {title} {Quantum-logic-based
  ${}^{25}$$\mathrm{Mg}$${}^{+}$-${}^{27}$$\mathrm{Al}$${}^{+}$ optical
  frequency standard for the redefinition of the si second},\ }\href
  {https://doi.org/10.1103/PhysRevApplied.21.044017} {\bibfield  {journal}
  {\bibinfo  {journal} {Phys. Rev. Appl.}\ }\textbf {\bibinfo {volume} {21}},\
  \bibinfo {pages} {044017} (\bibinfo {year} {2024})}\BibitemShut {NoStop}%
\bibitem [{\citenamefont {Li}\ \emph {et~al.}(2024)\citenamefont {Li},
  \citenamefont {Cui}, \citenamefont {Jia}, \citenamefont {Kong}, \citenamefont
  {Yu}, \citenamefont {Zhu}, \citenamefont {Liu}, \citenamefont {Wang},
  \citenamefont {Zhang}, \citenamefont {Huang} \emph
  {et~al.}}]{li2024strontium}%
  \BibitemOpen
  \bibfield  {author} {\bibinfo {author} {\bibfnamefont {J.}~\bibnamefont
  {Li}}, \bibinfo {author} {\bibfnamefont {X.-Y.}\ \bibnamefont {Cui}},
  \bibinfo {author} {\bibfnamefont {Z.-P.}\ \bibnamefont {Jia}}, \bibinfo
  {author} {\bibfnamefont {D.-Q.}\ \bibnamefont {Kong}}, \bibinfo {author}
  {\bibfnamefont {H.-W.}\ \bibnamefont {Yu}}, \bibinfo {author} {\bibfnamefont
  {X.-Q.}\ \bibnamefont {Zhu}}, \bibinfo {author} {\bibfnamefont {X.-Y.}\
  \bibnamefont {Liu}}, \bibinfo {author} {\bibfnamefont {D.-Z.}\ \bibnamefont
  {Wang}}, \bibinfo {author} {\bibfnamefont {X.}~\bibnamefont {Zhang}},
  \bibinfo {author} {\bibfnamefont {X.-Y.}\ \bibnamefont {Huang}}, \emph
  {et~al.},\ }\bibfield  {title} {\bibinfo {title} {A strontium lattice clock
  with both stability and uncertainty below},\ }\href@noop {} {\bibfield
  {journal} {\bibinfo  {journal} {Metrologia}\ }\textbf {\bibinfo {volume}
  {61}},\ \bibinfo {pages} {015006} (\bibinfo {year} {2024})}\BibitemShut
  {NoStop}%
\bibitem [{\citenamefont {Zeng}\ \emph {et~al.}(2023)\citenamefont {Zeng},
  \citenamefont {Huang}, \citenamefont {Zhang}, \citenamefont {Hao},
  \citenamefont {Ma}, \citenamefont {Hu}, \citenamefont {Zhang}, \citenamefont
  {Chen}, \citenamefont {Wang}, \citenamefont {Guan},\ and\ \citenamefont
  {Gao}}]{PhysRevApplied.19.064004}%
  \BibitemOpen
  \bibfield  {author} {\bibinfo {author} {\bibfnamefont {M.}~\bibnamefont
  {Zeng}}, \bibinfo {author} {\bibfnamefont {Y.}~\bibnamefont {Huang}},
  \bibinfo {author} {\bibfnamefont {B.}~\bibnamefont {Zhang}}, \bibinfo
  {author} {\bibfnamefont {Y.}~\bibnamefont {Hao}}, \bibinfo {author}
  {\bibfnamefont {Z.}~\bibnamefont {Ma}}, \bibinfo {author} {\bibfnamefont
  {R.}~\bibnamefont {Hu}}, \bibinfo {author} {\bibfnamefont {H.}~\bibnamefont
  {Zhang}}, \bibinfo {author} {\bibfnamefont {Z.}~\bibnamefont {Chen}},
  \bibinfo {author} {\bibfnamefont {M.}~\bibnamefont {Wang}}, \bibinfo {author}
  {\bibfnamefont {H.}~\bibnamefont {Guan}},\ and\ \bibinfo {author}
  {\bibfnamefont {K.}~\bibnamefont {Gao}},\ }\bibfield  {title} {\bibinfo
  {title} {Toward a transportable ${\mathrm{ca}}^{+}$ optical clock with a
  systematic uncertainty of 4.8 \ifmmode\times\else\texttimes\fi{}
  10\ensuremath{-}18},\ }\href
  {https://doi.org/10.1103/PhysRevApplied.19.064004} {\bibfield  {journal}
  {\bibinfo  {journal} {Phys. Rev. Appl.}\ }\textbf {\bibinfo {volume} {19}},\
  \bibinfo {pages} {064004} (\bibinfo {year} {2023})}\BibitemShut {NoStop}%
\bibitem [{\citenamefont {Hohensee}\ \emph
  {et~al.}(2013{\natexlab{b}})\citenamefont {Hohensee}, \citenamefont
  {M\"uller},\ and\ \citenamefont {Wiringa}}]{PhysRevLett.111.151102}%
  \BibitemOpen
  \bibfield  {author} {\bibinfo {author} {\bibfnamefont {M.~A.}\ \bibnamefont
  {Hohensee}}, \bibinfo {author} {\bibfnamefont {H.}~\bibnamefont {M\"uller}},\
  and\ \bibinfo {author} {\bibfnamefont {R.~B.}\ \bibnamefont {Wiringa}},\
  }\bibfield  {title} {\bibinfo {title} {Equivalence principle and bound
  kinetic energy},\ }\href {https://doi.org/10.1103/PhysRevLett.111.151102}
  {\bibfield  {journal} {\bibinfo  {journal} {Phys. Rev. Lett.}\ }\textbf
  {\bibinfo {volume} {111}},\ \bibinfo {pages} {151102} (\bibinfo {year}
  {2013}{\natexlab{b}})}\BibitemShut {NoStop}%
\bibitem [{\citenamefont {Hohensee}\ \emph {et~al.}(2011)\citenamefont
  {Hohensee}, \citenamefont {Chu}, \citenamefont {Peters},\ and\ \citenamefont
  {M\"uller}}]{PhysRevLett.106.151102}%
  \BibitemOpen
  \bibfield  {author} {\bibinfo {author} {\bibfnamefont {M.~A.}\ \bibnamefont
  {Hohensee}}, \bibinfo {author} {\bibfnamefont {S.}~\bibnamefont {Chu}},
  \bibinfo {author} {\bibfnamefont {A.}~\bibnamefont {Peters}},\ and\ \bibinfo
  {author} {\bibfnamefont {H.}~\bibnamefont {M\"uller}},\ }\bibfield  {title}
  {\bibinfo {title} {Equivalence principle and gravitational redshift},\ }\href
  {https://doi.org/10.1103/PhysRevLett.106.151102} {\bibfield  {journal}
  {\bibinfo  {journal} {Phys. Rev. Lett.}\ }\textbf {\bibinfo {volume} {106}},\
  \bibinfo {pages} {151102} (\bibinfo {year} {2011})}\BibitemShut {NoStop}%
\bibitem [{\citenamefont {Dzuba}\ and\ \citenamefont
  {Flambaum}(2017)}]{PhysRevD.95.015019}%
  \BibitemOpen
  \bibfield  {author} {\bibinfo {author} {\bibfnamefont {V.~A.}\ \bibnamefont
  {Dzuba}}\ and\ \bibinfo {author} {\bibfnamefont {V.~V.}\ \bibnamefont
  {Flambaum}},\ }\bibfield  {title} {\bibinfo {title} {Limits on gravitational
  einstein equivalence principle violation from monitoring atomic clock
  frequencies during a year},\ }\href
  {https://doi.org/10.1103/PhysRevD.95.015019} {\bibfield  {journal} {\bibinfo
  {journal} {Phys. Rev. D}\ }\textbf {\bibinfo {volume} {95}},\ \bibinfo
  {pages} {015019} (\bibinfo {year} {2017})}\BibitemShut {NoStop}%
\bibitem [{\citenamefont {Kosteleck\'y}\ and\ \citenamefont
  {Tasson}(2011)}]{PhysRevD.83.016013}%
  \BibitemOpen
  \bibfield  {author} {\bibinfo {author} {\bibfnamefont {V.~A.}\ \bibnamefont
  {Kosteleck\'y}}\ and\ \bibinfo {author} {\bibfnamefont {J.~D.}\ \bibnamefont
  {Tasson}},\ }\bibfield  {title} {\bibinfo {title} {Matter-gravity couplings
  and lorentz violation},\ }\href {https://doi.org/10.1103/PhysRevD.83.016013}
  {\bibfield  {journal} {\bibinfo  {journal} {Phys. Rev. D}\ }\textbf {\bibinfo
  {volume} {83}},\ \bibinfo {pages} {016013} (\bibinfo {year}
  {2011})}\BibitemShut {NoStop}%
\bibitem [{\citenamefont {Colladay}\ and\ \citenamefont
  {Kosteleck\'y}(1997)}]{PhysRevD.55.6760}%
  \BibitemOpen
  \bibfield  {author} {\bibinfo {author} {\bibfnamefont {D.}~\bibnamefont
  {Colladay}}\ and\ \bibinfo {author} {\bibfnamefont {V.~A.}\ \bibnamefont
  {Kosteleck\'y}},\ }\bibfield  {title} {\bibinfo {title} {$\mathrm{CPT}$
  violation and the standard model},\ }\href
  {https://doi.org/10.1103/PhysRevD.55.6760} {\bibfield  {journal} {\bibinfo
  {journal} {Phys. Rev. D}\ }\textbf {\bibinfo {volume} {55}},\ \bibinfo
  {pages} {6760} (\bibinfo {year} {1997})}\BibitemShut {NoStop}%
\bibitem [{\citenamefont {Takamoto}\ \emph {et~al.}(2020)\citenamefont
  {Takamoto}, \citenamefont {Ushijima}, \citenamefont {Ohmae}, \citenamefont
  {Yahagi}, \citenamefont {Kokado}, \citenamefont {Shinkai},\ and\
  \citenamefont {Katori}}]{takamoto2020test}%
  \BibitemOpen
  \bibfield  {author} {\bibinfo {author} {\bibfnamefont {M.}~\bibnamefont
  {Takamoto}}, \bibinfo {author} {\bibfnamefont {I.}~\bibnamefont {Ushijima}},
  \bibinfo {author} {\bibfnamefont {N.}~\bibnamefont {Ohmae}}, \bibinfo
  {author} {\bibfnamefont {T.}~\bibnamefont {Yahagi}}, \bibinfo {author}
  {\bibfnamefont {K.}~\bibnamefont {Kokado}}, \bibinfo {author} {\bibfnamefont
  {H.}~\bibnamefont {Shinkai}},\ and\ \bibinfo {author} {\bibfnamefont
  {H.}~\bibnamefont {Katori}},\ }\bibfield  {title} {\bibinfo {title} {Test of
  general relativity by a pair of transportable optical lattice clocks},\
  }\href@noop {} {\bibfield  {journal} {\bibinfo  {journal} {Nature photonics}\
  }\textbf {\bibinfo {volume} {14}},\ \bibinfo {pages} {411} (\bibinfo {year}
  {2020})}\BibitemShut {NoStop}%
\bibitem [{\citenamefont {Delva}\ \emph {et~al.}(2018)\citenamefont {Delva},
  \citenamefont {Puchades}, \citenamefont {Sch\"onemann}, \citenamefont
  {Dilssner}, \citenamefont {Courde}, \citenamefont {Bertone}, \citenamefont
  {Gonzalez}, \citenamefont {Hees}, \citenamefont {Le~Poncin-Lafitte},
  \citenamefont {Meynadier}, \citenamefont {Prieto-Cerdeira}, \citenamefont
  {Sohet}, \citenamefont {Ventura-Traveset},\ and\ \citenamefont
  {Wolf}}]{PhysRevLett.121.231101}%
  \BibitemOpen
  \bibfield  {author} {\bibinfo {author} {\bibfnamefont {P.}~\bibnamefont
  {Delva}}, \bibinfo {author} {\bibfnamefont {N.}~\bibnamefont {Puchades}},
  \bibinfo {author} {\bibfnamefont {E.}~\bibnamefont {Sch\"onemann}}, \bibinfo
  {author} {\bibfnamefont {F.}~\bibnamefont {Dilssner}}, \bibinfo {author}
  {\bibfnamefont {C.}~\bibnamefont {Courde}}, \bibinfo {author} {\bibfnamefont
  {S.}~\bibnamefont {Bertone}}, \bibinfo {author} {\bibfnamefont
  {F.}~\bibnamefont {Gonzalez}}, \bibinfo {author} {\bibfnamefont
  {A.}~\bibnamefont {Hees}}, \bibinfo {author} {\bibfnamefont {C.}~\bibnamefont
  {Le~Poncin-Lafitte}}, \bibinfo {author} {\bibfnamefont {F.}~\bibnamefont
  {Meynadier}}, \bibinfo {author} {\bibfnamefont {R.}~\bibnamefont
  {Prieto-Cerdeira}}, \bibinfo {author} {\bibfnamefont {B.}~\bibnamefont
  {Sohet}}, \bibinfo {author} {\bibfnamefont {J.}~\bibnamefont
  {Ventura-Traveset}},\ and\ \bibinfo {author} {\bibfnamefont {P.}~\bibnamefont
  {Wolf}},\ }\bibfield  {title} {\bibinfo {title} {Gravitational redshift test
  using eccentric galileo satellites},\ }\href
  {https://doi.org/10.1103/PhysRevLett.121.231101} {\bibfield  {journal}
  {\bibinfo  {journal} {Phys. Rev. Lett.}\ }\textbf {\bibinfo {volume} {121}},\
  \bibinfo {pages} {231101} (\bibinfo {year} {2018})}\BibitemShut {NoStop}%
\bibitem [{\citenamefont {Herrmann}\ \emph {et~al.}(2018)\citenamefont
  {Herrmann}, \citenamefont {Finke}, \citenamefont {L{\"u}lf}, \citenamefont
  {Kichakova}, \citenamefont {Puetzfeld}, \citenamefont {Knickmann},
  \citenamefont {List}, \citenamefont {Rievers}, \citenamefont {Giorgi},
  \citenamefont {G{\"u}nther} \emph {et~al.}}]{herrmann2018test}%
  \BibitemOpen
  \bibfield  {author} {\bibinfo {author} {\bibfnamefont {S.}~\bibnamefont
  {Herrmann}}, \bibinfo {author} {\bibfnamefont {F.}~\bibnamefont {Finke}},
  \bibinfo {author} {\bibfnamefont {M.}~\bibnamefont {L{\"u}lf}}, \bibinfo
  {author} {\bibfnamefont {O.}~\bibnamefont {Kichakova}}, \bibinfo {author}
  {\bibfnamefont {D.}~\bibnamefont {Puetzfeld}}, \bibinfo {author}
  {\bibfnamefont {D.}~\bibnamefont {Knickmann}}, \bibinfo {author}
  {\bibfnamefont {M.}~\bibnamefont {List}}, \bibinfo {author} {\bibfnamefont
  {B.}~\bibnamefont {Rievers}}, \bibinfo {author} {\bibfnamefont
  {G.}~\bibnamefont {Giorgi}}, \bibinfo {author} {\bibfnamefont
  {C.}~\bibnamefont {G{\"u}nther}}, \emph {et~al.},\ }\bibfield  {title}
  {\bibinfo {title} {Test of the gravitational redshift with galileo satellites
  in an eccentric orbit},\ }\href@noop {} {\bibfield  {journal} {\bibinfo
  {journal} {Physical review letters}\ }\textbf {\bibinfo {volume} {121}},\
  \bibinfo {pages} {231102} (\bibinfo {year} {2018})}\BibitemShut {NoStop}%
\bibitem [{\citenamefont {Savalle}\ \emph {et~al.}(2019)\citenamefont
  {Savalle}, \citenamefont {Guerlin}, \citenamefont {Delva}, \citenamefont
  {Meynadier}, \citenamefont {le~Poncin-Lafitte},\ and\ \citenamefont
  {Wolf}}]{savalle2019gravitational}%
  \BibitemOpen
  \bibfield  {author} {\bibinfo {author} {\bibfnamefont {E.}~\bibnamefont
  {Savalle}}, \bibinfo {author} {\bibfnamefont {C.}~\bibnamefont {Guerlin}},
  \bibinfo {author} {\bibfnamefont {P.}~\bibnamefont {Delva}}, \bibinfo
  {author} {\bibfnamefont {F.}~\bibnamefont {Meynadier}}, \bibinfo {author}
  {\bibfnamefont {C.}~\bibnamefont {le~Poncin-Lafitte}},\ and\ \bibinfo
  {author} {\bibfnamefont {P.}~\bibnamefont {Wolf}},\ }\bibfield  {title}
  {\bibinfo {title} {Gravitational redshift test with the future aces
  mission},\ }\href@noop {} {\bibfield  {journal} {\bibinfo  {journal}
  {Classical and Quantum Gravity}\ }\textbf {\bibinfo {volume} {36}},\ \bibinfo
  {pages} {245004} (\bibinfo {year} {2019})}\BibitemShut {NoStop}%
\bibitem [{\citenamefont {Qin}\ \emph {et~al.}(2024)\citenamefont {Qin},
  \citenamefont {Liu}, \citenamefont {Dai}, \citenamefont {Guo}, \citenamefont
  {Huang}, \citenamefont {Liu}, \citenamefont {Tan},\ and\ \citenamefont
  {Shao}}]{qin2024preliminary}%
  \BibitemOpen
  \bibfield  {author} {\bibinfo {author} {\bibfnamefont {C.-G.}\ \bibnamefont
  {Qin}}, \bibinfo {author} {\bibfnamefont {T.}~\bibnamefont {Liu}}, \bibinfo
  {author} {\bibfnamefont {X.-Y.}\ \bibnamefont {Dai}}, \bibinfo {author}
  {\bibfnamefont {P.-B.}\ \bibnamefont {Guo}}, \bibinfo {author} {\bibfnamefont
  {W.}~\bibnamefont {Huang}}, \bibinfo {author} {\bibfnamefont {X.-P.}\
  \bibnamefont {Liu}}, \bibinfo {author} {\bibfnamefont {Y.-J.}\ \bibnamefont
  {Tan}},\ and\ \bibinfo {author} {\bibfnamefont {C.-G.}\ \bibnamefont
  {Shao}},\ }\bibfield  {title} {\bibinfo {title} {Preliminary sensitivity
  study for a gravitational redshift measurement with china’s lunar
  exploration project},\ }\href@noop {} {\bibfield  {journal} {\bibinfo
  {journal} {Classical and Quantum Gravity}\ }\textbf {\bibinfo {volume}
  {41}},\ \bibinfo {pages} {135006} (\bibinfo {year} {2024})}\BibitemShut
  {NoStop}%
\bibitem [{\citenamefont {Shen}\ \emph {et~al.}(2023)\citenamefont {Shen},
  \citenamefont {Zhang}, \citenamefont {Shen}, \citenamefont {Xu},
  \citenamefont {Sun}, \citenamefont {Ashry}, \citenamefont {Ruby},
  \citenamefont {Xu}, \citenamefont {Wu}, \citenamefont {Wu}, \citenamefont
  {Ning}, \citenamefont {Wang}, \citenamefont {Li},\ and\ \citenamefont
  {Cai}}]{PhysRevD.108.064031}%
  \BibitemOpen
  \bibfield  {author} {\bibinfo {author} {\bibfnamefont {W.}~\bibnamefont
  {Shen}}, \bibinfo {author} {\bibfnamefont {P.}~\bibnamefont {Zhang}},
  \bibinfo {author} {\bibfnamefont {Z.}~\bibnamefont {Shen}}, \bibinfo {author}
  {\bibfnamefont {R.}~\bibnamefont {Xu}}, \bibinfo {author} {\bibfnamefont
  {X.}~\bibnamefont {Sun}}, \bibinfo {author} {\bibfnamefont {M.}~\bibnamefont
  {Ashry}}, \bibinfo {author} {\bibfnamefont {A.}~\bibnamefont {Ruby}},
  \bibinfo {author} {\bibfnamefont {W.}~\bibnamefont {Xu}}, \bibinfo {author}
  {\bibfnamefont {K.}~\bibnamefont {Wu}}, \bibinfo {author} {\bibfnamefont
  {Y.}~\bibnamefont {Wu}}, \bibinfo {author} {\bibfnamefont {A.}~\bibnamefont
  {Ning}}, \bibinfo {author} {\bibfnamefont {L.}~\bibnamefont {Wang}}, \bibinfo
  {author} {\bibfnamefont {L.}~\bibnamefont {Li}},\ and\ \bibinfo {author}
  {\bibfnamefont {C.}~\bibnamefont {Cai}},\ }\bibfield  {title} {\bibinfo
  {title} {Testing gravitational redshift based on microwave frequency links
  onboard the china space station},\ }\href
  {https://doi.org/10.1103/PhysRevD.108.064031} {\bibfield  {journal} {\bibinfo
   {journal} {Phys. Rev. D}\ }\textbf {\bibinfo {volume} {108}},\ \bibinfo
  {pages} {064031} (\bibinfo {year} {2023})}\BibitemShut {NoStop}%
\bibitem [{\citenamefont {Derevianko}\ \emph {et~al.}(2022)\citenamefont
  {Derevianko}, \citenamefont {Gibble}, \citenamefont {Hollberg}, \citenamefont
  {Newbury}, \citenamefont {Oates}, \citenamefont {Safronova}, \citenamefont
  {Sinclair},\ and\ \citenamefont {Yu}}]{derevianko2022fundamental}%
  \BibitemOpen
  \bibfield  {author} {\bibinfo {author} {\bibfnamefont {A.}~\bibnamefont
  {Derevianko}}, \bibinfo {author} {\bibfnamefont {K.}~\bibnamefont {Gibble}},
  \bibinfo {author} {\bibfnamefont {L.}~\bibnamefont {Hollberg}}, \bibinfo
  {author} {\bibfnamefont {N.~R.}\ \bibnamefont {Newbury}}, \bibinfo {author}
  {\bibfnamefont {C.}~\bibnamefont {Oates}}, \bibinfo {author} {\bibfnamefont
  {M.~S.}\ \bibnamefont {Safronova}}, \bibinfo {author} {\bibfnamefont {L.~C.}\
  \bibnamefont {Sinclair}},\ and\ \bibinfo {author} {\bibfnamefont
  {N.}~\bibnamefont {Yu}},\ }\bibfield  {title} {\bibinfo {title} {Fundamental
  physics with a state-of-the-art optical clock in space},\ }\href@noop {}
  {\bibfield  {journal} {\bibinfo  {journal} {Quantum Science and Technology}\
  }\textbf {\bibinfo {volume} {7}},\ \bibinfo {pages} {044002} (\bibinfo {year}
  {2022})}\BibitemShut {NoStop}%
\bibitem [{\citenamefont {Gu\'ena}\ \emph {et~al.}(2012)\citenamefont
  {Gu\'ena}, \citenamefont {Abgrall}, \citenamefont {Rovera}, \citenamefont
  {Rosenbusch}, \citenamefont {Tobar}, \citenamefont {Laurent}, \citenamefont
  {Clairon},\ and\ \citenamefont {Bize}}]{PhysRevLett.109.080801}%
  \BibitemOpen
  \bibfield  {author} {\bibinfo {author} {\bibfnamefont {J.}~\bibnamefont
  {Gu\'ena}}, \bibinfo {author} {\bibfnamefont {M.}~\bibnamefont {Abgrall}},
  \bibinfo {author} {\bibfnamefont {D.}~\bibnamefont {Rovera}}, \bibinfo
  {author} {\bibfnamefont {P.}~\bibnamefont {Rosenbusch}}, \bibinfo {author}
  {\bibfnamefont {M.~E.}\ \bibnamefont {Tobar}}, \bibinfo {author}
  {\bibfnamefont {P.}~\bibnamefont {Laurent}}, \bibinfo {author} {\bibfnamefont
  {A.}~\bibnamefont {Clairon}},\ and\ \bibinfo {author} {\bibfnamefont
  {S.}~\bibnamefont {Bize}},\ }\bibfield  {title} {\bibinfo {title} {Improved
  tests of local position invariance using $^{87}\mathrm{Rb}$ and
  $^{133}\mathrm{Cs}$ fountains},\ }\href
  {https://doi.org/10.1103/PhysRevLett.109.080801} {\bibfield  {journal}
  {\bibinfo  {journal} {Phys. Rev. Lett.}\ }\textbf {\bibinfo {volume} {109}},\
  \bibinfo {pages} {080801} (\bibinfo {year} {2012})}\BibitemShut {NoStop}%
\bibitem [{\citenamefont {Peil}\ \emph {et~al.}(2013)\citenamefont {Peil},
  \citenamefont {Crane}, \citenamefont {Hanssen}, \citenamefont {Swanson},\
  and\ \citenamefont {Ekstrom}}]{PhysRevA.87.010102}%
  \BibitemOpen
  \bibfield  {author} {\bibinfo {author} {\bibfnamefont {S.}~\bibnamefont
  {Peil}}, \bibinfo {author} {\bibfnamefont {S.}~\bibnamefont {Crane}},
  \bibinfo {author} {\bibfnamefont {J.~L.}\ \bibnamefont {Hanssen}}, \bibinfo
  {author} {\bibfnamefont {T.~B.}\ \bibnamefont {Swanson}},\ and\ \bibinfo
  {author} {\bibfnamefont {C.~R.}\ \bibnamefont {Ekstrom}},\ }\bibfield
  {title} {\bibinfo {title} {Tests of local position invariance using
  continuously running atomic clocks},\ }\href
  {https://doi.org/10.1103/PhysRevA.87.010102} {\bibfield  {journal} {\bibinfo
  {journal} {Phys. Rev. A}\ }\textbf {\bibinfo {volume} {87}},\ \bibinfo
  {pages} {010102} (\bibinfo {year} {2013})}\BibitemShut {NoStop}%
\bibitem [{\citenamefont {Ashby}\ \emph {et~al.}(2018)\citenamefont {Ashby},
  \citenamefont {Parker},\ and\ \citenamefont {Patla}}]{ashby2018null}%
  \BibitemOpen
  \bibfield  {author} {\bibinfo {author} {\bibfnamefont {N.}~\bibnamefont
  {Ashby}}, \bibinfo {author} {\bibfnamefont {T.~E.}\ \bibnamefont {Parker}},\
  and\ \bibinfo {author} {\bibfnamefont {B.~R.}\ \bibnamefont {Patla}},\
  }\bibfield  {title} {\bibinfo {title} {A null test of general relativity
  based on a long-term comparison of atomic transition frequencies},\
  }\href@noop {} {\bibfield  {journal} {\bibinfo  {journal} {Nature Physics}\
  }\textbf {\bibinfo {volume} {14}},\ \bibinfo {pages} {822} (\bibinfo {year}
  {2018})}\BibitemShut {NoStop}%
\bibitem [{\citenamefont {Schwarz}\ \emph {et~al.}(2020)\citenamefont
  {Schwarz}, \citenamefont {D\"orscher}, \citenamefont {Al-Masoudi},
  \citenamefont {Benkler}, \citenamefont {Legero}, \citenamefont {Sterr},
  \citenamefont {Weyers}, \citenamefont {Rahm}, \citenamefont {Lipphardt},\
  and\ \citenamefont {Lisdat}}]{PhysRevResearch.2.033242}%
  \BibitemOpen
  \bibfield  {author} {\bibinfo {author} {\bibfnamefont {R.}~\bibnamefont
  {Schwarz}}, \bibinfo {author} {\bibfnamefont {S.}~\bibnamefont {D\"orscher}},
  \bibinfo {author} {\bibfnamefont {A.}~\bibnamefont {Al-Masoudi}}, \bibinfo
  {author} {\bibfnamefont {E.}~\bibnamefont {Benkler}}, \bibinfo {author}
  {\bibfnamefont {T.}~\bibnamefont {Legero}}, \bibinfo {author} {\bibfnamefont
  {U.}~\bibnamefont {Sterr}}, \bibinfo {author} {\bibfnamefont
  {S.}~\bibnamefont {Weyers}}, \bibinfo {author} {\bibfnamefont
  {J.}~\bibnamefont {Rahm}}, \bibinfo {author} {\bibfnamefont {B.}~\bibnamefont
  {Lipphardt}},\ and\ \bibinfo {author} {\bibfnamefont {C.}~\bibnamefont
  {Lisdat}},\ }\bibfield  {title} {\bibinfo {title} {Long term measurement of
  the $^{87}\mathrm{Sr}$ clock frequency at the limit of primary cs clocks},\
  }\href {https://doi.org/10.1103/PhysRevResearch.2.033242} {\bibfield
  {journal} {\bibinfo  {journal} {Phys. Rev. Res.}\ }\textbf {\bibinfo {volume}
  {2}},\ \bibinfo {pages} {033242} (\bibinfo {year} {2020})}\BibitemShut
  {NoStop}%
\bibitem [{\citenamefont {Fortier}\ \emph {et~al.}(2007)\citenamefont
  {Fortier}, \citenamefont {Ashby}, \citenamefont {Bergquist}, \citenamefont
  {Delaney}, \citenamefont {Diddams}, \citenamefont {Heavner}, \citenamefont
  {Hollberg}, \citenamefont {Itano}, \citenamefont {Jefferts}, \citenamefont
  {Kim}, \citenamefont {Levi}, \citenamefont {Lorini}, \citenamefont {Oskay},
  \citenamefont {Parker}, \citenamefont {Shirley},\ and\ \citenamefont
  {Stalnaker}}]{PhysRevLett.98.070801}%
  \BibitemOpen
  \bibfield  {author} {\bibinfo {author} {\bibfnamefont {T.~M.}\ \bibnamefont
  {Fortier}}, \bibinfo {author} {\bibfnamefont {N.}~\bibnamefont {Ashby}},
  \bibinfo {author} {\bibfnamefont {J.~C.}\ \bibnamefont {Bergquist}}, \bibinfo
  {author} {\bibfnamefont {M.~J.}\ \bibnamefont {Delaney}}, \bibinfo {author}
  {\bibfnamefont {S.~A.}\ \bibnamefont {Diddams}}, \bibinfo {author}
  {\bibfnamefont {T.~P.}\ \bibnamefont {Heavner}}, \bibinfo {author}
  {\bibfnamefont {L.}~\bibnamefont {Hollberg}}, \bibinfo {author}
  {\bibfnamefont {W.~M.}\ \bibnamefont {Itano}}, \bibinfo {author}
  {\bibfnamefont {S.~R.}\ \bibnamefont {Jefferts}}, \bibinfo {author}
  {\bibfnamefont {K.}~\bibnamefont {Kim}}, \bibinfo {author} {\bibfnamefont
  {F.}~\bibnamefont {Levi}}, \bibinfo {author} {\bibfnamefont {L.}~\bibnamefont
  {Lorini}}, \bibinfo {author} {\bibfnamefont {W.~H.}\ \bibnamefont {Oskay}},
  \bibinfo {author} {\bibfnamefont {T.~E.}\ \bibnamefont {Parker}}, \bibinfo
  {author} {\bibfnamefont {J.}~\bibnamefont {Shirley}},\ and\ \bibinfo {author}
  {\bibfnamefont {J.~E.}\ \bibnamefont {Stalnaker}},\ }\bibfield  {title}
  {\bibinfo {title} {Precision atomic spectroscopy for improved limits on
  variation of the fine structure constant and local position invariance},\
  }\href {https://doi.org/10.1103/PhysRevLett.98.070801} {\bibfield  {journal}
  {\bibinfo  {journal} {Phys. Rev. Lett.}\ }\textbf {\bibinfo {volume} {98}},\
  \bibinfo {pages} {070801} (\bibinfo {year} {2007})}\BibitemShut {NoStop}%
\bibitem [{\citenamefont {Goti}\ \emph {et~al.}(2023)\citenamefont {Goti},
  \citenamefont {Condio}, \citenamefont {Clivati}, \citenamefont {Risaro},
  \citenamefont {Gozzelino}, \citenamefont {Costanzo}, \citenamefont {Levi},
  \citenamefont {Calonico},\ and\ \citenamefont
  {Pizzocaro}}]{goti2023absolute}%
  \BibitemOpen
  \bibfield  {author} {\bibinfo {author} {\bibfnamefont {I.}~\bibnamefont
  {Goti}}, \bibinfo {author} {\bibfnamefont {S.}~\bibnamefont {Condio}},
  \bibinfo {author} {\bibfnamefont {C.}~\bibnamefont {Clivati}}, \bibinfo
  {author} {\bibfnamefont {M.}~\bibnamefont {Risaro}}, \bibinfo {author}
  {\bibfnamefont {M.}~\bibnamefont {Gozzelino}}, \bibinfo {author}
  {\bibfnamefont {G.~A.}\ \bibnamefont {Costanzo}}, \bibinfo {author}
  {\bibfnamefont {F.}~\bibnamefont {Levi}}, \bibinfo {author} {\bibfnamefont
  {D.}~\bibnamefont {Calonico}},\ and\ \bibinfo {author} {\bibfnamefont
  {M.}~\bibnamefont {Pizzocaro}},\ }\bibfield  {title} {\bibinfo {title}
  {Absolute frequency measurement of a yb optical clock at the limit of the cs
  fountain},\ }\href@noop {} {\bibfield  {journal} {\bibinfo  {journal}
  {Metrologia}\ }\textbf {\bibinfo {volume} {60}},\ \bibinfo {pages} {035002}
  (\bibinfo {year} {2023})}\BibitemShut {NoStop}%
\bibitem [{\citenamefont {McGrew}\ \emph {et~al.}(2019)\citenamefont {McGrew},
  \citenamefont {Zhang}, \citenamefont {Leopardi}, \citenamefont {Fasano},
  \citenamefont {Nicolodi}, \citenamefont {Beloy}, \citenamefont {Yao},
  \citenamefont {Sherman}, \citenamefont {Schaeffer}, \citenamefont {Savory}
  \emph {et~al.}}]{mcgrew2019towards}%
  \BibitemOpen
  \bibfield  {author} {\bibinfo {author} {\bibfnamefont {W.~F.}\ \bibnamefont
  {McGrew}}, \bibinfo {author} {\bibfnamefont {X.}~\bibnamefont {Zhang}},
  \bibinfo {author} {\bibfnamefont {H.}~\bibnamefont {Leopardi}}, \bibinfo
  {author} {\bibfnamefont {R.}~\bibnamefont {Fasano}}, \bibinfo {author}
  {\bibfnamefont {D.}~\bibnamefont {Nicolodi}}, \bibinfo {author}
  {\bibfnamefont {K.}~\bibnamefont {Beloy}}, \bibinfo {author} {\bibfnamefont
  {J.}~\bibnamefont {Yao}}, \bibinfo {author} {\bibfnamefont {J.~A.}\
  \bibnamefont {Sherman}}, \bibinfo {author} {\bibfnamefont {S.~A.}\
  \bibnamefont {Schaeffer}}, \bibinfo {author} {\bibfnamefont {J.}~\bibnamefont
  {Savory}}, \emph {et~al.},\ }\bibfield  {title} {\bibinfo {title} {Towards
  the optical second: verifying optical clocks at the si limit},\ }\href@noop
  {} {\bibfield  {journal} {\bibinfo  {journal} {Optica}\ }\textbf {\bibinfo
  {volume} {6}},\ \bibinfo {pages} {448} (\bibinfo {year} {2019})}\BibitemShut
  {NoStop}%
\bibitem [{\citenamefont {Huntemann}\ \emph {et~al.}(2014)\citenamefont
  {Huntemann}, \citenamefont {Lipphardt}, \citenamefont {Tamm}, \citenamefont
  {Gerginov}, \citenamefont {Weyers},\ and\ \citenamefont
  {Peik}}]{PhysRevLett.113.210802}%
  \BibitemOpen
  \bibfield  {author} {\bibinfo {author} {\bibfnamefont {N.}~\bibnamefont
  {Huntemann}}, \bibinfo {author} {\bibfnamefont {B.}~\bibnamefont
  {Lipphardt}}, \bibinfo {author} {\bibfnamefont {C.}~\bibnamefont {Tamm}},
  \bibinfo {author} {\bibfnamefont {V.}~\bibnamefont {Gerginov}}, \bibinfo
  {author} {\bibfnamefont {S.}~\bibnamefont {Weyers}},\ and\ \bibinfo {author}
  {\bibfnamefont {E.}~\bibnamefont {Peik}},\ }\bibfield  {title} {\bibinfo
  {title} {Improved limit on a temporal variation of ${m}_{p}/{m}_{e}$ from
  comparisons of ${\mathrm{yb}}^{+}$ and cs atomic clocks},\ }\href
  {https://doi.org/10.1103/PhysRevLett.113.210802} {\bibfield  {journal}
  {\bibinfo  {journal} {Phys. Rev. Lett.}\ }\textbf {\bibinfo {volume} {113}},\
  \bibinfo {pages} {210802} (\bibinfo {year} {2014})}\BibitemShut {NoStop}%
\bibitem [{\citenamefont {Rosenband}\ \emph {et~al.}(2008)\citenamefont
  {Rosenband}, \citenamefont {Hume}, \citenamefont {Schmidt}, \citenamefont
  {Chou}, \citenamefont {Brusch}, \citenamefont {Lorini}, \citenamefont
  {Oskay}, \citenamefont {Drullinger}, \citenamefont {Fortier}, \citenamefont
  {Stalnaker} \emph {et~al.}}]{rosenband2008frequency}%
  \BibitemOpen
  \bibfield  {author} {\bibinfo {author} {\bibfnamefont {T.}~\bibnamefont
  {Rosenband}}, \bibinfo {author} {\bibfnamefont {D.}~\bibnamefont {Hume}},
  \bibinfo {author} {\bibfnamefont {P.}~\bibnamefont {Schmidt}}, \bibinfo
  {author} {\bibfnamefont {C.-W.}\ \bibnamefont {Chou}}, \bibinfo {author}
  {\bibfnamefont {A.}~\bibnamefont {Brusch}}, \bibinfo {author} {\bibfnamefont
  {L.}~\bibnamefont {Lorini}}, \bibinfo {author} {\bibfnamefont
  {W.}~\bibnamefont {Oskay}}, \bibinfo {author} {\bibfnamefont {R.~E.}\
  \bibnamefont {Drullinger}}, \bibinfo {author} {\bibfnamefont {T.~M.}\
  \bibnamefont {Fortier}}, \bibinfo {author} {\bibfnamefont {J.~E.}\
  \bibnamefont {Stalnaker}}, \emph {et~al.},\ }\bibfield  {title} {\bibinfo
  {title} {Frequency ratio of al+ and hg+ single-ion optical clocks; metrology
  at the 17th decimal place},\ }\href@noop {} {\bibfield  {journal} {\bibinfo
  {journal} {Science}\ }\textbf {\bibinfo {volume} {319}},\ \bibinfo {pages}
  {1808} (\bibinfo {year} {2008})}\BibitemShut {NoStop}%
\bibitem [{\citenamefont {Lange}\ \emph
  {et~al.}(2021{\natexlab{b}})\citenamefont {Lange}, \citenamefont {Huntemann},
  \citenamefont {Rahm}, \citenamefont {Sanner}, \citenamefont {Shao},
  \citenamefont {Lipphardt}, \citenamefont {Tamm}, \citenamefont {Weyers},\
  and\ \citenamefont {Peik}}]{PhysRevLett.126.011102}%
  \BibitemOpen
  \bibfield  {author} {\bibinfo {author} {\bibfnamefont {R.}~\bibnamefont
  {Lange}}, \bibinfo {author} {\bibfnamefont {N.}~\bibnamefont {Huntemann}},
  \bibinfo {author} {\bibfnamefont {J.~M.}\ \bibnamefont {Rahm}}, \bibinfo
  {author} {\bibfnamefont {C.}~\bibnamefont {Sanner}}, \bibinfo {author}
  {\bibfnamefont {H.}~\bibnamefont {Shao}}, \bibinfo {author} {\bibfnamefont
  {B.}~\bibnamefont {Lipphardt}}, \bibinfo {author} {\bibfnamefont
  {C.}~\bibnamefont {Tamm}}, \bibinfo {author} {\bibfnamefont {S.}~\bibnamefont
  {Weyers}},\ and\ \bibinfo {author} {\bibfnamefont {E.}~\bibnamefont {Peik}},\
  }\bibfield  {title} {\bibinfo {title} {Improved limits for violations of
  local position invariance from atomic clock comparisons},\ }\href
  {https://doi.org/10.1103/PhysRevLett.126.011102} {\bibfield  {journal}
  {\bibinfo  {journal} {Phys. Rev. Lett.}\ }\textbf {\bibinfo {volume} {126}},\
  \bibinfo {pages} {011102} (\bibinfo {year} {2021}{\natexlab{b}})}\BibitemShut
  {NoStop}%
\bibitem [{\citenamefont {Filzinger}\ \emph {et~al.}(2023)\citenamefont
  {Filzinger}, \citenamefont {D\"orscher}, \citenamefont {Lange}, \citenamefont
  {Klose}, \citenamefont {Steinel}, \citenamefont {Benkler}, \citenamefont
  {Peik}, \citenamefont {Lisdat},\ and\ \citenamefont
  {Huntemann}}]{PhysRevLett.130.253001}%
  \BibitemOpen
  \bibfield  {author} {\bibinfo {author} {\bibfnamefont {M.}~\bibnamefont
  {Filzinger}}, \bibinfo {author} {\bibfnamefont {S.}~\bibnamefont
  {D\"orscher}}, \bibinfo {author} {\bibfnamefont {R.}~\bibnamefont {Lange}},
  \bibinfo {author} {\bibfnamefont {J.}~\bibnamefont {Klose}}, \bibinfo
  {author} {\bibfnamefont {M.}~\bibnamefont {Steinel}}, \bibinfo {author}
  {\bibfnamefont {E.}~\bibnamefont {Benkler}}, \bibinfo {author} {\bibfnamefont
  {E.}~\bibnamefont {Peik}}, \bibinfo {author} {\bibfnamefont {C.}~\bibnamefont
  {Lisdat}},\ and\ \bibinfo {author} {\bibfnamefont {N.}~\bibnamefont
  {Huntemann}},\ }\bibfield  {title} {\bibinfo {title} {Improved limits on the
  coupling of ultralight bosonic dark matter to photons from optical atomic
  clock comparisons},\ }\href {https://doi.org/10.1103/PhysRevLett.130.253001}
  {\bibfield  {journal} {\bibinfo  {journal} {Phys. Rev. Lett.}\ }\textbf
  {\bibinfo {volume} {130}},\ \bibinfo {pages} {253001} (\bibinfo {year}
  {2023})}\BibitemShut {NoStop}%
\bibitem [{\citenamefont {Godun}\ \emph {et~al.}(2014)\citenamefont {Godun},
  \citenamefont {Nisbet-Jones}, \citenamefont {Jones}, \citenamefont {King},
  \citenamefont {Johnson}, \citenamefont {Margolis}, \citenamefont {Szymaniec},
  \citenamefont {Lea}, \citenamefont {Bongs},\ and\ \citenamefont
  {Gill}}]{PhysRevLett.113.210801}%
  \BibitemOpen
  \bibfield  {author} {\bibinfo {author} {\bibfnamefont {R.~M.}\ \bibnamefont
  {Godun}}, \bibinfo {author} {\bibfnamefont {P.~B.~R.}\ \bibnamefont
  {Nisbet-Jones}}, \bibinfo {author} {\bibfnamefont {J.~M.}\ \bibnamefont
  {Jones}}, \bibinfo {author} {\bibfnamefont {S.~A.}\ \bibnamefont {King}},
  \bibinfo {author} {\bibfnamefont {L.~A.~M.}\ \bibnamefont {Johnson}},
  \bibinfo {author} {\bibfnamefont {H.~S.}\ \bibnamefont {Margolis}}, \bibinfo
  {author} {\bibfnamefont {K.}~\bibnamefont {Szymaniec}}, \bibinfo {author}
  {\bibfnamefont {S.~N.}\ \bibnamefont {Lea}}, \bibinfo {author} {\bibfnamefont
  {K.}~\bibnamefont {Bongs}},\ and\ \bibinfo {author} {\bibfnamefont
  {P.}~\bibnamefont {Gill}},\ }\bibfield  {title} {\bibinfo {title} {Frequency
  ratio of two optical clock transitions in $^{171}{\mathrm{yb}}^{+}$ and
  constraints on the time variation of fundamental constants},\ }\href
  {https://doi.org/10.1103/PhysRevLett.113.210801} {\bibfield  {journal}
  {\bibinfo  {journal} {Phys. Rev. Lett.}\ }\textbf {\bibinfo {volume} {113}},\
  \bibinfo {pages} {210801} (\bibinfo {year} {2014})}\BibitemShut {NoStop}%
\bibitem [{\citenamefont {Le~Targat}\ \emph {et~al.}(2013)\citenamefont
  {Le~Targat}, \citenamefont {Lorini}, \citenamefont {Le~Coq}, \citenamefont
  {Zawada}, \citenamefont {Gu{\'e}na}, \citenamefont {Abgrall}, \citenamefont
  {Gurov}, \citenamefont {Rosenbusch}, \citenamefont {Rovera}, \citenamefont
  {Nag{\'o}rny} \emph {et~al.}}]{le2013experimental}%
  \BibitemOpen
  \bibfield  {author} {\bibinfo {author} {\bibfnamefont {R.}~\bibnamefont
  {Le~Targat}}, \bibinfo {author} {\bibfnamefont {L.}~\bibnamefont {Lorini}},
  \bibinfo {author} {\bibfnamefont {Y.}~\bibnamefont {Le~Coq}}, \bibinfo
  {author} {\bibfnamefont {M.}~\bibnamefont {Zawada}}, \bibinfo {author}
  {\bibfnamefont {J.}~\bibnamefont {Gu{\'e}na}}, \bibinfo {author}
  {\bibfnamefont {M.}~\bibnamefont {Abgrall}}, \bibinfo {author} {\bibfnamefont
  {M.}~\bibnamefont {Gurov}}, \bibinfo {author} {\bibfnamefont
  {P.}~\bibnamefont {Rosenbusch}}, \bibinfo {author} {\bibfnamefont
  {D.}~\bibnamefont {Rovera}}, \bibinfo {author} {\bibfnamefont
  {B.}~\bibnamefont {Nag{\'o}rny}}, \emph {et~al.},\ }\bibfield  {title}
  {\bibinfo {title} {Experimental realization of an optical second with
  strontium lattice clocks},\ }\href@noop {} {\bibfield  {journal} {\bibinfo
  {journal} {Nature communications}\ }\textbf {\bibinfo {volume} {4}},\
  \bibinfo {pages} {2109} (\bibinfo {year} {2013})}\BibitemShut {NoStop}%
\bibitem [{\citenamefont {Falke}\ \emph {et~al.}(2014)\citenamefont {Falke},
  \citenamefont {Lemke}, \citenamefont {Grebing}, \citenamefont {Lipphardt},
  \citenamefont {Weyers}, \citenamefont {Gerginov}, \citenamefont {Huntemann},
  \citenamefont {Hagemann}, \citenamefont {Al-Masoudi}, \citenamefont
  {H{\"a}fner} \emph {et~al.}}]{falke2014strontium}%
  \BibitemOpen
  \bibfield  {author} {\bibinfo {author} {\bibfnamefont {S.}~\bibnamefont
  {Falke}}, \bibinfo {author} {\bibfnamefont {N.}~\bibnamefont {Lemke}},
  \bibinfo {author} {\bibfnamefont {C.}~\bibnamefont {Grebing}}, \bibinfo
  {author} {\bibfnamefont {B.}~\bibnamefont {Lipphardt}}, \bibinfo {author}
  {\bibfnamefont {S.}~\bibnamefont {Weyers}}, \bibinfo {author} {\bibfnamefont
  {V.}~\bibnamefont {Gerginov}}, \bibinfo {author} {\bibfnamefont
  {N.}~\bibnamefont {Huntemann}}, \bibinfo {author} {\bibfnamefont
  {C.}~\bibnamefont {Hagemann}}, \bibinfo {author} {\bibfnamefont
  {A.}~\bibnamefont {Al-Masoudi}}, \bibinfo {author} {\bibfnamefont
  {S.}~\bibnamefont {H{\"a}fner}}, \emph {et~al.},\ }\bibfield  {title}
  {\bibinfo {title} {A strontium lattice clock with inaccuracy and its
  frequency},\ }\href@noop {} {\bibfield  {journal} {\bibinfo  {journal} {New
  Journal of Physics}\ }\textbf {\bibinfo {volume} {16}},\ \bibinfo {pages}
  {073023} (\bibinfo {year} {2014})}\BibitemShut {NoStop}%
\bibitem [{\citenamefont {Hill}\ \emph {et~al.}(2016)\citenamefont {Hill},
  \citenamefont {Hobson}, \citenamefont {Bowden}, \citenamefont {Bridge},
  \citenamefont {Donnellan}, \citenamefont {Curtis},\ and\ \citenamefont
  {Gill}}]{hill2016low}%
  \BibitemOpen
  \bibfield  {author} {\bibinfo {author} {\bibfnamefont {I.~R.}\ \bibnamefont
  {Hill}}, \bibinfo {author} {\bibfnamefont {R.}~\bibnamefont {Hobson}},
  \bibinfo {author} {\bibfnamefont {W.}~\bibnamefont {Bowden}}, \bibinfo
  {author} {\bibfnamefont {E.~M.}\ \bibnamefont {Bridge}}, \bibinfo {author}
  {\bibfnamefont {S.}~\bibnamefont {Donnellan}}, \bibinfo {author}
  {\bibfnamefont {E.~A.}\ \bibnamefont {Curtis}},\ and\ \bibinfo {author}
  {\bibfnamefont {P.}~\bibnamefont {Gill}},\ }\bibfield  {title} {\bibinfo
  {title} {A low maintenance sr optical lattice clock},\ }in\ \href@noop {}
  {\emph {\bibinfo {booktitle} {Journal of Physics: Conference Series}}},\
  Vol.\ \bibinfo {volume} {723}\ (\bibinfo {organization} {IOP Publishing},\
  \bibinfo {year} {2016})\ p.\ \bibinfo {pages} {012019}\BibitemShut {NoStop}%
\bibitem [{\citenamefont {Lodewyck}\ \emph {et~al.}(2016)\citenamefont
  {Lodewyck}, \citenamefont {Bilicki}, \citenamefont {Bookjans}, \citenamefont
  {Robyr}, \citenamefont {Shi}, \citenamefont {Vallet}, \citenamefont
  {Le~Targat}, \citenamefont {Nicolodi}, \citenamefont {Le~Coq}, \citenamefont
  {Gu{\'e}na} \emph {et~al.}}]{lodewyck2016optical}%
  \BibitemOpen
  \bibfield  {author} {\bibinfo {author} {\bibfnamefont {J.}~\bibnamefont
  {Lodewyck}}, \bibinfo {author} {\bibfnamefont {S.}~\bibnamefont {Bilicki}},
  \bibinfo {author} {\bibfnamefont {E.}~\bibnamefont {Bookjans}}, \bibinfo
  {author} {\bibfnamefont {J.-L.}\ \bibnamefont {Robyr}}, \bibinfo {author}
  {\bibfnamefont {C.}~\bibnamefont {Shi}}, \bibinfo {author} {\bibfnamefont
  {G.}~\bibnamefont {Vallet}}, \bibinfo {author} {\bibfnamefont
  {R.}~\bibnamefont {Le~Targat}}, \bibinfo {author} {\bibfnamefont
  {D.}~\bibnamefont {Nicolodi}}, \bibinfo {author} {\bibfnamefont
  {Y.}~\bibnamefont {Le~Coq}}, \bibinfo {author} {\bibfnamefont
  {J.}~\bibnamefont {Gu{\'e}na}}, \emph {et~al.},\ }\bibfield  {title}
  {\bibinfo {title} {Optical to microwave clock frequency ratios with a nearly
  continuous strontium optical lattice clock},\ }\href@noop {} {\bibfield
  {journal} {\bibinfo  {journal} {Metrologia}\ }\textbf {\bibinfo {volume}
  {53}},\ \bibinfo {pages} {1123} (\bibinfo {year} {2016})}\BibitemShut
  {NoStop}%
\bibitem [{\citenamefont {Delva}\ \emph {et~al.}(2017)\citenamefont {Delva},
  \citenamefont {Lodewyck}, \citenamefont {Bilicki}, \citenamefont {Bookjans},
  \citenamefont {Vallet}, \citenamefont {Le~Targat}, \citenamefont {Pottie},
  \citenamefont {Guerlin}, \citenamefont {Meynadier}, \citenamefont
  {Le~Poncin-Lafitte}, \citenamefont {Lopez}, \citenamefont {Amy-Klein},
  \citenamefont {Lee}, \citenamefont {Quintin}, \citenamefont {Lisdat},
  \citenamefont {Al-Masoudi}, \citenamefont {D\"orscher}, \citenamefont
  {Grebing}, \citenamefont {Grosche}, \citenamefont {Kuhl}, \citenamefont
  {Raupach}, \citenamefont {Sterr}, \citenamefont {Hill}, \citenamefont
  {Hobson}, \citenamefont {Bowden}, \citenamefont {Kronj\"ager}, \citenamefont
  {Marra}, \citenamefont {Rolland}, \citenamefont {Baynes}, \citenamefont
  {Margolis},\ and\ \citenamefont {Gill}}]{PhysRevLett.118.221102}%
  \BibitemOpen
  \bibfield  {author} {\bibinfo {author} {\bibfnamefont {P.}~\bibnamefont
  {Delva}}, \bibinfo {author} {\bibfnamefont {J.}~\bibnamefont {Lodewyck}},
  \bibinfo {author} {\bibfnamefont {S.}~\bibnamefont {Bilicki}}, \bibinfo
  {author} {\bibfnamefont {E.}~\bibnamefont {Bookjans}}, \bibinfo {author}
  {\bibfnamefont {G.}~\bibnamefont {Vallet}}, \bibinfo {author} {\bibfnamefont
  {R.}~\bibnamefont {Le~Targat}}, \bibinfo {author} {\bibfnamefont {P.-E.}\
  \bibnamefont {Pottie}}, \bibinfo {author} {\bibfnamefont {C.}~\bibnamefont
  {Guerlin}}, \bibinfo {author} {\bibfnamefont {F.}~\bibnamefont {Meynadier}},
  \bibinfo {author} {\bibfnamefont {C.}~\bibnamefont {Le~Poncin-Lafitte}},
  \bibinfo {author} {\bibfnamefont {O.}~\bibnamefont {Lopez}}, \bibinfo
  {author} {\bibfnamefont {A.}~\bibnamefont {Amy-Klein}}, \bibinfo {author}
  {\bibfnamefont {W.-K.}\ \bibnamefont {Lee}}, \bibinfo {author} {\bibfnamefont
  {N.}~\bibnamefont {Quintin}}, \bibinfo {author} {\bibfnamefont
  {C.}~\bibnamefont {Lisdat}}, \bibinfo {author} {\bibfnamefont
  {A.}~\bibnamefont {Al-Masoudi}}, \bibinfo {author} {\bibfnamefont
  {S.}~\bibnamefont {D\"orscher}}, \bibinfo {author} {\bibfnamefont
  {C.}~\bibnamefont {Grebing}}, \bibinfo {author} {\bibfnamefont
  {G.}~\bibnamefont {Grosche}}, \bibinfo {author} {\bibfnamefont
  {A.}~\bibnamefont {Kuhl}}, \bibinfo {author} {\bibfnamefont {S.}~\bibnamefont
  {Raupach}}, \bibinfo {author} {\bibfnamefont {U.}~\bibnamefont {Sterr}},
  \bibinfo {author} {\bibfnamefont {I.~R.}\ \bibnamefont {Hill}}, \bibinfo
  {author} {\bibfnamefont {R.}~\bibnamefont {Hobson}}, \bibinfo {author}
  {\bibfnamefont {W.}~\bibnamefont {Bowden}}, \bibinfo {author} {\bibfnamefont
  {J.}~\bibnamefont {Kronj\"ager}}, \bibinfo {author} {\bibfnamefont
  {G.}~\bibnamefont {Marra}}, \bibinfo {author} {\bibfnamefont
  {A.}~\bibnamefont {Rolland}}, \bibinfo {author} {\bibfnamefont {F.~N.}\
  \bibnamefont {Baynes}}, \bibinfo {author} {\bibfnamefont {H.~S.}\
  \bibnamefont {Margolis}},\ and\ \bibinfo {author} {\bibfnamefont
  {P.}~\bibnamefont {Gill}},\ }\bibfield  {title} {\bibinfo {title} {Test of
  special relativity using a fiber network of optical clocks},\ }\href
  {https://doi.org/10.1103/PhysRevLett.118.221102} {\bibfield  {journal}
  {\bibinfo  {journal} {Phys. Rev. Lett.}\ }\textbf {\bibinfo {volume} {118}},\
  \bibinfo {pages} {221102} (\bibinfo {year} {2017})}\BibitemShut {NoStop}%
\bibitem [{\citenamefont {Schioppo}\ \emph {et~al.}(2022)\citenamefont
  {Schioppo}, \citenamefont {Kronjaeger}, \citenamefont {Silva}, \citenamefont
  {Ilieva}, \citenamefont {Paterson}, \citenamefont {Baynham}, \citenamefont
  {Bowden}, \citenamefont {Hill}, \citenamefont {Hobson}, \citenamefont
  {Vianello} \emph {et~al.}}]{schioppo2022comparing}%
  \BibitemOpen
  \bibfield  {author} {\bibinfo {author} {\bibfnamefont {M.}~\bibnamefont
  {Schioppo}}, \bibinfo {author} {\bibfnamefont {J.}~\bibnamefont
  {Kronjaeger}}, \bibinfo {author} {\bibfnamefont {A.}~\bibnamefont {Silva}},
  \bibinfo {author} {\bibfnamefont {R.}~\bibnamefont {Ilieva}}, \bibinfo
  {author} {\bibfnamefont {J.}~\bibnamefont {Paterson}}, \bibinfo {author}
  {\bibfnamefont {C.}~\bibnamefont {Baynham}}, \bibinfo {author} {\bibfnamefont
  {W.}~\bibnamefont {Bowden}}, \bibinfo {author} {\bibfnamefont
  {I.}~\bibnamefont {Hill}}, \bibinfo {author} {\bibfnamefont {R.}~\bibnamefont
  {Hobson}}, \bibinfo {author} {\bibfnamefont {A.}~\bibnamefont {Vianello}},
  \emph {et~al.},\ }\bibfield  {title} {\bibinfo {title} {Comparing ultrastable
  lasers at 7$\times$ 10- 17 fractional frequency instability through a 2220 km
  optical fibre network},\ }\href@noop {} {\bibfield  {journal} {\bibinfo
  {journal} {Nature communications}\ }\textbf {\bibinfo {volume} {13}},\
  \bibinfo {pages} {212} (\bibinfo {year} {2022})}\BibitemShut {NoStop}%
\bibitem [{\citenamefont {Roberts}\ \emph {et~al.}(2024)\citenamefont
  {Roberts}, \citenamefont {Filzinger}, \citenamefont {Caddell}, \citenamefont
  {Jani}, \citenamefont {Steinel}, \citenamefont {Giani},\ and\ \citenamefont
  {Huntemann}}]{roberts2024ultralight}%
  \BibitemOpen
  \bibfield  {author} {\bibinfo {author} {\bibfnamefont {B.}~\bibnamefont
  {Roberts}}, \bibinfo {author} {\bibfnamefont {M.}~\bibnamefont {Filzinger}},
  \bibinfo {author} {\bibfnamefont {A.}~\bibnamefont {Caddell}}, \bibinfo
  {author} {\bibfnamefont {D.}~\bibnamefont {Jani}}, \bibinfo {author}
  {\bibfnamefont {M.}~\bibnamefont {Steinel}}, \bibinfo {author} {\bibfnamefont
  {L.}~\bibnamefont {Giani}},\ and\ \bibinfo {author} {\bibfnamefont
  {N.}~\bibnamefont {Huntemann}},\ }\bibfield  {title} {\bibinfo {title}
  {Ultralight dark matter search with space-time separated atomic clocks and
  cavities},\ }\href@noop {} {\bibfield  {journal} {\bibinfo  {journal}
  {Bulletin of the American Physical Society}\ } (\bibinfo {year}
  {2024})}\BibitemShut {NoStop}%
\bibitem [{\citenamefont {Kopeikin}\ and\ \citenamefont
  {Kaplan}(2024)}]{PhysRevD.110.084047}%
  \BibitemOpen
  \bibfield  {author} {\bibinfo {author} {\bibfnamefont {S.~M.}\ \bibnamefont
  {Kopeikin}}\ and\ \bibinfo {author} {\bibfnamefont {G.~H.}\ \bibnamefont
  {Kaplan}},\ }\bibfield  {title} {\bibinfo {title} {Lunar time in general
  relativity},\ }\href {https://doi.org/10.1103/PhysRevD.110.084047} {\bibfield
   {journal} {\bibinfo  {journal} {Phys. Rev. D}\ }\textbf {\bibinfo {volume}
  {110}},\ \bibinfo {pages} {084047} (\bibinfo {year} {2024})}\BibitemShut
  {NoStop}%
\bibitem [{\citenamefont {Ashby}\ and\ \citenamefont
  {Patla}(2024)}]{ashby2024relativistic}%
  \BibitemOpen
  \bibfield  {author} {\bibinfo {author} {\bibfnamefont {N.}~\bibnamefont
  {Ashby}}\ and\ \bibinfo {author} {\bibfnamefont {B.~R.}\ \bibnamefont
  {Patla}},\ }\bibfield  {title} {\bibinfo {title} {A relativistic framework to
  estimate clock rates on the moon},\ }\href@noop {} {\bibfield  {journal}
  {\bibinfo  {journal} {The Astronomical Journal}\ }\textbf {\bibinfo {volume}
  {168}},\ \bibinfo {pages} {112} (\bibinfo {year} {2024})}\BibitemShut
  {NoStop}%
\end{thebibliography}
%

\end{document}